\documentclass[]{jfm}
\usepackage[]{graphicx}
\usepackage{epstopdf, epsfig}
\usepackage{textcomp}
\usepackage[usenames,dvipsnames]{color} 
\usepackage{physics}
\usepackage[version=3]{mhchem}
\usepackage{multirow}
\usepackage[normalem]{ulem}
\useunder{\uline}{\ul}{}
\usepackage{rotating}
\usepackage{float}
\usepackage{setspace}
\usepackage[font=small,skip=0pt]{caption}
\usepackage{caption}
\usepackage[labelformat=simple]{subcaption}

\shorttitle{RMI of a Thermal, Isotope, and Species Interface in a two-fluid plasma}
\shortauthor{K. C. Tapinou, V. Wheatley, D. Bond, and Ingo Jahn}

\title{The Richtmyer-Meshkov Instability of Thermal, Isotope, and Species Interfaces in a five-moment multi-fluid plasma}

\author{K. C. Tapinou\aff{1}
		\corresp{\email{kyriakos.tapinou@uq.net.au}},
		V. Wheatley\aff{1},
		D. Bond\aff{1}, 
		\and Ingo Jahn\aff{1}
		}

\affiliation{\aff{1}Centre for Hypersonics, School of Mechanical and Mining Engineering, The University of Queensland,St Lucia Qld 4072, Australia} 

\begin{document}
\maketitle
\begin{abstract}
The Richtmyer-Meshkov instability (RMI) results from the impulsive acceleration of a density interface where either it or the acceleration is perturbed.
Density interfaces may arise due to a change in gas species, isotope, temperature or a combination of these. We computationally investigate the effect of interface type on the plasma RMI, which is relevant for a range of applications, including inertial confinement fusion. We simulate the evolution of single-mode perturbed thermal, species and isotope interfaces in an ideal ion-electron plasma using the multi-fluid plasma (MFP) model. 

We find that in the MFP model, the evolution of different types of interface differ significantly, in contrast to single-fluid models where they behave similarly if the Atwood number is matched. The thermal and species interfaces produce the most severe response to shock acceleration, experiencing the secondary instabilities and enhanced primary mode growth. The isotope interface evolution is restrained in comparison to the former cases, resembling the response predicted by single-fluid models. The determining factor in the severity of the MFP RMI is the density ratio across the initial interface in the electron fluid, which is unity for an isotope interface. We observe that as the density ratio across the electron interface decreases, so do the magnitudes of the self-generated fields and consequently the severity of the growth amplification. Generally, the evolution of the RMI with different types of interface becomes more similar as the level of coupling between the ion and electron fluids is increased, characterised by reducing the plasma non-dimensional skin depth.


\end{abstract}


\section{Introduction}
Research in thermonuclear fusion has been pursued for over a generation in the hopes of developing a carbon-neutral and plentiful power source that would solve several existential challenges facing humanity.
Significant progress towards power generation has been made with many milestones passed and others on the near horizon. Inertial confinement fusion (ICF) is one fusion technology which is predicted to be the first to achieve the next milestone, ``ignition''. During ignition, self sustaining fusion reactions consume all inertially confined fuel in the target, this has never been achieved. ICF experiments have made significant progress towards this goal of ignition, notably achieving a ``burning'' plasmas \citep{zylstra2022burning} where more heat is released from fusion reactions than is input to the fuel target. There are many obstacles to achieving the ignition condition but the presence of hydrodynamic instabilities within the fuel target is a leading problem that must be overcome \citep{lindl2014review,nagel2017platform,remington2019rayleigh}.

The most advanced fusion experiments, in the context of progress towards ignition, are conducted by the National Ignitiion Facility (NIF). Experiments at the NIF operate in the region between a burning plasma, where fusion has begun but is not self sustaining, and the ignition point. The ablative shock in these experiments propagates with implosion velocities of 300-400 km/s \citep{betti2015alpha} and produces ablation front pressures in excess of 1 TPa. Peak pressures experienced in the hot-spot due to heating and spherical convergence can reach $1-10 \times10^{15}$ Pa \citep{remington2019rayleigh}. The implosion results in a final fuel assembly comprised of a hot-spot density and temperature of 30-100 g/cc and 3-7 keV (34.8-81.2 MK), respectively, and a compressed fuel-shell density and temperature of 300-1000 g/cc and 200-400 eV (2.32-4.64 MK) \citep{betti2015alpha}. Peak self generated magnetic fields are of the order of $10^3$ T \citep{srinivasan2012mechanism}, which is strong enough to affect the electron heat conduction. Any physical imperfections within the fuel assembly can act as an initiation point for the Richtmyer-Meshkov instability (RMI), the instability of an impulsively accelerated density-interface (DI) where the DI and/or flow fields are perturbed \citep{meshkov1969instability,richtmyer1960taylor}. In ICF the characteristic length scales of hydrodynamic instabilities are on the order to 1-1000 \textmu{m} \citep{remington2019rayleigh}.

The highly energetic and violent processes created during the ICF implosion make it difficult to receive telemetry on many of the parameters of interest within the fuel assembly. The problem is further complicated by the requirement for non-invasive diagnostic techniques, as not to disrupt the implosion symmetry and fusion process. The burn-averaged total areal density, neutron-averaged hot-spot ion temperature, and neutron yield are typical measurable quantities \citep{zhou2008measurable}. The areal density is a measure of the amount of compression achieved and is a fundamental ICF parameter given from the Lawson criteria. All other parameters are related to the measurable parameters, assuming that the physics of energy confinement and losses is well understood and correctly described by existing theories and simulation tools. The significant limitations on experimental diagnostics mean that researchers have turned to numerical simulations to gain insights otherwise out of reach. 

The hydrodynamically equivalent physics (HEP) program \citep{lindl2004physics} studied the hydrodynamic instabilities in preparation for the first NIF ICF ignition campaign. The study concluded from early experiments on planar, cylindrical, and hemispherical geometries that ``ICF ignition capsules remain in the linear or weakly nonlinear [growth rate] regime'' \cite{lindl2004physics}. \citet{lindl2004physics} concluded that the ablation front of the target fuel pellet is Rayleigh-Taylor (RT) unstable and interface perturbations can seed the Richtmyer-Meshkov instability (RMI) within the capsule interior during acceleration and then deceleration of non-ablated material. These instabilities, if severe enough, ``can cause ablator material to mix into the core and radiatively cool the hot spot, decreasing the hot spot temperature and nuclear yield'' \citet{park2014high}.

Despite the awareness of hydrodynamic instabilities as an obstacle to fusion experiments, their significance was underestimated in the lead up to the national ignition campaign (NIC) of 2009-2012. The instabilities were modelled using simplified analytical theory combined with empirical relations based on experimental data from previous investigations. A review of the first three years of ignition experiments concluded that ``Current evidence points to low-mode asymmetry and hydrodynamic instability as key areas of research to improve the performance of ignition experiments on the NIF and are a central focus of the Ignition Program going forward'' \citep{lindl2014review}. Since this time, considerable effort has been made to suppress the RTI instability and to a lesser extent the RMI, with a focus on the laser pulse characteristics.

The hydrodynamic instabilities, which are well understood for conventional fluids, are complicated by EM effects in the plasma which constitutes the ICF fuel target after the laser drive commences. The RMI is one of the primary instabilities that plagues ICF efforts. The literature on the plasma RMI is limited and lacks a consensus on how to appropriately model the ICF RMI. The common models for the plasma RMI are the ideal magnetohydrodynamic model (MHD) \citep{wheatley2005regular,wheatley2012magnetohydrodynamic,wheatley2013magnetohydrodynamic,mostert2015effects,mostert2017magnetohydrodynamic}, the Hall magnetohydrodynamic (HMHD) model \citep{shen2019impulse,srinivasan2012mechanism}, and the multi-fluid plasma (MFP) model \citep{bond2017richtmyer}. The motivation for using a MFP model is its superior grasp of the physics involved, compared to the ideal and Hall MHD models which are simplified in their treatment of the physics. In essence, the MFP theory captures more accurately the effects of charge separation, Lorentz forces, self generated EM fields, fluid interactions (electron fluid exciting ion fluid), and high frequency phenomena. Furthermore, it allows for investigation of the fundamental types of interfaces present in a plasma RMI which cannot be distinguished by MHD and HMHD due to their simplifications. 

The first theoretical and experimental characterisations of the RMI was by \citet{richtmyer1960taylor} and \citet{meshkov1969instability}, respectively, giving the instability its namesake. At the time of writing, the applications driving new research on the RMI are the mitigation of hydrodynamic instabilities in ICF \citep{lindl2014review,bond2017richtmyer,bond2017electron,hohenberger2012inertial,nagel2017platform,remington2019rayleigh,bender2021simulation}, and enhancing mixing in supersonic combustion \citep{yang2014richtmyer,yang1993applications}. Furthermore, as indicated by the reviews of RMI (and RTI) by \citet{zhou2017rayleigh} and \citet{brouillette2002richtmyer}, the RMI is important in many other natural and engineered formats. These formats include: many astrophysical phenomena \citep{arnett1989supernova,arnett2000role}, atmospheric sonic boom propagation \citep{davy1971measurements}, driver gas contamination in reflected shock tunnels \citep{stalker1978driver,brouillette1999experiments}, combustion wave deflagration-to-detonation transition \citep{khokhlov1999ainteraction,khokhlov1999bnumerical}, laser-material interactions including but not limited to micro fluid dynamics \citep{lugomer2007micro} and micron-scale fragment ejection \citep{buttler2012unstable}, high energy density turbulent mixing \citep{bender2021simulation}, and many more fundamental studies investigating solid-liquid and solid-solid media interactions with lasers and fluid flows. 

\citet{bond2017richtmyer} investigated the multi-fluid plasma RMI of a thermal interface for a single-mode perturbation. Simulation results showed the MFP thermal RMI (TRMI) differed significantly from the single-fluid hydrodynamic case, which in the absence of an applied magnetic field represents the same solution of an ideal MHD simulation. This investigation is the precursor to the work presented in the current paper with the same methods implemented by \citet{bond2017richtmyer} reproduced here but for differing applications. 

The key phenomena observed in the investigation were
\begin{itemize}
	\item formation of a precursor electron shock that accelerates the electron interface prior to arrival of the ion shock, 
	\item Lorentz force bulk oscillations, 
	\item electromagnetically driven RTI, 
	\item electron fluid exciting the ion-fluid interface, 
	\item and decreasing severity of MFP effects for reduced skin depth. 
\end{itemize}
The oscillation of the electron fluid about the ion interface is caused by the electron precursor shock. An EM force attempts to neutralise the plasma charge distribution, however, the electron fluid continually overshoots the distribution that would neutralise the plasma because it lacks physical dissipation. The overshoot generates EM fields, dominated by the electric field, that drive a reversal in motion and further drive the oscillations. At early times the oscillation is dominant and drives a variable acceleration RTI that alternates between stable and unstable configurations. 

The electron fluid, owing to its lesser particle mass, experiences orders of magnitude greater acceleration that produce high wave-number instabilities much sooner than the fluid comprised of the more massive ion particles. The passage of the ion shock at later times produces a region of positive charge between the shock and the interface that results in EM forces driving the electromagnetically drive Rayleigh-Taylor Instability in the unstable direction until simulation end. The RTI in the ion fluid results in growth of the low-mode ion interface perturbation to substantially enhance the interface growth. The distortion of the ion shock as it traverses the interface in conjunction with rapidly propagating electron vortices results in triple points.  

The works by \citet{bond2017richtmyer} varied the plasma Debye length ($\lambda_D$), a measure of the coupling between the ion and electron fluid. Upon refining the Debye length, many of the MFP effects are reduced. However, a result of the tighter coupling is an increase in high-wave number features on the ion interface. All cases of Debye length exhibit self-generated EM fields, the source of the electromagnetically driven RTI and Lorentz force bulk fluid oscillations. These EM fields are more intense in the smaller Debye length case (tighter coupling). \citet{bond2017richtmyer} concludes that the MFP effects have the potential to produce significant deviations from the behaviour of the plasma RMI reported by MHD models.

The methodology employed in the current study is driven by an effort to simplify the ICF RMI to a point where its effects can be understood without the superposition of other physical phenomena confusing the investigation. The ``other'' physical phenomena we refer to are the radiation transport, laser interactions, multi-phase modelling to include the dynamics of the shell, nuclear reactions, converging geometry, ablation, and the process of ionisation. The isolation of the RMI dramatically reduces the complexity of the problem, investigative tools, and analyses required for the works proposed. The removal of other physics allows the computational resources on hand to be directed to the physics of the plasma RMI. Therefore, the far more physically accurate, but also more computationally expensive, multi-fluid plasma model is used instead of the ideal and Hall MHD models to investigate the plasma RMI. 

This work studies the three fundamental types of DIs, within an ideal mutli-fluid plasma, that can host the RMI and compares them to each other. Section \ref{sec:Fundamental_interfaces} will discuss the relevance of the three interface types to the ICF problem. 
Section \ref{sec:MFP} introduces the ideal multi-fluid plasma model and section \ref{sec:NumericalSolver} outlines the numerical tool and non-dimensionalisation used for implementation of the model. Section \ref{sec:ProblemDescription} specifies the exact scenarios simulated in this study. The code is validated for the hydronyamic RMI in section \ref{sec:HRMIvalid} Some key phenomena which occur in the two-dimensional (2D) multi-fluid plasma RMI base flows will be isolated in the discussion of the analogous one-dimensional (1D) base flows explored in \ref{sec:1D_base_flow}. A comparison of the 2D planar RMI interface types will be made in section \ref{sec:comparison_interfaces} and with a discussion of the key parameters influencing the RMI evolution in section \ref{sec:parameter_dicsussion} and some comparisons to the hydrodynamic RMI in section \ref{sec:qualitativePlasmaHydroRMI}. A conclusion of the works completed is presented in section \ref{sec:Conclusion}. 

\section{Problem definition and methodology}

\subsection{Fundamental interfaces}\label{sec:Fundamental_interfaces}
\begin{figure}
	\centering
	\includegraphics[width=\textwidth]{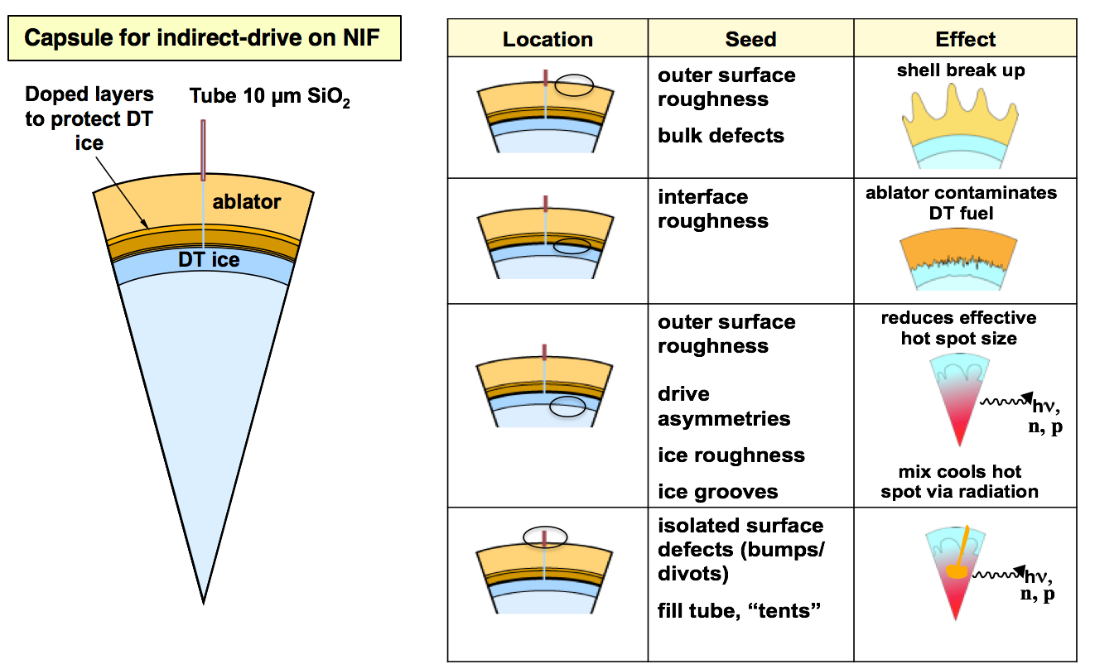}
	\caption{Sources of hydrodynamic instabilities for cryogenic ICF targets of deuterium-tritium (DT) fuel, reproduced with permission from \citet{smalyuk2017mix}.}
	\label{fig:hydrodynamic_instability_seeds} 
\end{figure}

The ICF fuel capsule is a complicated engineered assembly of ablator, dopant, and fuel layers that provide many seeds for hydrodynamic instabilities and general asymmetries. Figure \ref{fig:hydrodynamic_instability_seeds} shows the key types of perturbations that may seed the RMI and RTI within the ICF fuel capsule. Native interface perturbations and bulk defects in the capsule material are consequences of the processing techniques used for fabrication e.g. the surface roughness due to inhomogeneous solidification of the deuterium-tritium (DT) ice layer. Energy coupling non-uniformities and asymmetries are caused by unequal irradiation (indirect or direct drive) of the fuel target. Intentional engineering features, for example the fill tube used in constructing the fuel target that remains embedded, are at significant risk of instability growth initiation. The varying sources of seeds can each excite instabilities at different locations within the fuel capsule. The situation is further complicated by the ionisation of the fuel capsule constituent materials during implosion as this process augments the perturbations. 

\begin{figure}
	\centering
	\includegraphics[width=\textwidth]{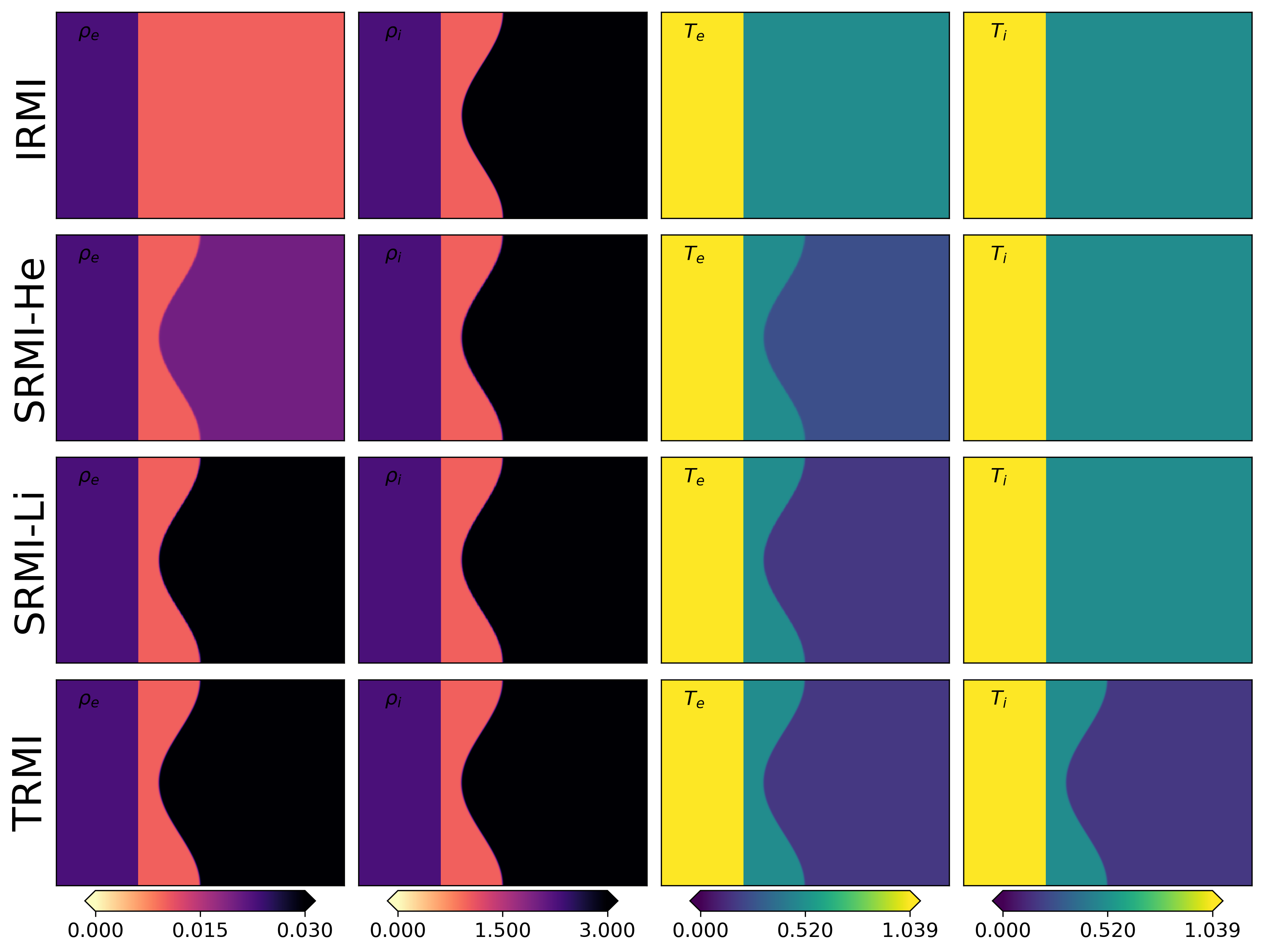}
	\caption{Illustration of the initial conditions for the four interface conditions investigated. Columns form left to right show the electron mass density, ion mass density, electron temperature, and ion temperature. }
	\label{fig:initial_conditions} 
\end{figure}

The fundamental types of interface that have unique responses to the RMI are the: (1) thermal, (2) species, and (3) isotope interfaces. Figure \ref{fig:initial_conditions} shows the non-dimensional initial conditions for all interface types studied. To elucidate the effect of each interface type we enforce mechanical equilibrium (pressure through $P=nkT$ and charge neutrality ($n_eq_e + n_iq_i$) across the interface. The (1) thermal interface possesses a \textit{discontinuity in thermal energy} within both the ion and electron fluids, producing a \textit{DI in both fluids}. The thermal RMI (TRMI) may occur at the solid-gas fuel interface, shown in figure \ref{fig:hydrodynamic_instability_seeds}, seeded by native surface roughness on the solid-fuel. The (2) species interface possesses a is \textit{discontinuity in ion particle mass and ion charge}, also producing a \textit{DI in both fluids}. The species RMI (SRMI)  exists at the interfaces formed by layers of ablator-dopant and dopant-solid fuel and can be especially detrimental to ignition by introduction of non-fusible material into the hot-spot. Finally, the (3) isotope interface possesses a \textit{discontinuity in the ion particle mass only}, i.e. different isotopes of an element, that results in a \textit{DI in only the ion fluid} since the isotopes on each side of the interface have the same number of electrons in their shells. An interface discontinuity in charge density and thermal energy, and whether these discontinuities exist in both or one of the ion and electron fluids, will produce unique responses to the RMI, despite having the same mass density ratio. The isotope RMI (IRMI) does not have an immediately obvious place in an ideal ICF fuel target, however, it provides insight into targets with inhomogeneous fuel layers, intended or defective, as well as providing completeness to elucidate key influences in 1) and 2). 

\subsection{The multi-fluid plasma model}\label{sec:MFP}
This study is conducted using the ideal multi-fluid plasma model, modelling the plasma as comprised of only the ion and electron fluids, neglecting any neutral  species. It is appropriate to use a fluid modelling approach for the ICF plasma because of the high density and temperatures at which the plasma exists \citep{bellan2008FundPlasma,chen2016introduction}. It is appropriate to apply the fluid description of plasma when the hydrodynamical time scale ($\tau_H$) is much slower than the thermal relaxation time scale ($\tau_e$, $\tau_i$, and $\tau_{eq}$) of all charged particles comprising the plasma \citep{goedbloed2004principles}. Furthermore, for ideal MFP to be valid we require the dissipative diffusion time scale ($\tau_D$) to be greater than the hydrodynamical time scale. For a typical ICF plasma, the characteristic time scales are $\tau_e \approx 10^{-17}<<\tau_i \approx 10^{-15} << \tau_{eq} \approx 10^{-14} << \tau_H \approx 10^{-11} << \tau_D \approx 10^{-7}$, and therefore the use of an ideal MFP model is satisfied. We note that, \textit{while not the case in this work}, the presence of an exceptionally strong magnetic field produces a perpendicular diffusion time scale that is of the same magnitude as the hydrodynamical timescale increasing the error in fluid description.

In the ideal MFP model, the fluid equations are derived by taking moments of the Vlasov equation, where the result ends as the familiar Euler equations with EM forcing terms. A set of fluid equations,  conservation of mass; momentum; and energy, exist for each fluid modelled. In our case we have one ion fluid and one electron fluid, yielding two sets of fluid equations. The following assumptions for the plasmas investigated are used unless otherwise noted. The ions and electrons are each in thermal equilibrium with themselves. The velocity distribution for each fluid is Maxwellian and this distribution gives an accurate representation of currents within each species. The plasmas, as per ideal MFP theory, neglect viscous and resistive terms in the momentum equation and conductive heat transfer in the energy equation. This reduction can be appropriate when the time scale of diffusive terms is much larger than that of the advective terms \cite{goedbloed2004principles}, which is generally the case in our problem of interest. Furthermore, the inclusion of these terms (that are often significant in effect) may obscure the ideal MFP phenomena that are the focus of the study. The thermal relaxation (resultant from collisions between species) are neglected for this reason, despite the expectation that they will produce a significant heat exchange between species.

We follow the non-dimensionalised ideal MFP model formulation of \citet{bond2017richtmyer} based on the work by \citet{loverich2003finite}. The non-dimensionalisation reduces the variance of the magnitudes of variables which must be resolved. This procedure is important when the disparate variables must be solved together in a single system. The chosen non-dimensionalisation results in a less stiff numerical system and conveniently allows the plasma regime to be set by only two parameters, the skin depth, $d_S$, and the plasma ratio of thermodynamic and magnetic pressure, $\beta$. The non-dimensionalisation is shown below, note the ``hat'' symbol and subscript zero indicate a non-dimensional and reference parameter, respectively. 
\begin{align}
\hat{n} & =\frac{n}{n_0} & \hat{m} & =\frac{m}{m_0} & \hat{q} & =  \frac{q}{q_0}  & \hat{\rho} &= \frac{\rho}{\rho_0} \nonumber \\
\hat{\mathbf{u}} & = \frac{\mathbf{u}}{u_0} & \hat{p} & = \frac{p}{n_0m_0u_0^2}& \hat{\varepsilon} & =\frac{\varepsilon}{n_0m_0u_0^2}  &  \hat{x} &= \frac{x}{x_0}\nonumber \\
\hat{c} & = \frac{c}{u_0} & \hat{t} & = \frac{t}{t_0} & \hat{\mathbf{B}} & =\frac{\mathbf{B}}{B_0}  &  \hat{\mathbf{E}} &= \frac{\mathbf{E}}{cB_0} \nonumber \\
\hat{\psi_E} & = \frac{\psi_E}{B_0} & \hat{\psi_B} & =\frac{\psi_B}{cB_0}  &  \hat{d_S} &= \frac{d_S}{x_0} \nonumber 
\end{align}

Where $n$ is the number density, $m$ is the particle mass, $q$ is the charge, $\rho$ is mass density, $\mathbf{u}$ is the velocity vector, $p$ is thermodynamic pressure, $\varepsilon$ is the thermal and kinetic specific energy, $x$ is a length, $c$ is the speed of light, $t$ is time, $\mathbf{B}$ is the magnetic field vector, $\mathbf{E}$ is the electric field vector, $\psi_E$ and $\psi_B$ represent the wave speed of the divergence cleaning. Further variables to be used temperature $T$, Boltzmann's constant $k_B$, atomic number of a species $Z$, ratio of specific heats $\gamma$, vacuum permittivity $\epsilon_0$, the permeability of free space $\mu_0$, the hydrodynamic Mach number of the propagating shock $M$. Note that the subscript $\alpha$ indicates some unique fluid i.e. ions or electrons.

\begin{table}
	\begin{center}
		\begin{tabular}{lccccccc}
			\cline{2-7}			
			\textbf{}                                         & \multicolumn{1}{l}{\textbf{\begin{tabular}[c]{c@{}c@{}} $T$ \\   $(K)$\end{tabular}}} & \multicolumn{1}{c}{\textbf{\begin{tabular}[c]{c@{}c@{}}$\rho$\\$(kg/m^3)$\end{tabular}}} & \multicolumn{1}{c}{\textbf{\begin{tabular}[c]{c@{}c@{}}$d_S$\\$(m)$\end{tabular}}} & \multicolumn{1}{c}{\textbf{\begin{tabular}[c]{c@{}c@{}}$v_{th,e}$\\$(m/s)$\end{tabular}}} & \multicolumn{1}{c}{\textbf{\begin{tabular}[c]{c@{}c@{}}$n$\\$(m^{-3})$\end{tabular}}} & \multicolumn{1}{c}{\textbf{\begin{tabular}[c]{c@{}c@{}}$U_{shock}$\\$(m/s)$\end{tabular}}} \\ \hline
			\multicolumn{1}{l}{\textbf{ICF conditions}}     & $10^6$     & $10^5$          & $10^{-9}$        & $10^7$              & $10^{31}$      & $10^5$               \\ \hline
			\multicolumn{1}{l}{\textbf{\begin{tabular}[c]{@{}l@{}}Current simulation, \\ $\hat{d}_S=0.1$, $\hat{c}=50$, \\$\hat{m}_i=100\hat{m}_e$\end{tabular}}} & $10^7$     & $10^2$          & $10^{-7}$      & $10^7$              & $10^{29}$      & $10^6$               \\ \hline
		\end{tabular}
		\caption{Comparison of typical ICF parameter orders of magnitude \citep{betti2015alpha} and those simulated in the present work. }
		\label{table:simulation_parameters}
	\end{center}
\end{table}

The choice of reference values for each property is problem specific. The default choices used in the works completed and proposed are as given next unless otherwise stated. The reference length $x_0$ is set to the typical wavelength of the RMI perturbation of about 1 \textmu{m}, reference charge is set to the elementary charge, reference velocity is set to the electron thermal velocity, and the reference number density, and reference mass are set to achieve the required $d_S$ and $\beta$. The reference magnetic field strength is set according to $\beta$ (equation \ref{eq:beta}), reference density is set to the product of reference number density and mass, reference time is set according to the ratio of reference length and reference velocity, and reference mass is set to a hundred times an electron mass --- an artificially large electron mass relaxes the problem stiffness allowing for more tractable solutions. A comparison of typical ICF conditions with the current simulation properties is given in table \ref{table:simulation_parameters}. The current conditions do not exactly mirror those of ICF but are close enough to give insight into the physical phenomena occurring. \textit{It is vital to understand that the purpose of these works is to provide detailed qualitative insight on the influence of the RMI and secondary instabilities which affect ICF efforts, not the exact quantitative extent of which these phenomena occur.}

The non-dimensional form of the fluid equations, with the hat symbol dropped for brevity, are given in equations \ref{eq:fluid1}, \ref{eq:fluid2}, and \ref{eq:fluid3}. The fluid equations are recognisable as the inviscid Euler equations with EM source terms included (gravity is negligible) and exist for each fluid. 
\begin{align}
\frac{\partial \rho_\alpha }{\partial t} + \div{\rho_{\alpha}\mathbf{u}_\alpha} &=0\label{eq:fluid1}\\
\frac{\partial \rho_\alpha \mathbf{u}_\alpha}{\partial t} + \div{\left(\rho_{\alpha}\mathbf{u}_\alpha\mathbf{u}_\alpha + p_\alpha \mathbf{I}\right)} &=\sqrt{\frac{2}{\beta_0}}\frac{n_\alpha q_\alpha}{d_S}\left(c\mathbf{E} + \mathbf{u}_\alpha\cross\mathbf{B}\right)\label{eq:fluid2}\\
\frac{\partial \varepsilon_\alpha }{\partial t} +\div{\left( \left( \varepsilon_\alpha +p_\alpha \right)\mathbf{u}_\alpha\right)} &= \sqrt{\frac{2}{\beta_0}}\frac{n_\alpha q_\alpha c}{d_S}\mathbf{E} \vdot \mathbf{u}_\alpha \label{eq:fluid3}
\end{align}
Here the subscript $\alpha \in (i,e)$ represents the species modelled. We define mass density, pressure, and energy density by
\begin{align}
\rho_{\alpha} = n_\alpha m_\alpha, \quad p_\alpha = n_\alpha k_B T_\alpha, \quad \textrm{and} \quad \varepsilon_\alpha = \frac{p_\alpha}{\gamma - 1} + \frac{\rho_\alpha|\mathbf{u}_\alpha|^2}{2}.
\end{align}

Maxwell's equations govern the evolution of the EM fields and are given in non-dimensional form
\begin{align}
\frac{\partial \mathbf{B}}{\partial t} + c\curl{\mathbf{E}} &= 0, \label{eq:farady}\\
\frac{\partial \mathbf{E}}{\partial t} - c\curl{\mathbf{B}} &= -\frac{c}{d_S}\sqrt[]{\frac{\beta_0}{2}}\sum_\alpha n_\alpha q_\alpha \mathbf{u}_\alpha,\label{eq:ampere} \\
c\div{\mathbf{E}} &= \frac{c^2}{d_S}\sqrt{\frac{\beta_0}{2}}\sum_\alpha n_\alpha q_\alpha, \label{eq:divE}\\
\text{and} \quad \div{\mathbf{B}} &= 0. \label{eq:divB}
\end{align}
Discretisation of the EM divergence constraints, equations \ref{eq:divE} and \ref{eq:divB}, introduces error, that is mitigated by implementation of the perfectly hyperbolic form of Maxwell's equations, as per \citet{munz2000three}, resulting in the new Maxwell's equations, 
\begin{align}
\frac{\partial \mathbf{B}}{\partial t} + c\curl{\mathbf{E}} + c\Gamma_B\grad{\psi_B}&= 0, \\
\frac{\partial \mathbf{E}}{\partial t} - c\curl{\mathbf{B}} +c\Gamma_E\grad{\psi_E} &= -\frac{c}{d_S}\sqrt[]{\frac{\beta_0}{2}}\sum_\alpha n_\alpha q_\alpha \mathbf{u}_\alpha,\label{eq:AmpereMaxwell} \\
\frac{\partial\psi_E}{\partial t} + c\div{\mathbf{E}} &= \frac{c^2\Gamma_E}{d_S}\sqrt{\frac{\beta_0}{2}}\sum_\alpha n_\alpha q_\alpha, \\
\text{and} \quad \frac{\partial \psi_B}{\partial t} + c\Gamma_B\div{\mathbf{B}} &= 0. 
\end{align}

We define the reference values of skin depth and $\beta$ ratio as 
\begin{align}
d_{S,0} = \frac{1}{q_0}\sqrt{\frac{m_0}{\mu_0 n_0}} \quad \textrm{and} \quad 
\beta_0 = \frac{2\mu_0 n_0 m_0 u_0^2}{B_0 ^2}\label{eq:beta}
\end{align}
The Lorentz force acting on a single particle is given by $\mathcal{L}=q_\alpha\left(\mathbf{E} + \mathbf{u}\cross \mathbf{B}\right)$, and the X and Y components (important for the 2D simulation results) are given by
\begin{align}
\mathcal{L}_x=q_\alpha\left(E_x + u_yB_z - u_zB_y\right) \quad \textrm{and} \quad \mathcal{L}_y=q_\alpha\left(E_y + u_xB_z - u_zB_x\right).
\end{align}


\subsection{Numerical solver}\label{sec:NumericalSolver}
\subsubsection{Description}
The numerical implementation used for this work was developed by \citet{bond2017richtmyer}. A finite volume method is used to allow for a development of a three-dimension modelling capability and to leverage existing knowledge in computational fluid dynamics of EM system, such as  \citet{loverich2003finite,abgrall2014robust}. The implementation uses a two-dimensional three-vector architecture, therefore no x- and y-magnetic field components are generated by the fluid motions in simulations, however, there is generation of the z-component of magnetic field. The solution requires a very high degree of spatial and temporal refinement due to the wide range of length scales and advective speeds associated with the plasma regimes modelled. The spatial refinement is satisfied by implementing the system of equations in the adaptive mesh frame work, AMReX \citep{zhang2019amrex}. The trigger for the cell refinement is the relative ion and electron density gradient,
\begin{align}
\frac{|\grad{\rho_\alpha}|\Delta x}{\rho_{\alpha}} > \frac{5}{100}. 
\end{align} 
We implement a general consideration for effective spatial resolution such that the smallest length scale is refined by at least two cells, $\delta x \leq \frac{\lambda_{min}}{2}$, though this is only enforced when the normalised  charge  density  exceeded  the  small  threshold  of  5/1000, $n_\alpha q_\alpha > \frac{5}{1000}$ (an emphasis is placed on high resolution in areas of EM features). The temporal refinement is satisfied in some part by manipulating the resolved value of speed of light in the non-dimensionalised  parameter space. The manipulation is in choosing a large value for the reference velocity so the non-dimensional speed of light, $\hat{c} = c/u_0$, is smaller, thereby, lessening the temporal refinement required by the CFL condition than what is typical in most plasmas. Care must be taken to ensure that the non-dimensional speed of light is still the greatest characteristic speed in the system, otherwise non-physical behaviour due to interaction of fluid and EM waves may occur. 

As discussed by \citet{loverich2003finite}, the homogeneous part of each of the three distinct systems of equations (ion fluid, electron fluid, and electromagnetic fields) can be solved independently because the systems are only coupled through the source terms. This allows the option, which is exercised by \citet{bond2017richtmyer}, to solve each system using its own Riemann solver. The source terms are then solved using an implicit solver. 
%
%
%
The time integration is done by a two-stage second-order accurate Runge-Kutta scheme. Piece-wise linear spatial reconstruction is used for both cell centred and face centred variables with limiting by a min-mod limiter in the primitive variables. As per the discussion above, the homogeneous solution of the fluid and EM system of equations is solved using HLLC \citep{einfeldt1991godunov,toro1994restoration} and HLLE \citep{harten1983upstream,einfeldt1988godunov} approximate Riemann solvers. The source terms are solved locally using an implicit method as per \citet{abgrall2014robust}. 

There exists no analytical test case and solution for the two-fluid model. There are, however, numerous plasma solvers that have been produced and documented in the literature, allowing for comparison of numerical experiments. The standard test for multi-fluid plasma solvers is the two-fluid electromagnetic plasma shock which is an extension of the Brio and Wu shock tube \citep{brio1988upwind}. Our numerical solver has been verified by comparing the solution of the EM plasma shock to \citet{loverich2003finite}. 

Verification of sufficient solution convergence is completed by comparing flow statistics of varying effective resolutions and performing a Richardson extrapolation to estimate convergence error. A two-dimensional refinement study was conducted in the preceding work by \citet{bond2017richtmyer}. Flow statistics are used instead of point-wise convergence because the latter is not possible without physical dissipation being modelled to set a minimum physical length scale. The chosen flow statistic is the ion interface width. The average L2 norm calculated for the relative error of the fine grid solution to the Richardson extrapolated solution showed that the calculated error for relevant length scales, $d_D=0.1$ and $d_D=0.01$) was significantly below 1 \% when using effective resolutions of 2048 and 4096, respectively. It was observed that these resolutions correspond to those required to resolve the high frequency plasma wave packets in the 1D base-flow. This is sufficient to capture the bulk flow characteristics of the interface evolution. One of the characteristic Debye lengths in the present study, $d_D=0.002$, is not covered by these results, but the 1D convergence study in section \ref{sec:ConvergenceStudy} indicates that the high frequency waves generated are well resolved. This meets the criteria for resolving the interface statistics from \citep{bond2019richtmyer} and gives the authors confidence that bulk flow characteristics are captured in 2D at these resolutions. Effective resolutions of 2048, 4096, and 4096 are used for the Debye lengths of 0.2, 0.02, and 0.002.

\subsubsection{Hydrodynamic RMI Validation}\label{sec:HRMIvalid}
\begin{figure}
	\centerline{\includegraphics[width=0.75\linewidth]{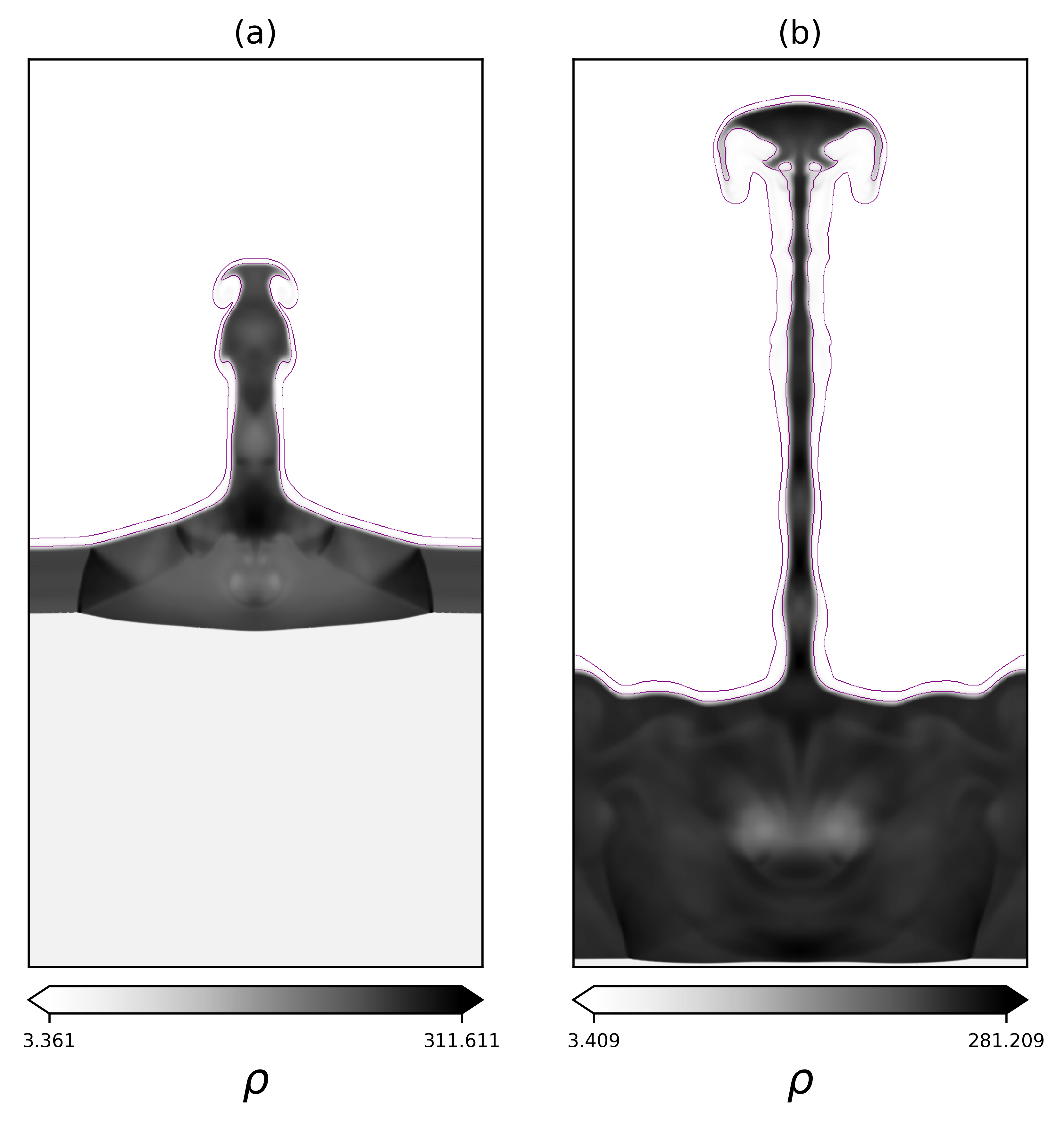}}
	\caption{Cerberus simulation RMI interface mass-density contours at dimensional times 0.67 ms and 1.77 ms for experimental conditions \citet{motl2009experimental} for the case 8 of a $M=1.95$ shock.}
	\label{fig:motl_fluid_case_8_simulation}
\end{figure}

\begin{figure}
	\centerline{\includegraphics[width=\linewidth]{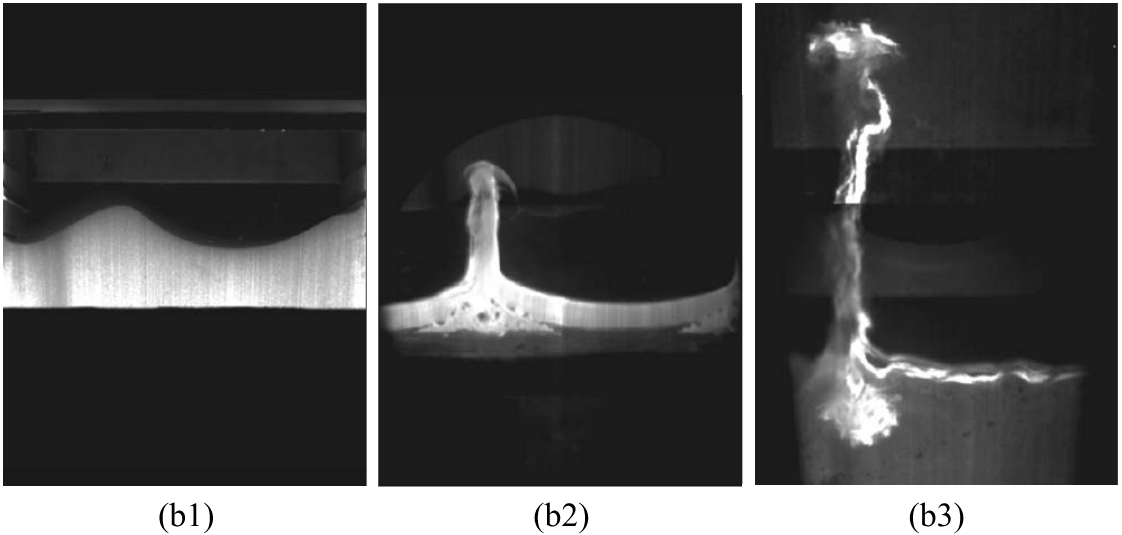}}
	\caption{Time-sequence of experimental images for He/$\textrm{SF}_6$, M=1.95 case (b1) initial condition $t\approx0.00$, (b2) $t\approx0.67$ ms, and (b3) $t\approx1.77$ ms. Reproduced from Motl, B., Oakley, J., Ranjan, D., Weber, C., Anderson, M. and Bonazza, R., 2009. Experimental validation of a Richtmyer–Meshkov scaling law over large density ratio and shock strength ranges. Physics of Fluids, 21(12), p.126102., with the permission of AIP Publishing. 
	}
	\label{fig:motl_fluid_case_8_experiment}
\end{figure}

\begin{figure}
	\centerline{\includegraphics[width=0.5\linewidth]{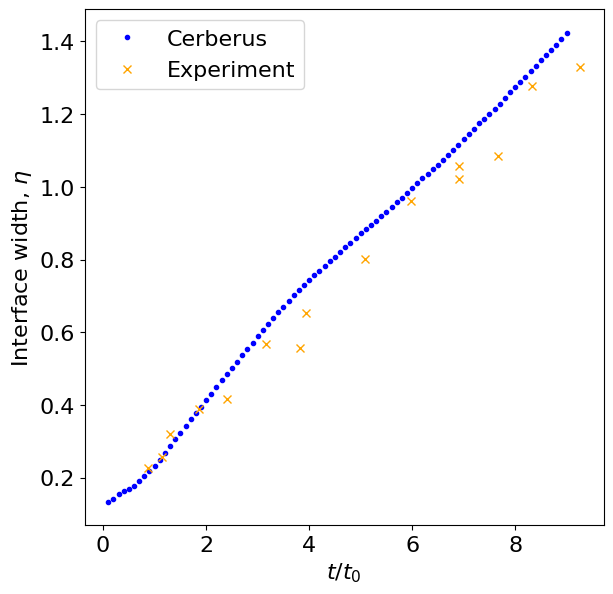}}
	\caption{Comparison of the interface width from \citet{motl2009experimental} case 8 and Cerberus simulation results for a $M=1.95$ shock in a He/$SF_6$ interface. }
	\label{fig:motl_eta_case_8}
\end{figure}

The experiments of \citet{motl2009experimental} were simulated with Cerberus, run hydrodynamically (gas is uncharged), to verify that the combination of numerical methods implemented is able to adequately capture the dynamics of the RMI. 
Good qualitative and quantitative agreement was found. 
The experimental conditions of the scenario (case 8 in the work by \citet{motl2009experimental}) simulated were an interface of helium and sulphur hexa-fluoride gasses at temperature $295$ K and pressure $98$ kPa, initial perturbation amplitude $1.36$ cm, perturbation wavelength $16.74$ cm, and Atwood number $\mathcal{A}=0.95$. 
Figure \ref{fig:motl_fluid_case_8_simulation} and \ref{fig:motl_fluid_case_8_experiment} show our simulation and experimental result by \citet{motl2009experimental}, respectively. 
The simulation reproduces all the qualitative features from the experiment, where both figures show a mushroom has developed at the head of the spike and behind the head, KHI rollers develop behind the interface, and a reverse jet is formed at the base of the spike. 
Figure \ref{fig:motl_eta_case_8} shows the comparison of the interface width, $\eta$, for the experiment and simulation, which show good agreement. 
There are two differences to note between simulation and experimental conditions (1) the vertical shock tube arrangement employed by \citet{motl2009experimental} and (2) and the lack of viscosity in the ideal hydrodynamic simulation. 
The ideal simulations did not model the gravitational force or viscosity which both retard the growth of the instability. 
These two differences account for the faster growth rate in the simulations. 
The comparison to the experiment shows Cerberus is able to accurately model the hydrodynamic RMI and secondary instabilities in a hydrodynamic fluid.

The scenarios from \citet{dell2017maximum} were simulated with Cerberus and compared to the simulation results and empirical relation by \citet{dell2017maximum}. 
\citet{dell2017maximum} uses a smoothed particle hydrodynamic numerical scheme \citep{stellingwerf1991smooth} and derives an empirical relation for the non-linear and linear initial growth rates. 
Our simulations reproduced the conditions from \citet{dell2017maximum} with shock $M=3$ and Atwood number $\mathcal{A}=0.8$. 
The initial perturbation wavelength was $\lambda=3.3\times 10^{-3}$ m, light and heavy densities were $\rho_l = 1\times 10^{-3}$ $kg/m^3$ and $\rho_h = 9\times 10^{-3}$ $kg/m^3$, respectively. 
The simulations were run with reference parameters of $x_0=3.3\times 10^{-3}$ m, $P_0=2.494\times 10^{6}$ Pa, $T_0 = 300$ K, and reference mass equal to mass of the light fluid particle. 
The particles in the light and heavy regions were assigned non-dimensional masses of $1$ and $9$, respectively, and both were ideal monoatomic gases with $\gamma = 5/3$. 
The domain length was $40$ in x and $2$ in y with zero-gradient and periodic boundary conditions in x and y, respectively.

The ratios of amplitude to wavelength, $a_0/\lambda$, simulated were $\{0.06$, $0.25$, $0.3$, $0.4$, $0.5$, $1.0\}$ all with the conditions specified above. 
The maximum grid refinement in each case (for the adaptive mesh refinement used by Cerberus) was set to ensure the interface amplitude was well resolved ($\Delta x \approx a_0/100$) 
, likewise, the time-step constraint was enforced (more restrictive than the CFL condition) such that the interaction of the initiating shock wave and interface was accurately resolved ($dt\approx a_0/10u_{shock}$). 
The initial growth rate was calculated as prescribed by \citet{dell2017maximum}, as the slope of the linear regression of interface amplitude data from the time the reflected and transmitted shocks are emitted from the bubble, to the time $0.3\tau$, where $\tau=\lambda/v_\infty$.

\begin{figure}
	\centerline{\includegraphics[width=0.75\linewidth]{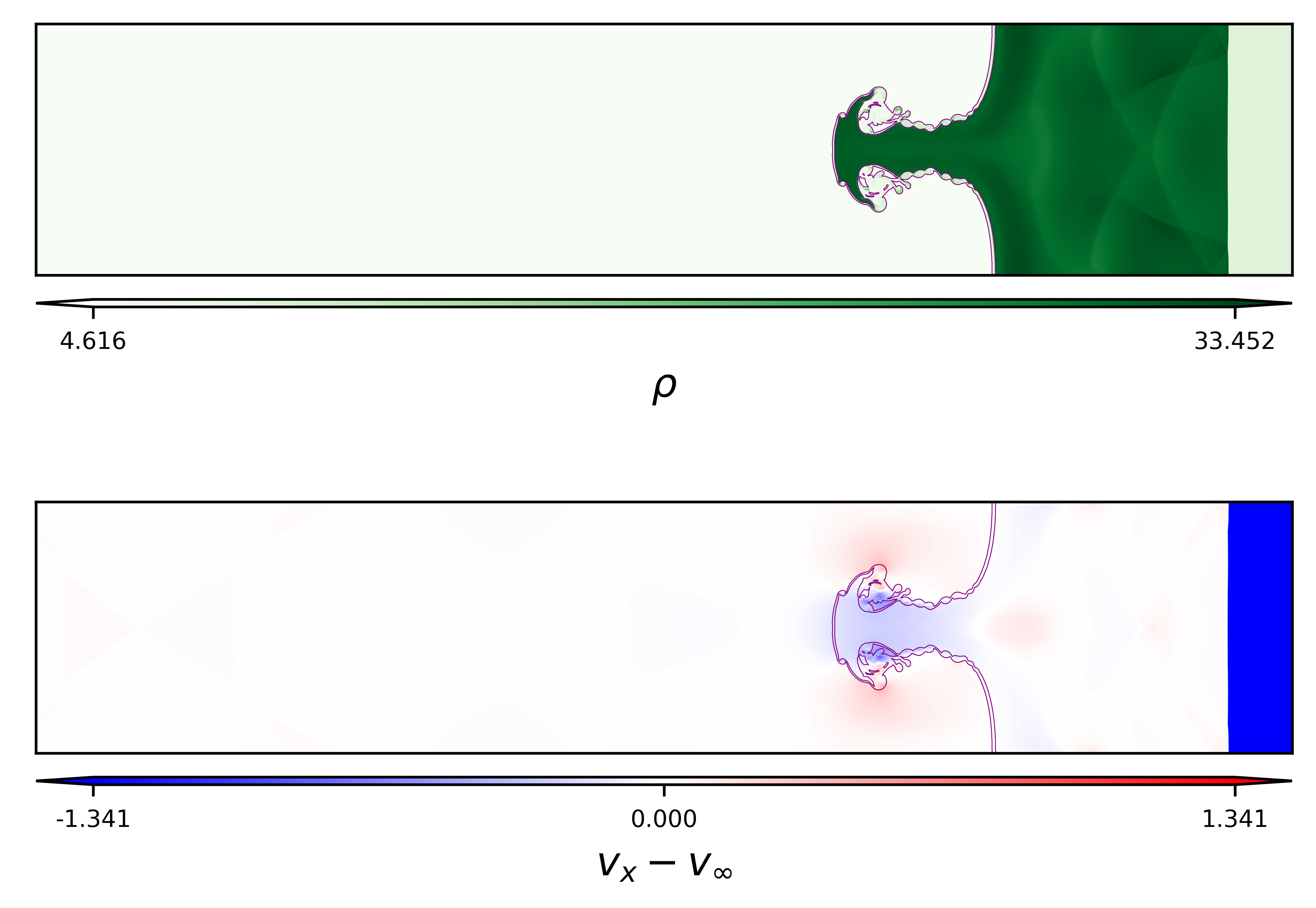}}
	\caption{The mass-density (upper) and x-velocity (lower), offset by the background motion $v_\infty$, for the $M=3$, $\mathcal{A}=0.8$, and $a_0/\lambda=0.06$ case. A contour line of the density-interface is shown in purple. }
	\label{fig:HRMI_aol_0p06}
\end{figure}
\begin{figure}
	\centerline{\includegraphics[width=0.75\linewidth]{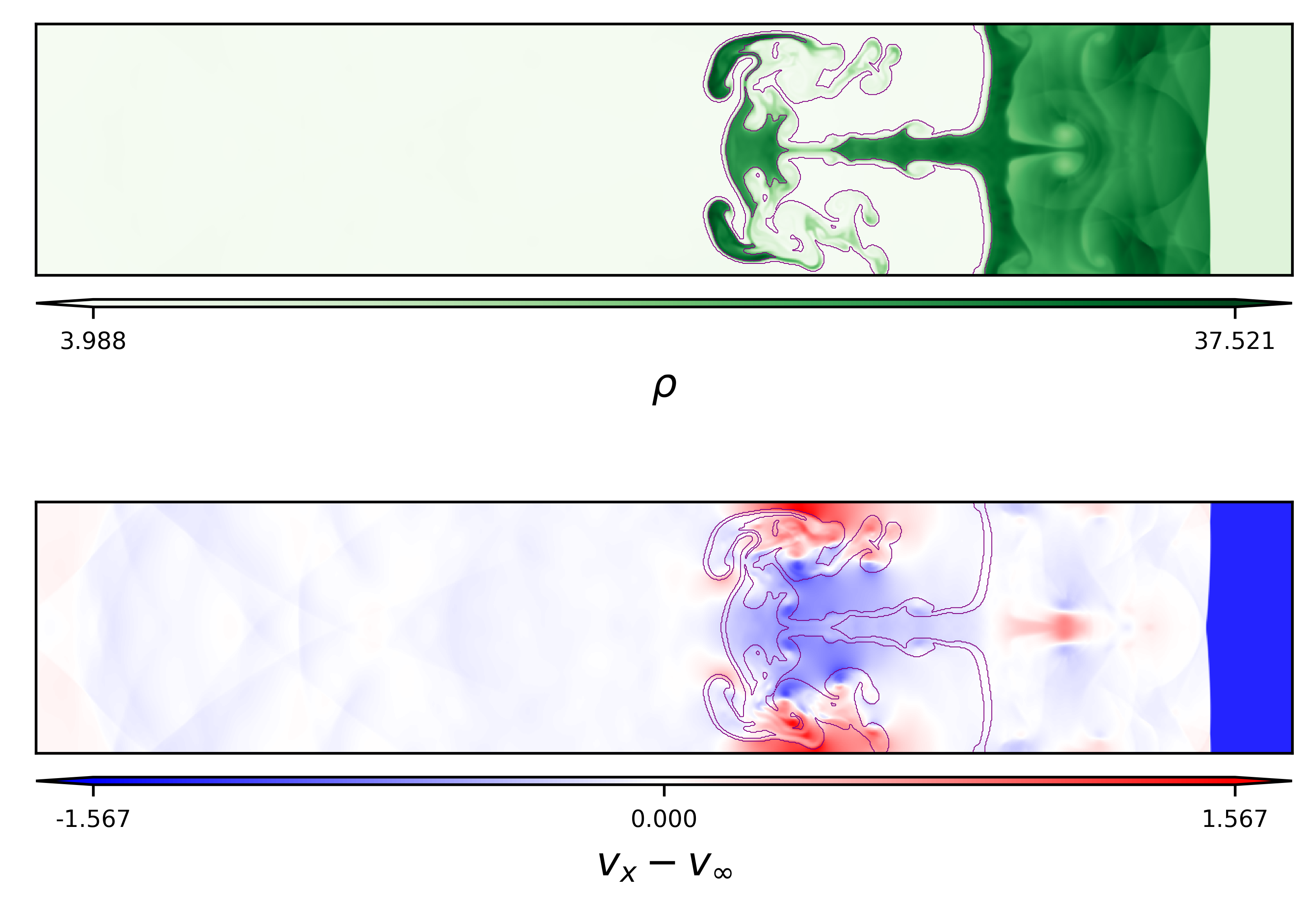}}
	\caption{The mass-density (upper) and x-velocity (lower), offset by the background motion $v_\infty$, for the $M=3$, $\mathcal{A}=0.8$, and $a_0/\lambda=0.25$ case. A contour line of the density-interface is shown in purple. }
	\label{fig:HRMI_aol_0p25}
\end{figure}
\begin{figure}
	\centerline{\includegraphics[width=0.75\linewidth]{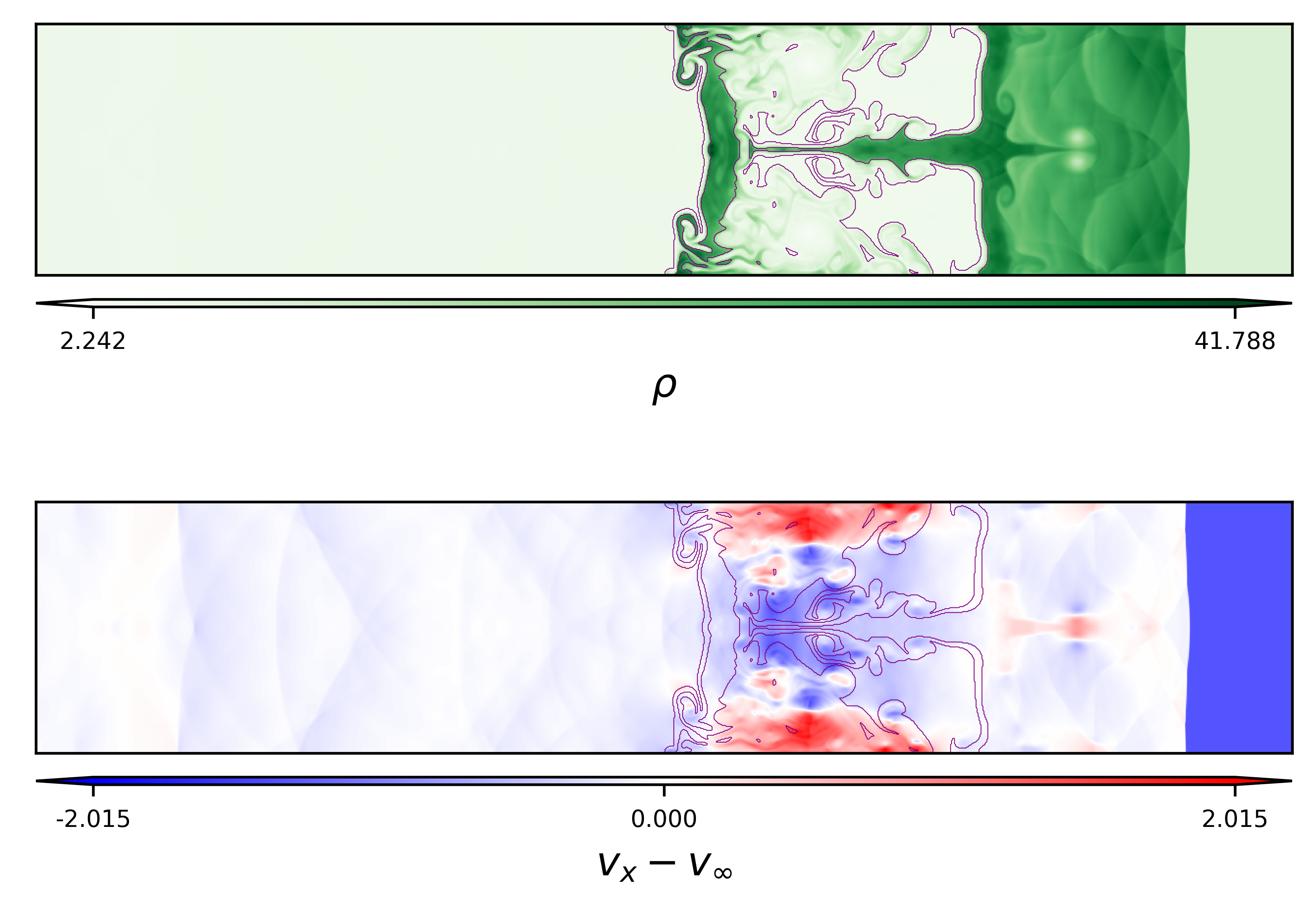}}
	\caption{The mass-density (upper) and x-velocity (lower), offset by the background motion $v_\infty$, for the $M=3$, $\mathcal{A}=0.8$, and $a_0/\lambda=0.4$ case. A contour line of the density-interface is shown in purple. }
	\label{fig:HRMI_aol_0p4}
\end{figure}

After shock traversal of the interface, the interface itself and surrounding fluid have a velocity that is characteristic of the problem. 
This velocity for a planar interface is $v_\infty$ and is used to define the background motion. 
It can be derived from linear theory \citep{wouchuk2001growth,richtmyer1960taylor} or simulations. 
Values for $v_\infty/c_l$, where $c_l$ is the sound speed in the light fluid, at the conditions described above from our simulation and that of \citet{dell2017maximum} are $1.0391$ and $1.0328$, respectively. 
The value of $v_\infty/c_l$ from linear theory \citep{wouchuk2001growth} is $1.0391$, showing excellent agreement with our simulation and good agreement with that of \citet{dell2017maximum}. 
Figures \ref{fig:HRMI_aol_0p06}, \ref{fig:HRMI_aol_0p25}, and \ref{fig:HRMI_aol_0p4} show the mass-density and x-velocity (offset by the background motion $v_\infty$) of the fluid at late time for $a_0/ \lambda=0.06$, $0.25$, and $0.4$. 
In the frame of reference moving with speed equal to $v_\infty$, theory \citep{abarzhi2003rayleigh,abarzhi2008review} predicts that the fluid motion away from the interface is small, there is significant fluid motion close to the interface, and the nonlinear RMI bubbles flatten at late times. 
We observe that for low values of $a_0/ \lambda$ the fluid motion away from the interface is very small but as $a_0/ \lambda$ increases, the increased presence of transverse waves disrupts the velocity field away from the interface more significantly, however, compared to the velocity field about the interface, it is still small. 
Our simulation results reproduce the qualitative prediction from the non-linear theory.

\begin{figure}
	\centerline{\includegraphics[width=\linewidth]{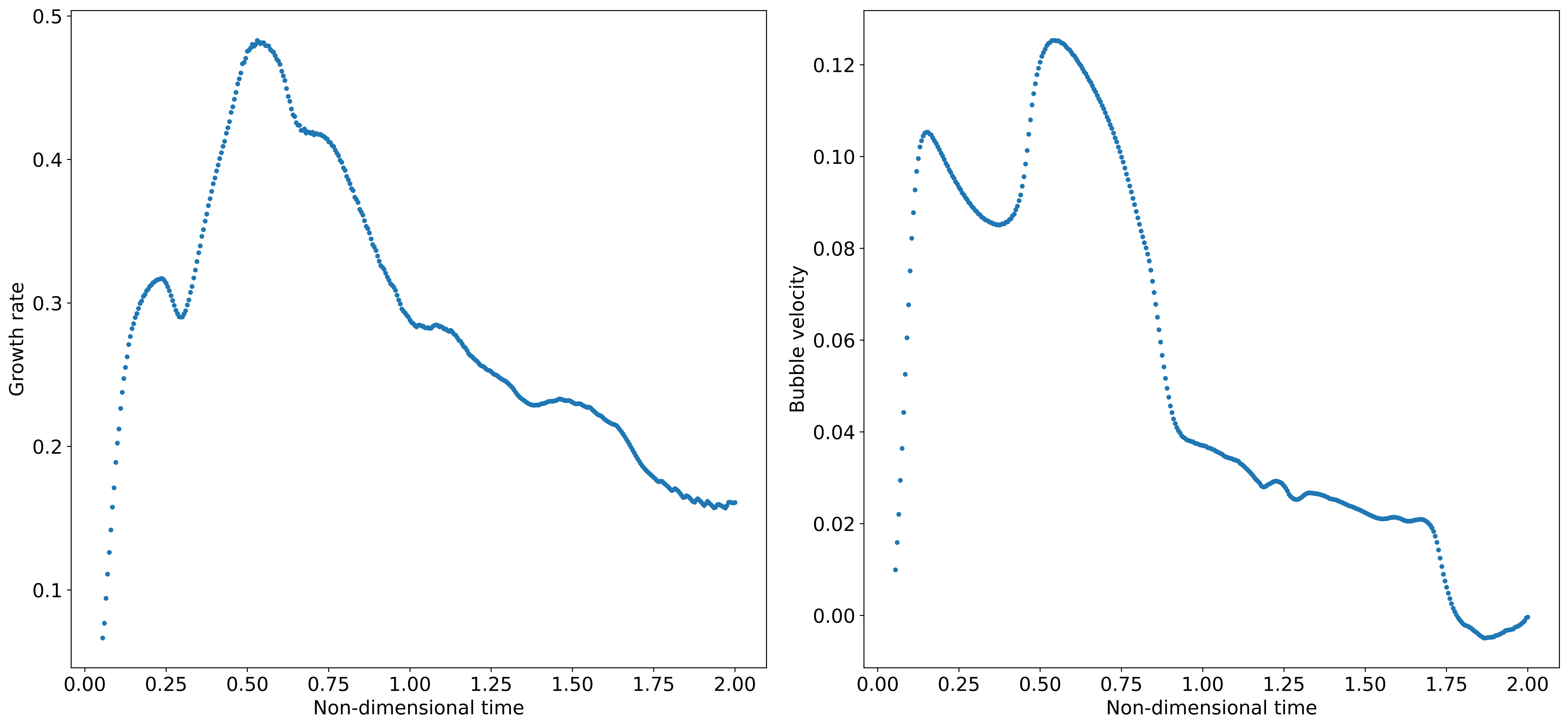}}
	\caption{The growth rate and bubble speed for the $M=3$, $\mathcal{A}=0.8$, and $a_0/\lambda=0.06$ case. The velocity is in the initiating shock direction, the reference frame of the background motion, and normalised by the background motion ($\left(v - v_\infty\right)/v_\infty$. }
	\label{fig:growth_spike_bubble_speed_aol_0p06}
\end{figure}

\begin{figure}
	\centerline{\includegraphics[width=\linewidth]{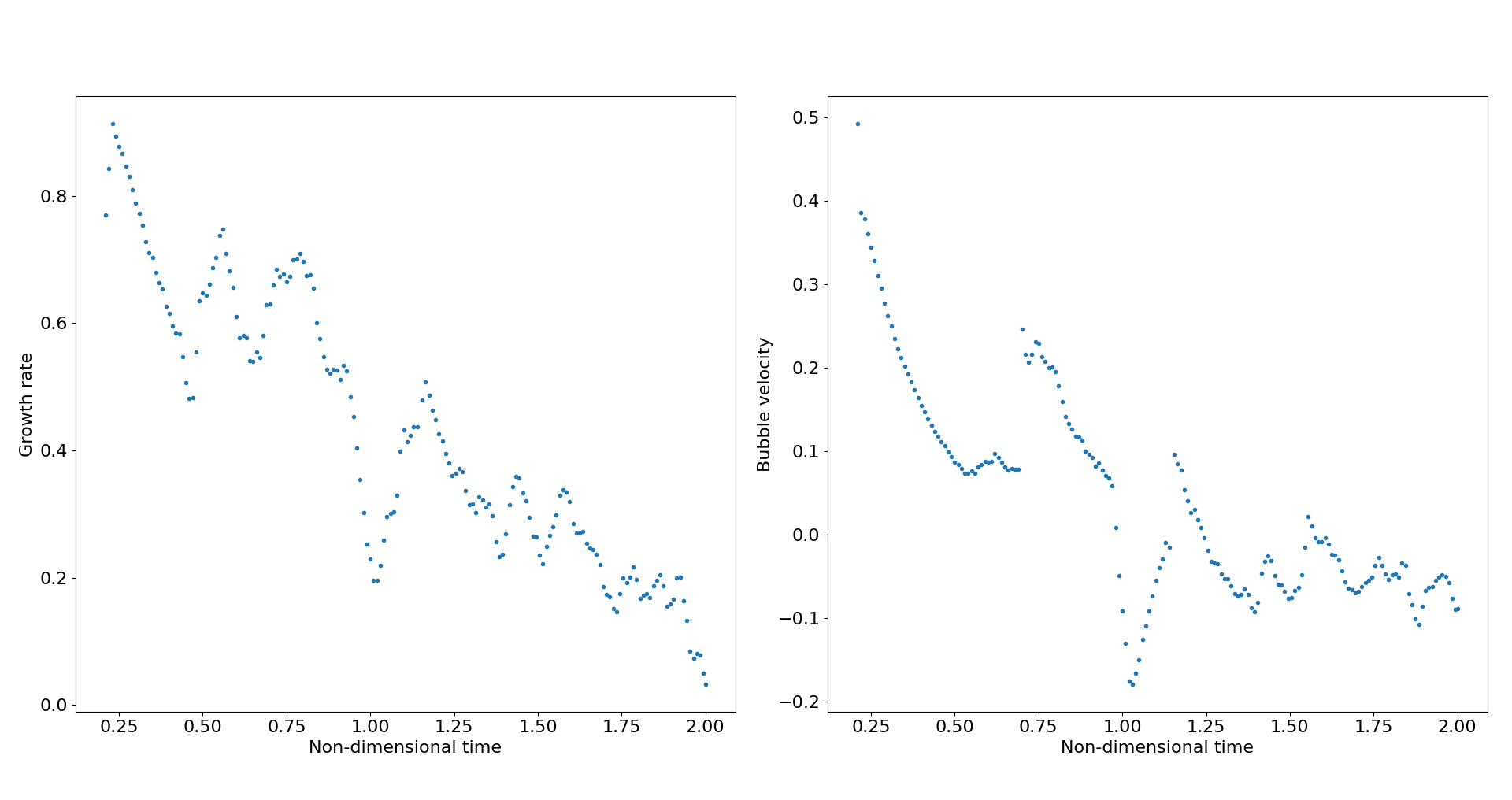}}
	\caption{The growth rate and bubble speed for the $M=3$, $\mathcal{A}=0.8$, and $a_0/\lambda=0.25$ case. The velocity is in the initiating shock direction, the reference frame of the background motion, and normalised by the background motion ($\left(v - v_\infty\right)/v_\infty$. }
	\label{fig:growth_spike_bubble_speed_aol_0p25}
\end{figure}

Our results indicate that as the ratio of initial amplitude to wavelength increases, the growth rate does not increase monotonically. 
We find that the growth rates increase, reaching a maximum between $a_0/\lambda=0.25\textrm{ and } 0.3$, then decreasing such that at $a_0/\lambda=1$ the growth rate is comparable to the $a_0/\lambda<0.1$, in qualitative agreement with the results from \citet{dell2017maximum}. 
The simulations that enter the non-linear regime exhibit the characteristic flattening of the bubble. 
This indicates that the pressure fluctuations around the interface are weak at the time of flattening \citep{pandian2017effect}. 
Results also capture the small scale dynamics of the problem, such as reverse jets which occur when two flows collide at small angle of attack \citep{pandian2017effect} and are fully immersed in the heavy fluid. 
Figure \ref{fig:growth_spike_bubble_speed_aol_0p06} and \ref{fig:growth_spike_bubble_speed_aol_0p25} shows the bubble speed (in the frame of reference of $v_\infty$ and normalised by $v_\infty$) for $a_0/\lambda=0.06$ and $0.25$. 
The figures show the bubble speeds and growth after the reflected and transmitted shocks pass the bubble. The bubble speed decreases but is periodically influenced by the passage of transverse reflected shocks. The influence of these transverse shocks is greater for increased $a_0/\lambda$ as inclination of the interface increases, aligning the reflected shocks more with the y-direction, therefore, interacting with the interface for longer time due to the periodic boundary conditions in y. 
The $a_0/\lambda=0.06$ case does not flatten during simulation run, but we see from figures \ref{fig:HRMI_aol_0p25} and \ref{fig:growth_spike_bubble_speed_aol_0p25} that the $a_0/\lambda=0.25$ case does, with a feature developing on the bubble. 
This feature forms in a similar way to the reverse jets at the base of the spike, except here the curvature of the fluid density profile (in the bulk of the fluid) is opposite, leading to a jet in the negative x-direction. 
The growth of this jet at the bubble is what produces the increase in bubble speed at late time $a_0/\lambda=0.06$ case, it is has already developed by the times shown in the $a_0/\lambda=0.25$ and $0.4$ cases. 
We see from figures \ref{fig:HRMI_aol_0p06}, \ref{fig:HRMI_aol_0p4}, and \ref{fig:HRMI_aol_0p25} that the x-velocity across the interface when the bubble has flattened is continuous but still discontinuous in the $a_0/\lambda=0.06$ where the bubble still has finite curvature, in agreement with the nonlinear theory for interface with no mass-flow \citep{abarzhi2003rayleigh}.
Overall, our simulation results show good qualitative agreement with the theory, reproducing all the expected features.

\begin{table}
	\begin{center}
		\begin{tabular}{lccccc}
			$a_0/\lambda$ & \begin{tabular}[c]{@{}l@{}}Effective \\ resolution\end{tabular} & \begin{tabular}[c]{@{}l@{}}Time \\ step\end{tabular} & \begin{tabular}[c]{@{}l@{}}$v_0/v_\infty$ \\ Cerberus\end{tabular} & \begin{tabular}[c]{@{}l@{}}Empirical \\ relation deviation\end{tabular} & \begin{tabular}[c]{@{}l@{}}Simulation \\ results deviation\end{tabular} \\ \hline
			0.06 & 2048 & $5\times 10^{-4}$ & 0.2043 & 0.09\% * & -28.53\% \\ 
			0.25 & 1024 & $1\times 10^{-3}$ & 0.5608 & -26.95\% & -6.69\% \\
			0.30 & 512  & $5\times 10^{-3}$ & 0.5043 & -8.49\% & 10.80\% \\ 
			0.40 & 512  & $5\times 10^{-3}$ & 0.4615 & 3.16\% & 0.00\% \\ 
			0.50 & 512  & $5\times 10^{-3}$ & 0.4305 & 6.02\% & -2.37\% \\ 
			1.00 & 512  & $5\times 10^{-3}$ & 0.1846 & 25.02\% & 18.19\% 
		\end{tabular}
	\end{center}
	\caption{Comparison of Cerberus results with the empirical relation and simulation results of \citet{dell2017maximum}. *In the case of $a_0/\lambda=0.06$, the empirical relation by \citet{dell2017maximum} for the initial growth rate $v_0$ was used, whereas the empirical relation for the non-linear $v_0$ was used for all others. }
	\label{table:dell_2017_comparison}
\end{table}

The results and comparison with \citet{dell2017maximum} are given in table 	\ref{table:dell_2017_comparison}. 
Some significant deviation from the empirical relation is found in the small $a_0/\lambda$ range and for the $a_0/\lambda=1.00$ case but good agreement is found for the remainder. 
The comparison to the simulation results is mostly in good agreement except for the $a_0/\lambda=0.06$ and $1.00$ cases.
The simulations results by \citet{dell2017maximum}, which form a subset of the group of simulation results that the empirical model was based on, exhibited deviations from the empirical model that is similar to our simulations. 
The empirical relations for the linear and non-linear regime \citep{dell2017maximum} were found to be a useful predictor of behaviour, despite some deviation, and is attractive due to its simplicity and universality, being dependent on $\mathcal{A}$ and $a_0/\lambda$, and being calibrated from simulation data.

Good qualitative and quantitative agreement was found when comparing to experiments \citep{motl2009experimental} and reasonable agreement was found when comparing to the simulation results and empirical relation derived from a different numerical scheme \citep{dell2017maximum}, giving the authors confidence that the Cerberus code accurately captures the RMI and secondary instabilities.

\subsubsection{1D convergence study}\label{sec:ConvergenceStudy}
\begin{figure}
	\centerline{\includegraphics[width=\linewidth]{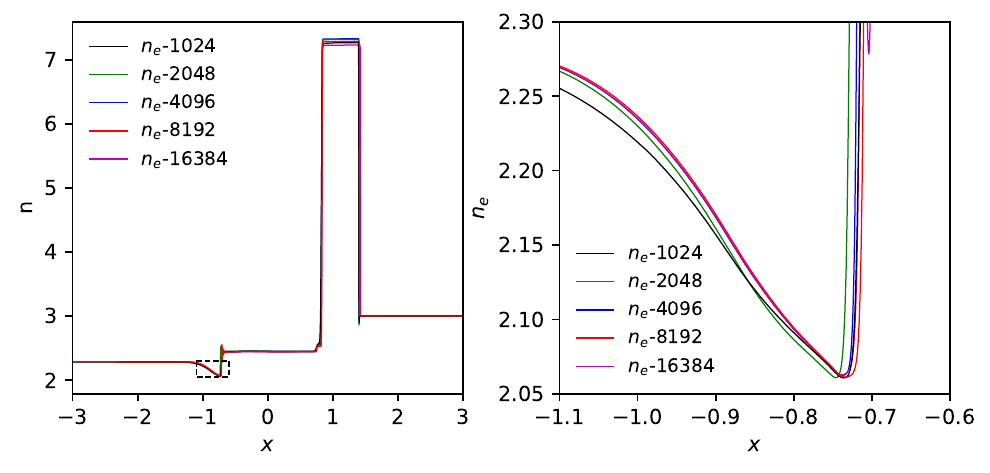}}
 replaced this with new corrected image 
	\caption{(a) Ion and electron number density from a one dimensional simulation of a discontinuous interface with incident shock at non-dimensional time t=0.3,  illustrating  grid  convergence.  (b) Zoomed view of results showing only the electron number density.}
	\label{fig:ConvergenceStudy}
\end{figure}
The results of a 1D test case have been used to determine an appropriate resolution in the stream wise direction. The ideal multi-fluid plasma model, lacking physical dissipation, does not converge to a single solution. As the grid is refined, progressively smaller flow features will emerge meaning no minimum length scale limit exists in this regard. Therefore, the criteria for appropriate resolution was taken to be the resolution of all plasma waves and to pragmatically minimise the relative change between simulation results. The results of successive refinements of resolution is shown in figure \ref{fig:ConvergenceStudy} for a skin depth of 0.1 (Debye length of 0.002) unit lengths. We see significant improvement between resolution of 1024 to 2048 cells per unit length and continued but diminished improvement between 2048 and 16384 cells per unit length. The expanded view especially shows that a reasonable resolution that captures the flow features well is the 4096 resolution.

\subsection{Problem Description}\label{sec:ProblemDescription}
\begin{figure}
	\includegraphics[width=\linewidth]{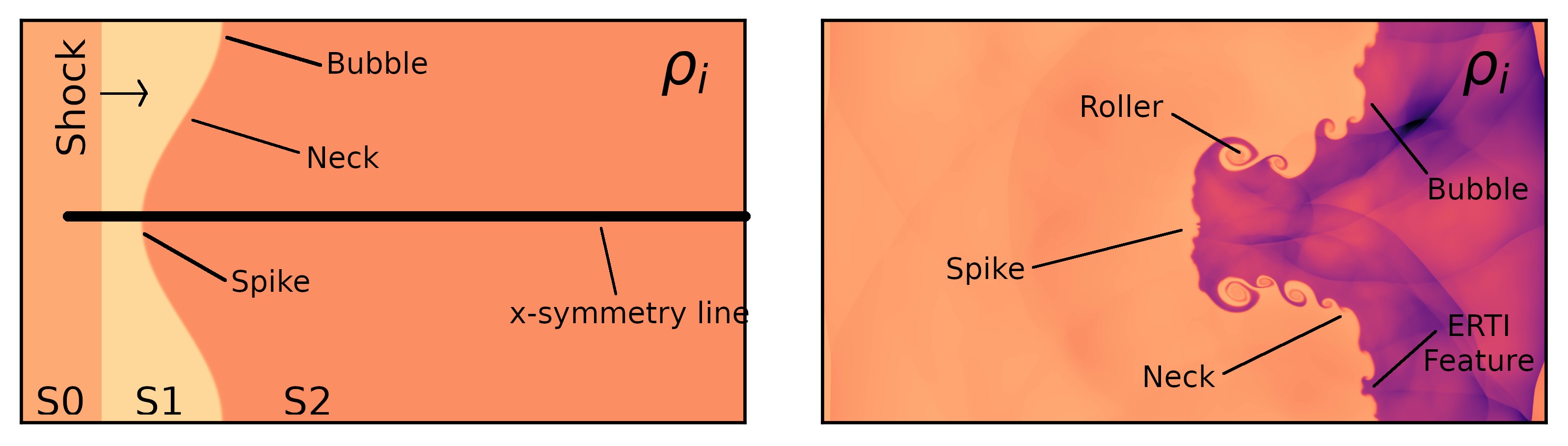}
	\caption{An example of initial the initial conditions and developed evolution of the RMI. }
	\label{fig:ProblemDescription}
\end{figure}
The generic problem follows that of \citet{bond2017richtmyer} which was designed to be comparable to previous RMI studies \citep{samtaney2003suppression,wheatley2005regular,wheatley2013magnetohydrodynamic}. Figure \ref{fig:ProblemDescription} shows the initial configuration that forms a three zone Riemann problem consisting of zones zero ($S_0$), one ($S_1$), and two ($S_2$) from the left boundary. The interface between $S_0$ and $S_1$ generates a shock that travels to the right where it encounters the interface formed between fluids in region $S_1$ and $S_2$. The perturbation of the interface between $S_1$ and $S_2$, is chosen to consist of a single period of a sinusoid with an amplitude of 0.1 non-dimensional length ($\frac{1}{10}$ domain width). The perturbation mean location in the x-dimension is 0.2 non-dimensional lengths from the initial shock. The density interface studied in this work is a material interface. The initial interface is a stationary contact discontinuity without any mass or heat flux (we do not consider, for example, the kind that would be involved in the RMI of a travelling flame front \citep{ilyin2022interface}). Furthermore, there are no phase changes associated with the interface in this work.

As an initial condition, charge neutrality is enforced everywhere and mechanical equilibrium is enforced between $S_1$ and $S_2$ as minimum requirements for initial stability and clarity of results. These two conditions ensure there is no movement of the interface $S_1$ to $S_2$ prior to the shock interaction due to EM or hydrodynamic effects, excluding the shocks involved in the simulation. The interface is not likely to be stable in actual ICF experiments due to an X-ray preheat preceding the shock, shown to exist in shock tube experiments \citep{keiter2002observation,yamada2019experimental}, and drive asymmetries. However, it is useful to model the interface as stable so we may elucidate the fundamental physical phenomena.

The boundary conditions are periodic in the y-dimension and zero gradient in the x-dimension. The domain is taken to be one reference length in the y-dimension, and in the x-dimension it is two times the distance travelled at the non-dimensional speed of light during simulation time, to avoid any possible numerical artefacts reflecting off the boundaries and interacting with the observed solution space. The expansive x-dimension of the simulation space is made practical by the adaptive mesh refinement used in the numerical solver, where a very coarse base grid of only 8 cells across the domain allows a grid to be inexpensively extended away from the region of interest. The boundary between $S_1$ and $S_2$ in figure \ref{fig:ProblemDescription} represents the RMI DI which is chosen to be in the light-to-heavy configuration to avoid complications from a phase inversion (heavy-to-light configuration). A hyperbolic tangent transition function is used between $S_1$ and $S_2$ to impose a smooth transition that makes the solution less susceptible to numerical artefacts and to ensure a consistent interface thickness at different resolutions. This function has the form
\begin{align}
f(x) = f_R + \frac{f_L-f_R}{2}\left(1+\tanh\left(\frac{2x}{\eta}\text{arctanh}\left[\frac{9f_R-10f_L}{10(f_L-f_R)}
\right]
\right)
\right)
\end{align}
where $f_L$ and $f_R$ are the variable of interest on the left and right of the interface, and $\eta$ is the width containing 90\% of the transition, chosen as 0.01 non-dimensional lengths. 

\subsection{MFP interface modelling}
A significant benefit of the MFP model is the ability to investigate fine details of the fluids that comprise the RMI interface types. Therefore, the subtle differences in the thermal, isotope, and species interface types can be distinctly simulated. \textit{If a single-fluid model, such as ideal MHD or HMHD were used, these details would not be captured and all three interfaces would behave similarly.} Consider figure \ref{fig:initial_conditions}, the unique differences across the cases presented would not be captured, in MHD only one homogeneous case of a density interface could be used to investigate all cases. The thermal case models a discontinuity in the temperature of elemental hydrogen, $\ce{^{1}_{1}H}$. The isotope case models a discontinuity in particle mass in the ion fluid that is produced by the interface of hydrogen, $\ce{^{1}_{1}H}$, and tritium, $\ce{^{3}_{1}H}$. The first species case is modelled with an interface of hydrogen, $\ce{^{1}_{1}H}$, with helium (isotope), $\ce{^{3}_{2}He}$, and the second species case is modelled with an interface of hydrogen with a fictitious isotope of lithium, $\ce{^{3}_{3}Li}$. We reiterate once more that the scenarios simulated are not accurate quantitative representations of the phenomena occurring in ICF, however, the physical phenomena which are elucidated through these simulations are representative of those which would be exhibited in an ICF experiment.

\subsubsection{Parameters}
The parameter space in the simulations is large, therefore a subset must be chosen to make the investigation tractable. The parameters used to define the simulated regime were taken from previous investigations \citep{samtaney2003suppression,wheatley2005regular,wheatley2013magnetohydrodynamic} and are the ion fluid species mass densities either side of the interface, the ion partial pressure for $S_1$ and $S_2$, species hydrodynamic and EM properties e.g. ratio of specific heats and particle charge, and the shock Mach number. Electron fluid parameters such as fluid mass densities, pressures, number densities, and temperatures (kT); are set according to the ideal gas equation of state, normal shock relations, and physical properties of the species involved. The following parameters were set 
\begin{align}
m_e &= 0.01, &m_{i0} =m_{i1} &=1.0, & \rho_{i1} &= 1, \\
q_e &= -1.0, &q_{i0} =q_{i1} &= 1.0, & \rho_{i2}&=3 ,\nonumber \\
\gamma_e &= 5/3,  &\gamma_{i0}=\gamma_{i1}=\gamma_{i2}&=5/3, &p_{i1} &= 0.5 \nonumber \\
\textrm{and} \quad  M_0 &= 2.0. \nonumber 
\end{align}
The non trivial relations are the normal shock relations and scalar pressure for a gas obeying a Maxwellian (ideal gas law), 
\begin{align}
\rho_{i0} &= \frac{\rho_{i1}}{1 - (2/(\gamma+1))(1 - 1/M_0^2)},\\
p_{i0} &= p_{i1}\left(1 + \frac{2\gamma}{\gamma+1}(M^2 - 1)\right),\nonumber\\
p_\alpha &= n_\alpha kT_\alpha \nonumber
\end{align}
and the requirements of charge neutrality resulting in the relations, 
\begin{align}
n_{i0} &= \frac{\rho_{i0}}{m_{\ce{^{1}_{1}H}}}, & n_{i1} &= \frac{\rho_{i1}}{m_{\ce{^{1}_{1}H}}}, & n_{i2} &= \frac{\rho_{i2}}{m_{i2}},\\
n_{e0} &= n_{i0}Z_{\ce{^{1}_{1}H}}, & n_{e1} &= n_{i1}Z_{\ce{^{1}_{1}H}},&n_{e2} &= n_{i2}Z_{{i2}}, \nonumber
\end{align}
All other parameters are set as a result of those above and the requirements of charge neutrality and mechanical equilibrium. The remaining relations are
\begin{align}
P_{i2} &= P_{i1}, \\
P_{e0} &= P_{i0}, &P_{e1} &= P_{i1}, &P_{e2} &= P_{i1},\nonumber\\
kT_{i0} &= \frac{P_{i0}}{n_{i0}}, &kT_{i1} &= \frac{P_{i1}}{n_{i1}}, &kT_{i2} &= \frac{P_{i2}}{n_{i2}}, \nonumber\\
kT_{e0} &= \frac{P_{e0}}{n_{i0}Z_{i1}}, &kT_{e1} &= \frac{P_{e1}}{n_{i1}Z_{i1}}, &\textrm{and} \quad kT_{e2} &=\frac{P_{e2}}{n_{i2}Z_{i2}}.\nonumber
\end{align}
The unique parameters for the three interface types are given in table \ref{table:interfaceParameters}. 

\begin{table}
	\begin{center}
		\begin{tabular}{lcccc}
		\textbf{Interface type} &   Thermal (TRMI) &  Isotope (IRMI) & Species (SRMI-He) & Species (SRMI-Li) \\ \hline
			$m_{i.2}$  & $1$     & $3$ & $3$ & $3$ \\ \hline
			$q_{i,2}$ & $1$      & $1$ & $2$ & $3$ \\ \hline
		\end{tabular}	
		\caption{ Distinguishing parameters for the thermal, isotope, and species interface cases. }
		\label{table:interfaceParameters}
	\end{center}
\end{table}
Ideally, a single shock-front would be maintained across both fluids until interaction with the DI. However, the greater sound speed and shock breakdown in the electron fluid mean an electron shock-wave co-incident to the ion shock cannot be produced and maintained after initialisation. The resulting general Riemann problem in the electron fluid propagates a single shock and multiple waves that traverse the interface prior to the arrival of the ion shock. 

The volume of fluid tracer quantity used to track the interface is also used to calculate the mixture properties of the ion fluid. 
The same hyperbolic tangent transition function is used for the tracer as for the conserved properties on the interface, and the tracer has a value varying $\varrho \in [0, 1]$ from left to right states. 
The single ion fluid modelled in this work is technically a mixture, where the tracer property is equivalent to a mixture fraction.  
The tracer value is conserved and is convected as a passive scalar. A mixture equation, shown below, is used to find the particle properties when the tracer value is between zero and one. In the equation below, $\varrho$ is the tracer value, $\phi$ is some property which varies as a linear mixture of the two values. 
\begin{align}
\phi_{effective} = (1-\varrho)\phi_0 + \varrho\phi_1.
\end{align}

\section{Results}

\subsection{1D Exemplar}\label{sec:1D_base_flow}
\begin{figure}
	\centering
	\begin{subfigure}[b]{0.45\linewidth}
		\centering
		\includegraphics[width=\linewidth]{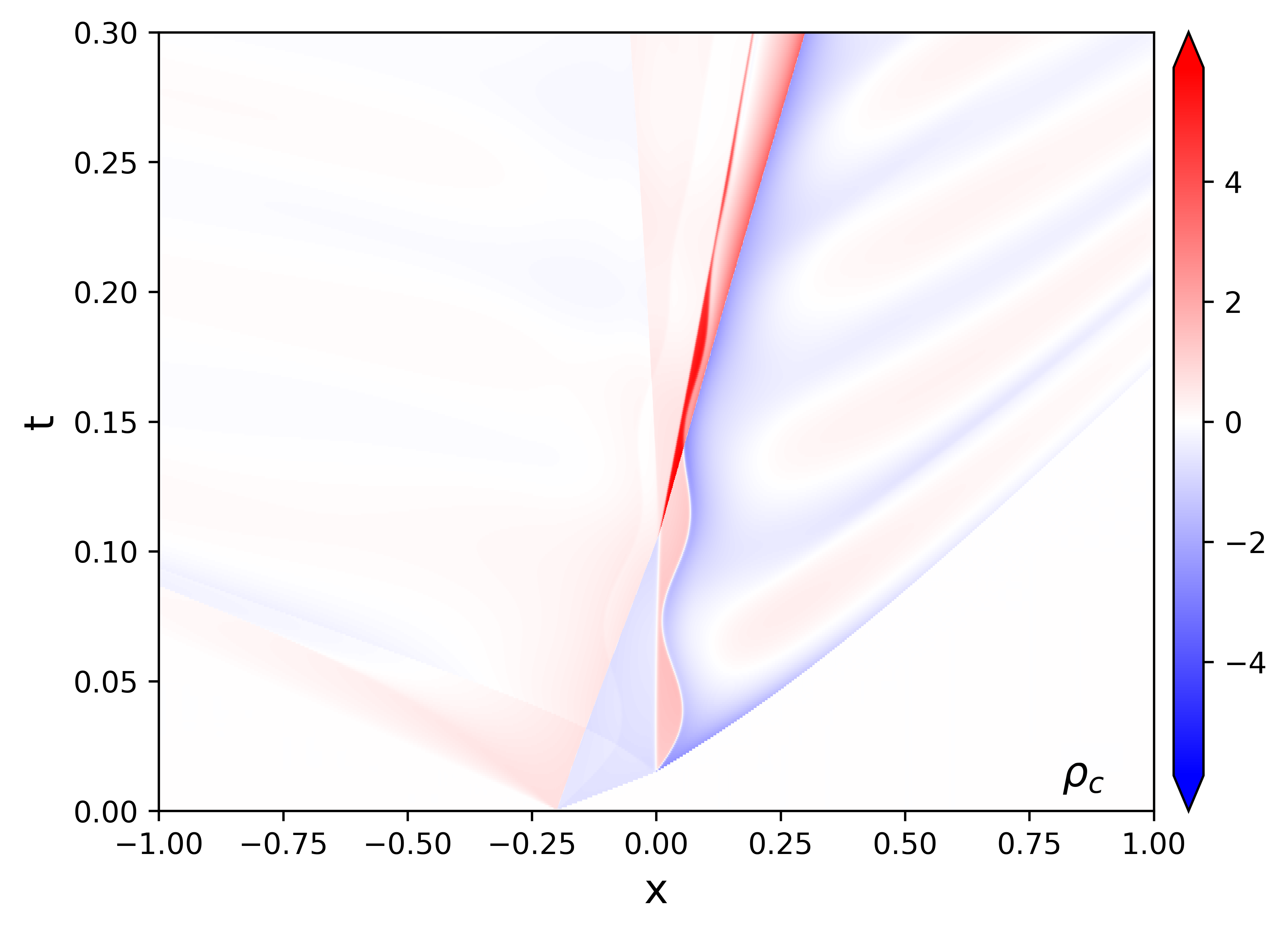}
		\caption{Thermal discontinuity}
		\label{fig:1D_Ther}
	\end{subfigure}
	\hfill
	\begin{subfigure}[b]{0.45\linewidth}
		\centering
		\includegraphics[width=\linewidth]{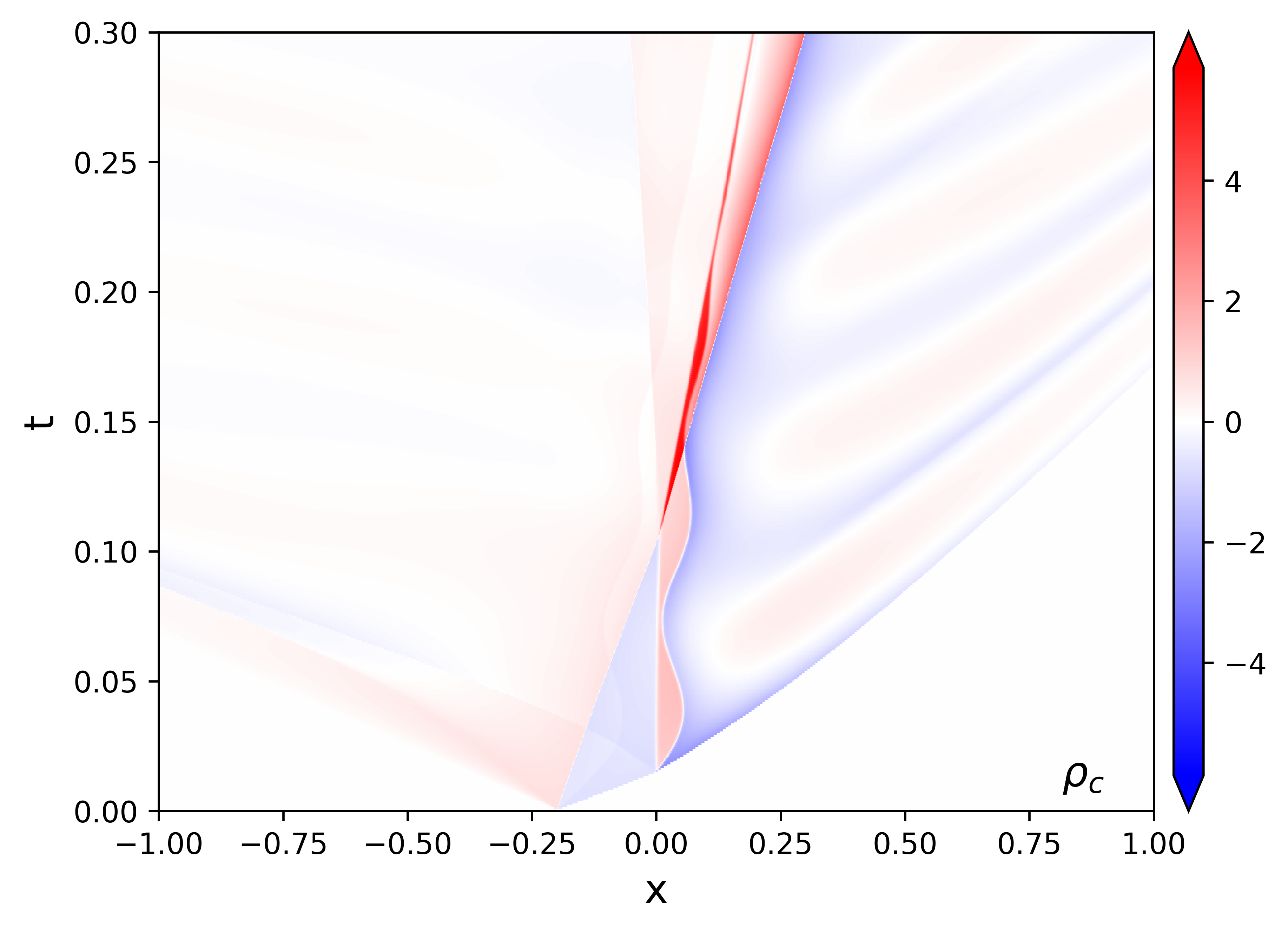}
		\caption{Species Li-3 discontinuity}
		\label{fig:1D_Spec_Li}
	\end{subfigure}
	\\
	\begin{subfigure}[b]{0.45\linewidth}
		\centering
		\includegraphics[width=\linewidth]{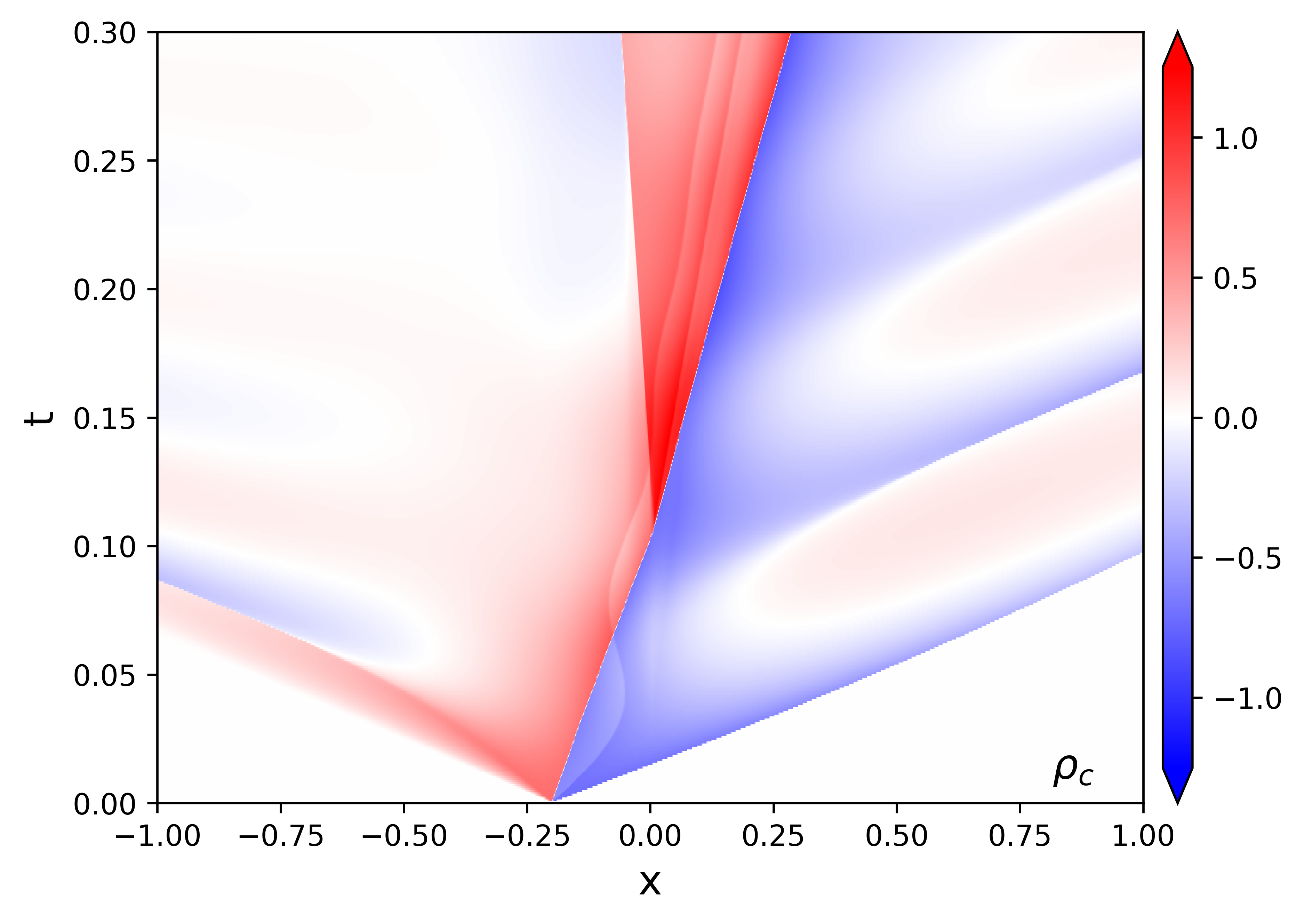}
		\caption{Isotope H-3 discontinuity}
		\label{fig:1D_Isot}
	\end{subfigure}
	\hfill
\begin{subfigure}[b]{0.45\linewidth}
	\centering
	\includegraphics[width=\linewidth]{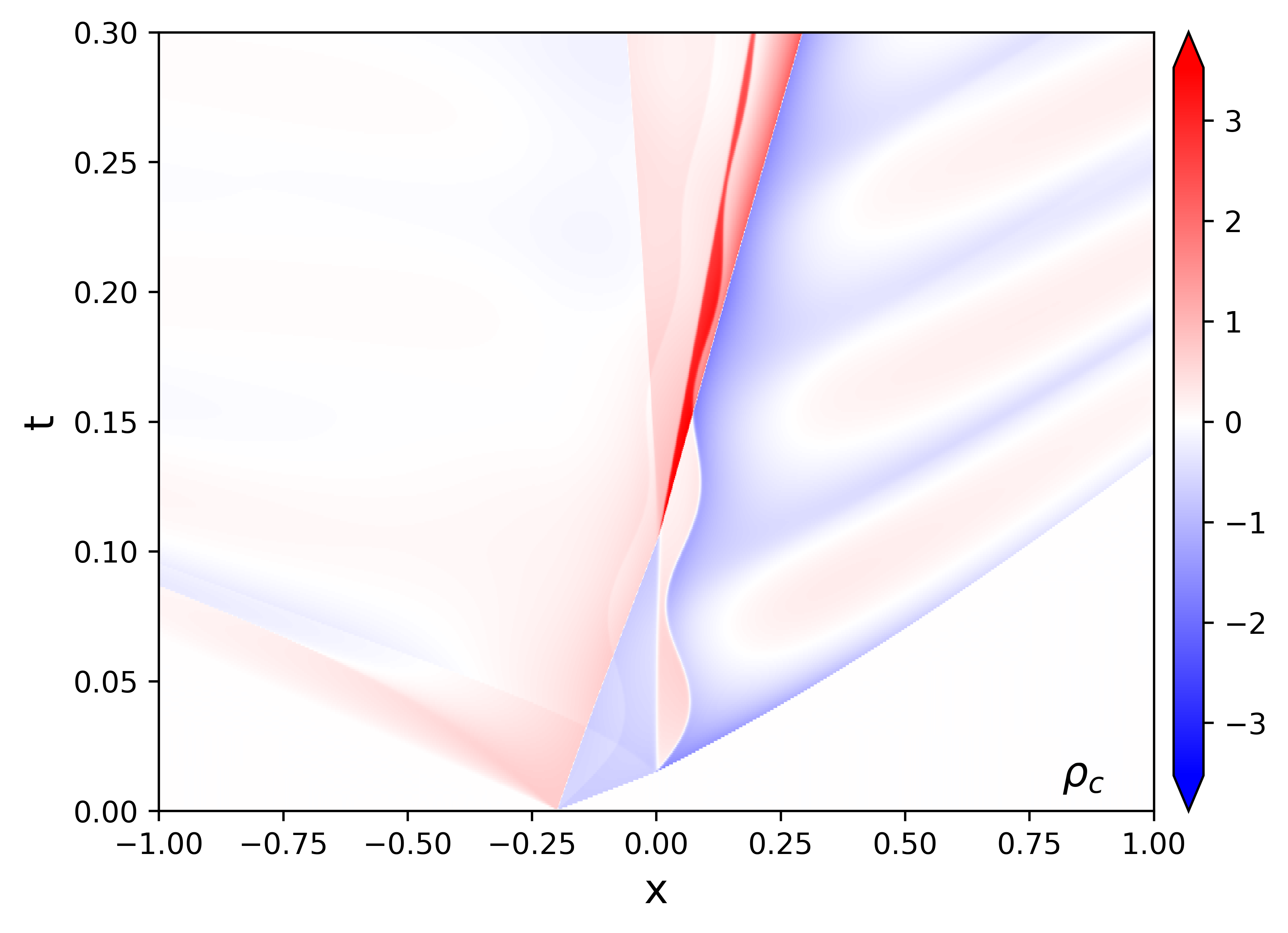}
	\caption{Species He-3 discontinuity}
	\label{fig:1D_Spec_He}
\end{subfigure}
	\caption{Charge density x-t diagrams for $d_S = 10$ scenarios.}
	\label{fig:1D_dS_10_xt_diagrams_charge}
\end{figure}

The 1D base flow simulation results for the case of an unperturbed thermal, species, and isotope interface, as per table \ref{table:interfaceParameters}, are useful for appreciating the simpler MFP mechanisms that exist in the 2D flows. Figure \ref{fig:1D_dS_10_xt_diagrams_charge}, charge density contour plots for displacement-time axis, shows the key phenomena observable from the 1D simulations which are the self-generation of EM field, the Lorentz force influence on the fluids, electron shock degradation, and oscillatory electron motion. Immediately obvious from figure \ref{fig:1D_dS_10_xt_diagrams_charge} is the breakdown of the general Riemann problem in the electron fluid, producing a backwards travelling rarefaction and forward travelling contact discontinuity, shock, and multiple waves that are not visible in the figure. The shock and multiple waves traverse the electron fluid density interface (EFDI), except for in the Isotope case which has no EFDI, prior to the arrival of the ion shock. The differential speed of the ion and electron shocks (and wave packets) produce a non-zero charge density that produces the first self-generated EM fields. As the electron shock processes the EFDI, the charge separation is increased as the denser side is compressed.

Following the EFDI compression, an oscillatory behaviour emerges in the electron fluid (viewed through the charge density). This oscillation is the result of the EFDI, compressed after the shock traversal, moving through the homogeneous and mostly undisturbed ion fluid background. As the EFDI moves to the right, it depletes the negative charge within its former location producing a net positive charge there. This deviation from charge neutrality generates an electric field that attracts the EFDI, opposing its shock imparted motion. The reciprocal of this process happens after the EFDI is accelerated to the left, causing the oscillation to the right. This oscillation only abates once the ion shock has processed the ion fluid density interface (IFDI). The DI experiences a positive x-electric field for a long duration which is produced from a line of non-zero charge density that results from the interaction of EFDI and IFDI.
This x-electric field endures until simulation end. In the two-dimensional case, this x-electric field contributes to a Lorentz force that drives the ERTI in an unstable configuration, contributing to growth.

The effects discussed above occur for all interface types with differences in the extent. The thermal interface, figure \ref{fig:1D_Ther}, and Li-3 (fictitious element for the purpose of comparison), figure \ref{fig:1D_Spec_Li}, show a very similar result with a minor difference in the interaction of the electron- and ion-fluid DI (EFDI and IFDI). After the initial shock traversal, the EFDI moves to the right with greater speed in the Li-3 species case than in the thermal case. However, the IFDI has the same speed in both, meaning that the thermal case experiences an interaction between the two fluid interfaces sooner. For the He-3 species case, figure \ref{fig:1D_Spec_He}, the severity of charge separation achieved is lesser than the thermal case and Li-3 species case scenarios because of the lower EFDI density ratio that the initiating shocks encounter. This reduction in charge density ultimately affects the magnitude of the self-generated fields in the plasma and driving of the secondary ERTI in the two-dimensional cases. The isotope case, figure \ref{fig:1D_Isot}, results highlight the significance of the initial EFDI. The cases with a lesser EFDI ratio (or none at all), figure \ref{fig:1D_Isot} and \ref{fig:1D_Spec_He}, produce weaker charge separation and consequentially less energetic responses in the fluid motions. This result is especially clear in the isotope case results, where the peak fluid velocities and electromagnetic field generation is almost an order of magnitude lower than the thermal and Li-3 species cases.

Lower magnitude EM fields are generated in the isotope case, and any cases with lower EFDI ratio such as the He-3 species case, because of reduced current densities. 
The relative motion between the ion and electron fluids combined with different charge density across the interface, within each fluid, will drive a varying current density depending on  the type of interface. 
The differing wave speeds in the fluids causes a significant relative motion at early times, see the small magnitude negative charge density at early time. Secondly, the oscillatory behaviour of the EFDI, being pulled back by the region of positive charge density in its wake after processing the EFDI at early times adds some complexity to the charge density. Thirdly and most significantly, different compression and acceleration of the DI in each fluid after processing by their respective shocks causes the greatest magnitude charge density and current densities. This is shown in figures \ref{fig:1D_Ther}, \ref{fig:1D_Spec_Li}, and \ref{fig:1D_Spec_He} by the bright red (high positive charge density) between the IFDI and EFDI after shock processing. Crucially, this region of high positive charge density is absent from the isotope case.

\begin{figure}
	\centering
	\begin{subfigure}[b]{0.475\linewidth}
		\centering
		\includegraphics[width=\linewidth]{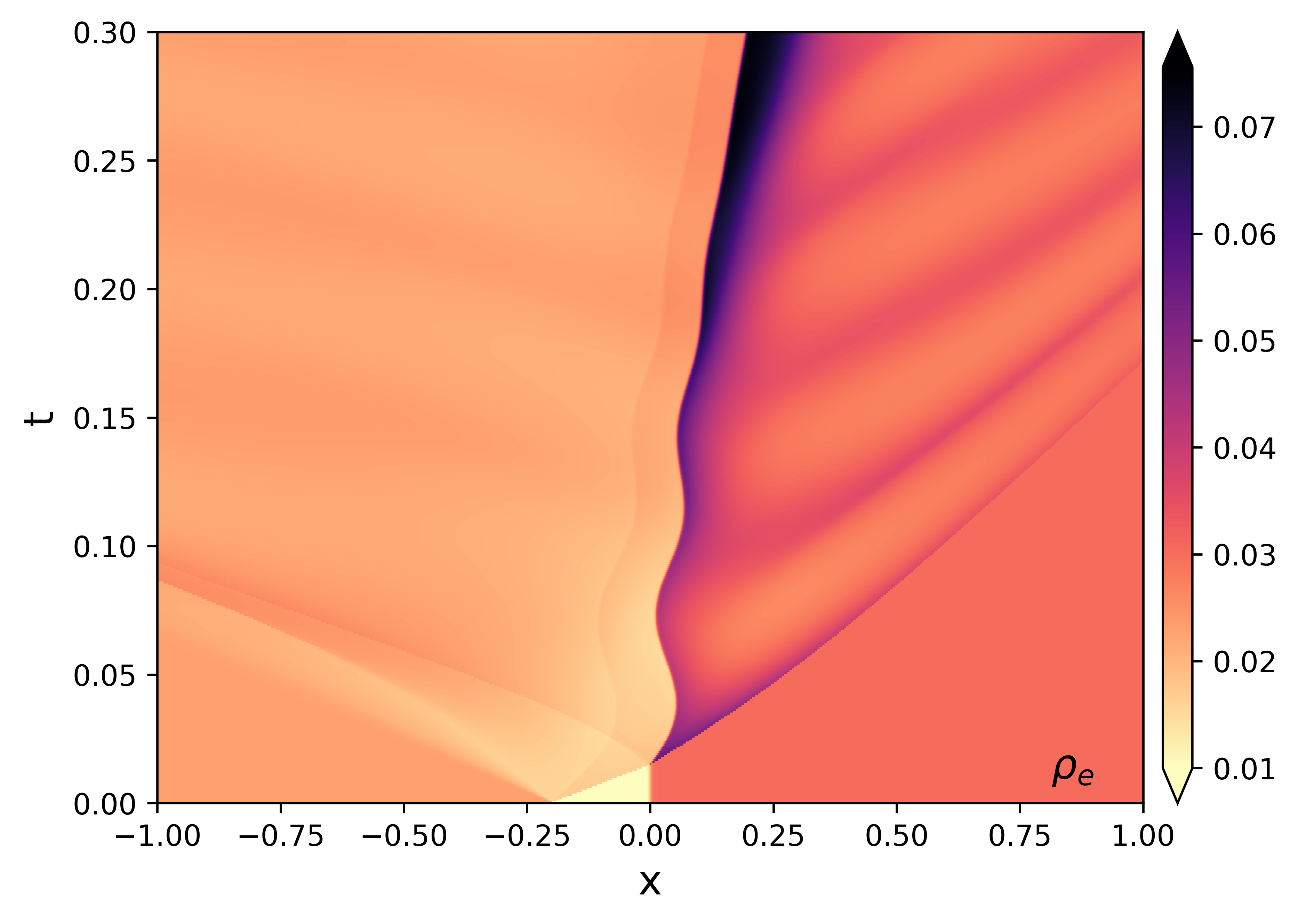}
		\caption{$d_S=10$}
		\label{fig:1D_Ther_10}
	\end{subfigure}
	\\
	\begin{subfigure}[b]{0.475\linewidth}
		\centering
		\includegraphics[width=\linewidth]{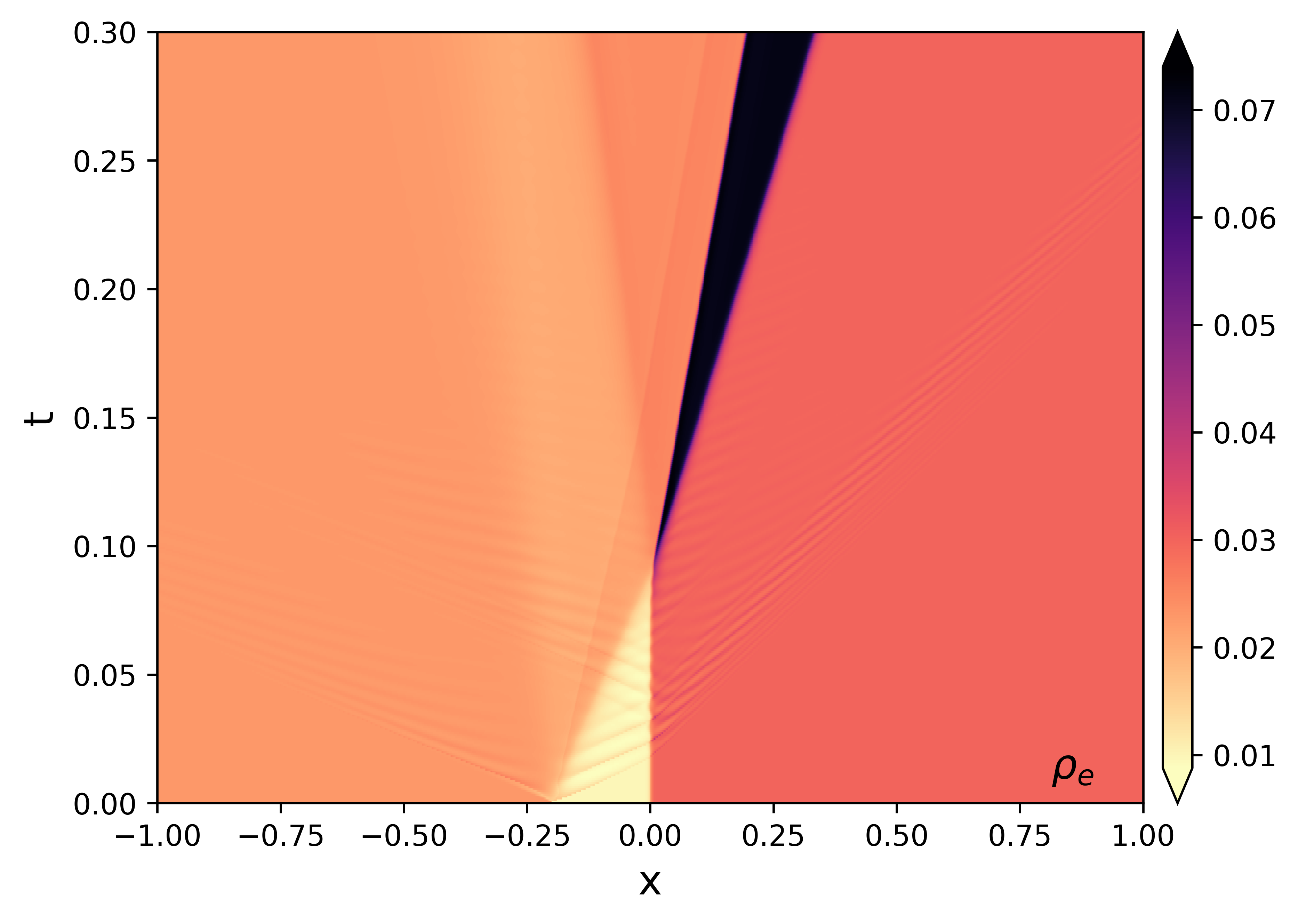}
		\caption{$d_S=1$}
		\label{fig:1D_Ther_1}
	\end{subfigure}
	\hfill
	\begin{subfigure}[b]{0.475\linewidth}
		\centering
		\includegraphics[width=\linewidth]{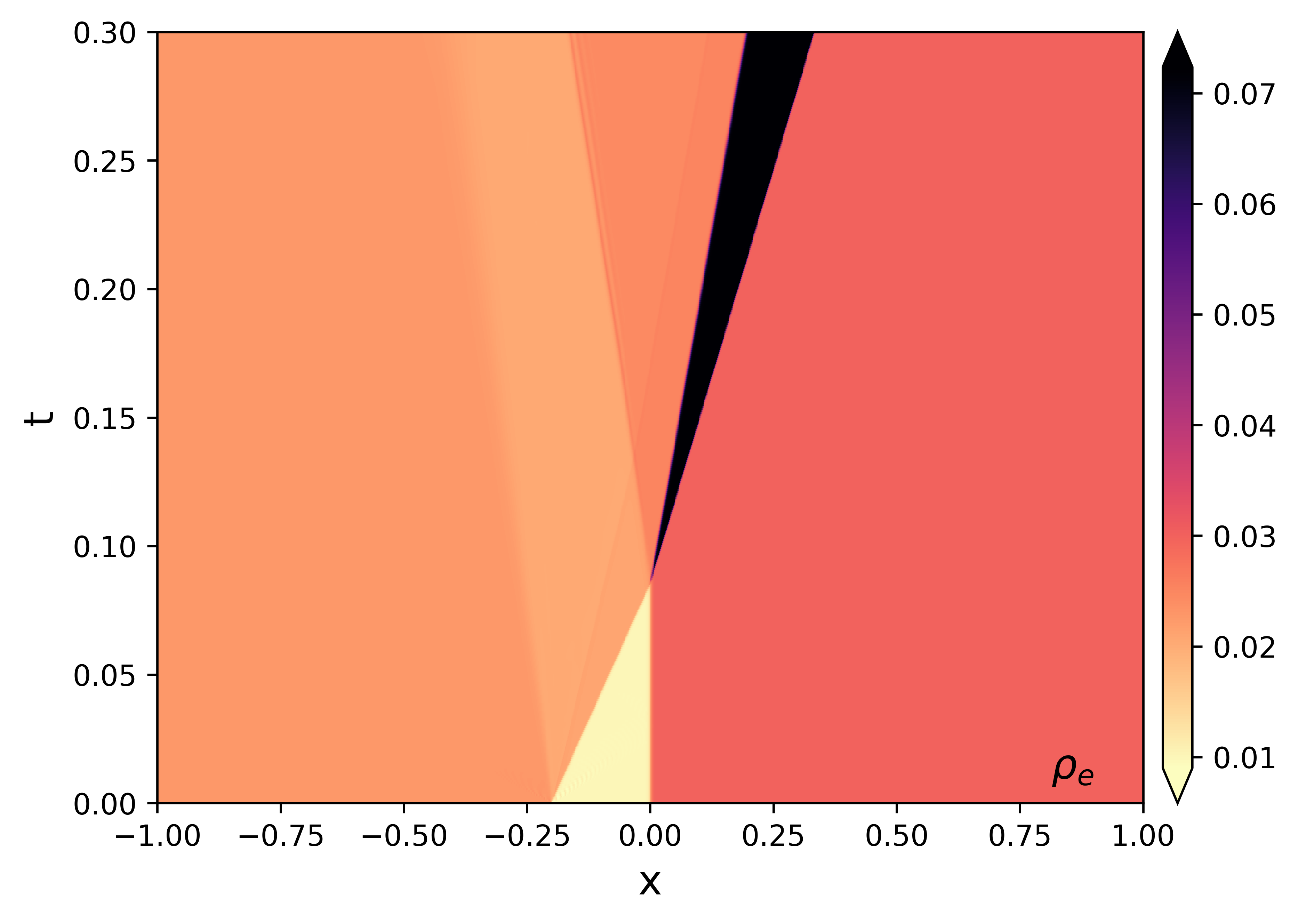}
		\caption{$d_S=0.1$}
		\label{fig:1D_Ther_0p1}
	\end{subfigure}
	\caption{Electron density distribution x-t diagram contour plot for Thermal for varying skin depths.}
	\label{fig:1D_Ther_xt_diagrams}
\end{figure}

Examining the Ther results for skin depths of 10, 1, and 0.1 (figure \ref{fig:1D_Ther_xt_diagrams}) demonstrates the increased coupling and subsequent reduction in multi-fluid effects for the 1D simulations. Figure \ref{fig:1D_Ther_10} corresponds to the charge density plot for the thermal case shown in figure \ref{fig:1D_Ther}. Figure \ref{fig:1D_Ther_1} shows the reduction in skin depth has caused a reduction in the oscillation amplitudes as the electron fluid is no longer able to achieve as large separation and consequent counteracting Lorentz force that would otherwise drive the greater oscillation amplitudes. The oscillations, though reduced in amplitude, increase in frequency, figure \ref{fig:1D_Ther_1}, as the electron fluid is pulled back sooner than it would be in the loosely coupled case. Within the $d_S = 0.1$ case, figure \ref{fig:1D_Ther_0p1}, no oscillation is visible. The reduction in skin depth allows the electron fluid to transfer a greater portion of its accumulated energy to the ions but the electrons accumulate less energy due to the reduced differential motion from the increased coupling. In 2D, however, the reduced skin depth cases produce more significant MFP effects than in 1D because (1) seeding of interface perturbations, by the electron fluid, triggers the KHI and (2) shock refraction at oblique angles produces destabilising features. 

\subsection{Comparison of 2D Fundamental interface types}\label{sec:comparison_interfaces}
This section compares the RMI simulation results for each of the fundamental interface types under varying values of non-dimensional skin depth. The key phenomena influencing the RMI are identified for each interface type. The MFP simulations are compared to the analogous single-fluid hydrodynamic simulations. In the absence of an applied magnetic field, as we simulate, single-fluid hydrodynamic and ideal MHD simulations are identical when the same density ratio is used. We provide two reference cases for the single-fluid hydrodynamic scenarios, pressures of $p=1$ and $p=0.5$. In the case of very loose coupling, the ion and electron fluids are permitted to experience large relative motions and act as if almost independent from each other, thus the partial pressure of 0.5 in the single fluid case is an appropriate comparison to the ion fluid. In the case of very tight coupling, the converse is true and the two fluids behave as one, thus a partial pressure of one in the hydrodynamic case is the most appropriate comparison. In this manner, the $p=1$ and $p=0.5$ hydrodynamic results create an envelope of theoretically expected behaviour for the RMI in the ion fluid.
This section closes with a discussion of the relative importance of the key mechanisms. 

\begin{figure}
	\centerline{\includegraphics[width=\linewidth]{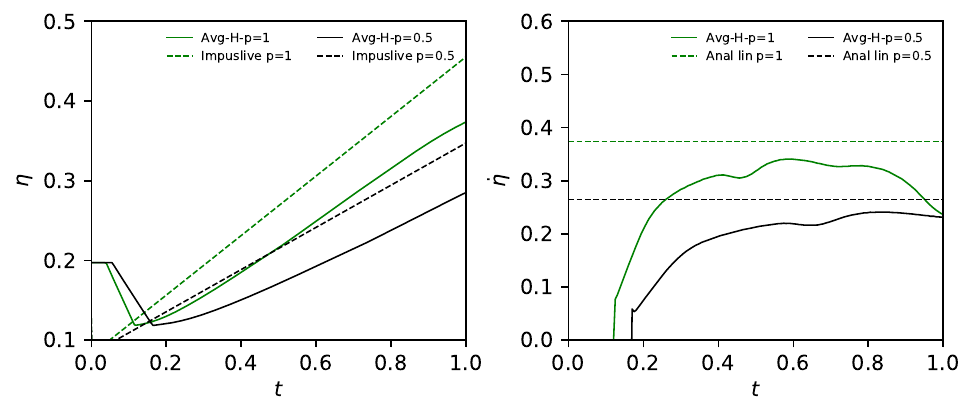}} 
	\caption{Hydrodynamic simulation results for a) cumulative circulation $\Gamma_Z$, b) total instantaneous generation of circulation $\dot{\Gamma_{Z, total}}$, c) interface width $\eta$, and d) growth rate $\dot{\eta}$, for all scenarios with $d_S = 10$ ($\lambda_D = 2\times10^{-1}$). Richtmyer's impulsive model \citep{richtmyer1960taylor} is provided for comparison. }
	\label{fig:hydro_analytical_interface_statistics_comparison}
\end{figure}

The simulation data is presented throughout this section with the format of figures \ref{fig:dD_0p2_density-ion-electron_all_interfaces} and \ref{fig:dD_0p2_interface_statistics_comparison_all}. The former shows the fluid mass density distributions at simulation end for each scenario and the latter shows time series of interface statistics. The interface statistics used are the circulation ($\Gamma_z$, on a half period of the interface), generation of circulation ($\dot{\Gamma_{Z, total}}$, total), and the ion fluid DI amplitude width ($\eta$) and growth rate ($\dot{\eta}$). The vorticity equation 
\begin{align}
\frac{\partial \mathbf{\omega}}{\partial t} = -\mathbf{\omega}\left(\div{\mathbf{u}}\right) -\left(\mathbf{u} \vdot \grad{}\right) \mathbf{\omega} + \frac{1}{\rho_\alpha^2}\left(\grad{\rho_\alpha}\cross\grad{p_\alpha}\right) + \curl{\left(\sqrt{\frac{2}{\beta}}\frac{q_\alpha}{d_S m_\alpha} \left(c\mathbf{E}+\mathbf{u}_\alpha \cross \mathbf{B}\right)\right)}\label{eq:vort_early}
\end{align}
shows the terms which combine to give the total time rate of change of vorticity, related to circulation by 
\begin{align}
\frac{\partial \Gamma } {\partial t} = \frac{\partial } {\partial t} \iint_S \vec{\omega}\cdot d\vec{S} = \iint_S \frac{\partial \vec{\omega}}{\partial t} \cdot d\vec{S}. 
\end{align}
For a small finite area in the x-y plane we have $\frac{\partial \Gamma_{Z}}{\partial t} = A\frac{\partial \mathbf{\omega_z}}{\partial t}$ where $A$ is the area of fluid enclosed by a loop with vorticity. If we take a cell to be the area enclosed by the enclosing loop, we can estimate the instantaneous circulation generated by multiplying terms from equation \ref{eq:vort_early} by the area of a cell. Summing the contributions from the cells that comprise the interface is how the circulation generation terms are calculated. Likewise, the circulation is found from the sum of vorticity multiplied by cell area for all cells comprising the interface.
\begin{align}
\frac{\partial \Gamma_{Z}}{\partial t} = \sum_i A_i\frac{\partial \omega_Z}{\partial t}, \text{ for } i \in \text{Interface cells}\\
\Gamma_{Z} = \sum_i A_i \omega_{Z,i}, \text{ for } i \in \text{Interface cells}
\end{align} 
The interface cells are determined by their volume of fluid tracer value, $\varrho$, that is convected with the flow. The interface width is calculated from the tracer average across the y-dimension where the X coordinates of $\varrho_y \in [0.05, 0.95]$ define the interface width.
\begin{align}
\varrho_y = \frac{\int \varrho dy}{\int dy}
\end{align}
The maximum and minimum transition points are then used to define the interface width. Numerical derivatives are then used to find the temporal growth rate of the interface from the interface width values.

The Canny shock detector proposed by \citet{fujimoto2019canny} was used to identify shocks travelling across the interface. The method is based off the Canny edge detection algorithm for image analysis. This detector was used rather than conventional methods because it relied on a simple principle that would be applicable despite strong EM influences. The algorithm steps through 1) computing pressure gradient magnitudes and direction, 2) collection of candidates --- local maximum search and filter, 3) and assessment of candidate points (comparing to shock jump conditions with an error tolerance). The Canny shock detector produced reasonable results that are used in the analysis for tracking the traversal of transverse shocks across the interface.

\begin{figure}
	\centerline{\includegraphics[width=\linewidth]{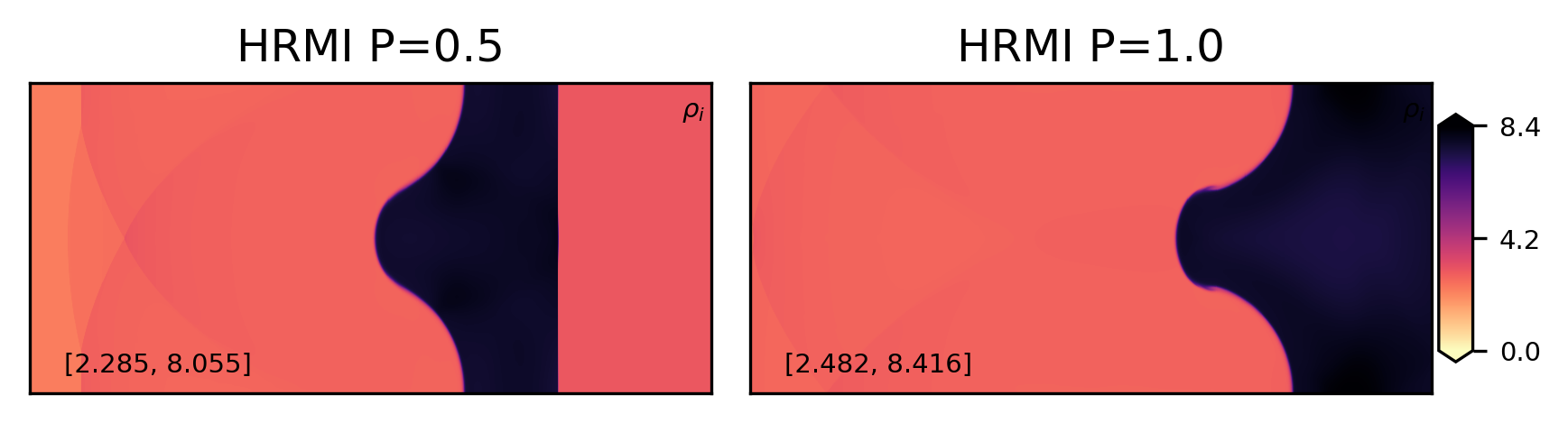}}
	\caption{Reference single-fluid hydrodynamic solution for the RMI showing the simulation end time ion density distribution. Bracketed tuples indicate the minimum and maximum each contour. }
	\label{fig:ref_HRMI_MHD_solution}
\end{figure}

The interface types represented on the time series plots are denoted as followed: H-p=1 --- single-fluid hydrodynamic reference solution with pressure of 1, H-p=0.5 --- single-fluid  hydrodynamic reference solution with pressure of 0.5, I-$d_S$-10 --- isotope RMI with skin-depth of 10, S-$d_S$-10 --- hydrogen-helium species RMI with skin-depth of 10, L-$d_S$-10 --- hydrogen-lithium species RMI with skin-depth of 10, and T-$d_S$-10 --- thermal RMI with skin-depth of 10. Figure \ref{fig:ref_HRMI_MHD_solution} shows the density contour plot for the reference single-fluid solutions as discussed in the preceding paragraph. 

When interpreting the results of the two-dimensional cases terminology introduced in figure \ref{fig:ProblemDescription} is used: (1) whenever the y-dimension is discussed in context of the DI, the lower x-symmetry plane is considered, (2) the spike refers to a region of high density fluid penetrating lower density, (3) the opposite case to point 2 is referred to as a bubble, (4) the region transitioning between the spike and bubble is referred to as the neck, and (5) the bottom left numbers within brackets on contour plots are the minimum and maximum values for that figure. 

It is useful to consider which regime the plasma RMI is in through the analogous hydrodynamic problem. The $p=1$ and $p=0.5$ hydrodynamic results for width and growth rate are compared to Richtmyer's impulsive model \citep{richtmyer1960taylor} in figure \ref{fig:hydro_analytical_interface_statistics_comparison}. Considering also figure \ref{fig:ref_HRMI_MHD_solution}, we expect the hydrodynamic cases and likely the plasma RMI will be transitioning between the linear and non-linear growth phases.

\subsubsection{Loosely coupled fluids $d_S=10$}\label{sec:dS10_discussion}
\begin{figure}
	\centerline{\includegraphics[width=\linewidth]{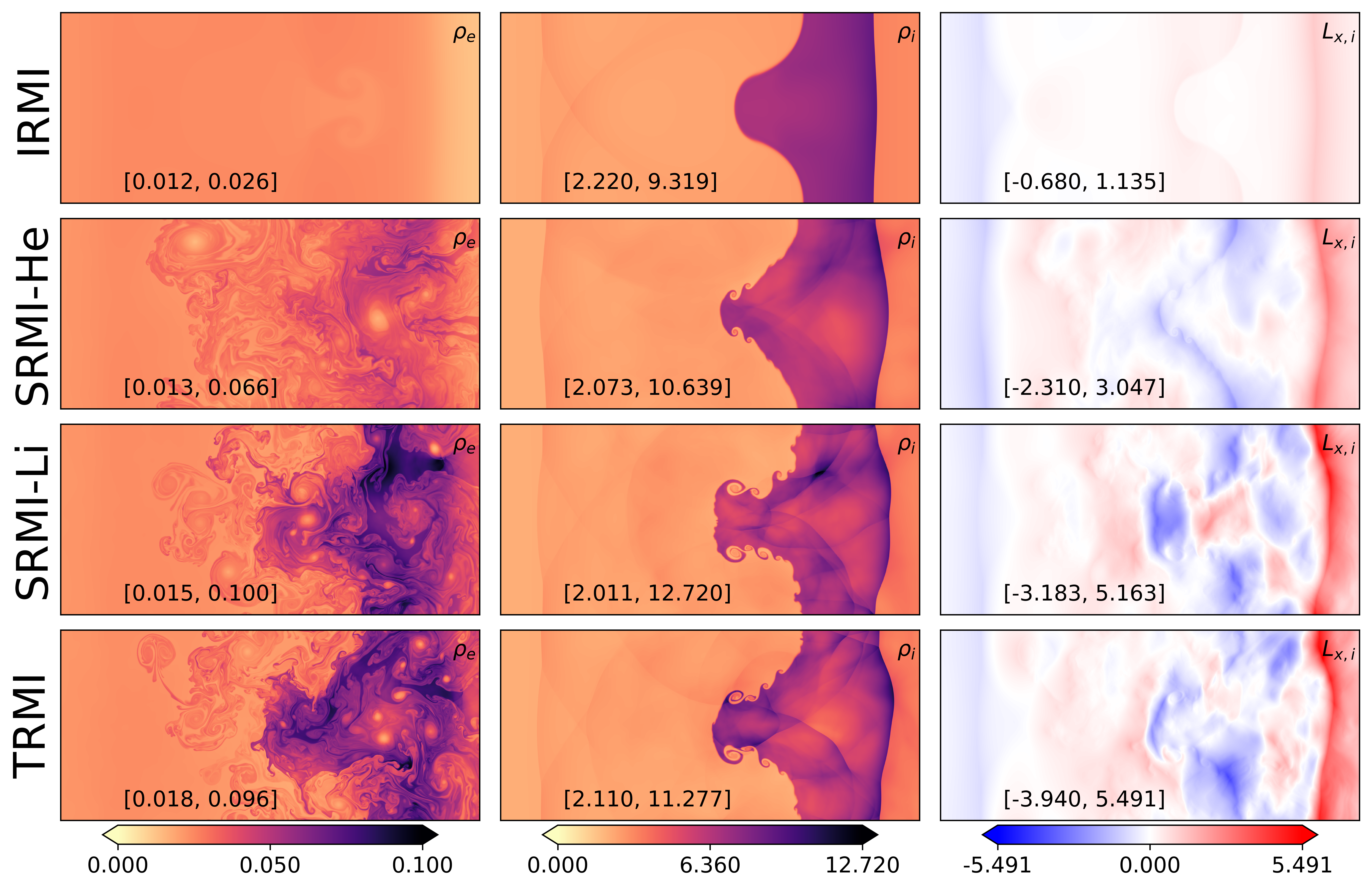}}
	\caption{Electron and ion fluid mass density and x-Lorentz force contours at simulation end time with Debye length ($\lambda_D = 2\times10^{-1}$). Bracketed tuples indicate the minimum and maximum values on each contour. }
	\label{fig:dD_0p2_density-ion-electron_all_interfaces}
\end{figure}

\begin{figure}
	\centerline{\includegraphics[width=\linewidth]{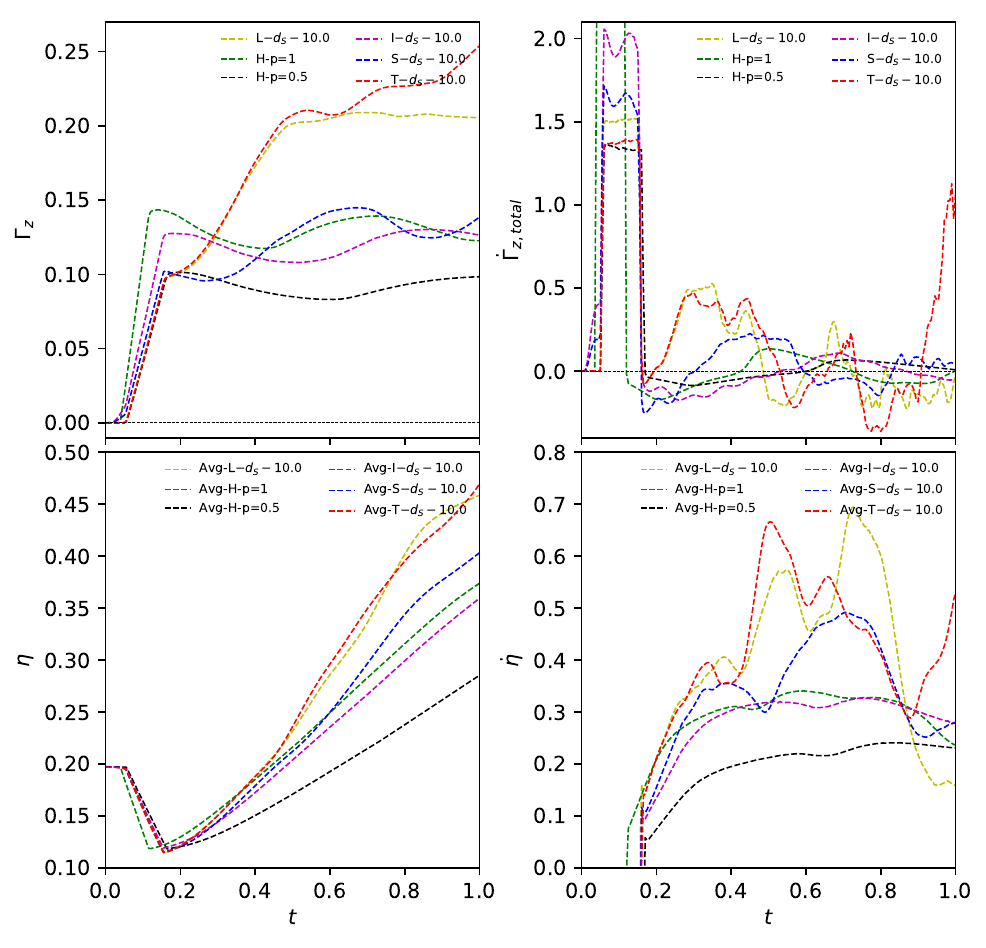}} 
	\caption{The evolution of the IFDI a) cumulative circulation $\Gamma_Z$, b) total instantaneous generation of circulation $\dot{\Gamma_{Z, total}}$, c) interface width $\eta$, and d) growth rate $\dot{\eta}$, for all scenarios with $d_S = 10$ ($\lambda_D = 2\times10^{-1}$). }
	\label{fig:dD_0p2_interface_statistics_comparison_all}
\end{figure}

A large non-dimensional skin depth indicates a loose coupling between the electron and ion fluids that ultimately enhances secondary instabilities. The EM forces couple the two fluids together but at this skin depth significant relative motion between them is permitted. This relative motion is critical in the development of secondary instabilities because it allows the evolution of expansive EM fields and allows the electron to perturb the ion interface. Figure \ref{fig:dD_0p2_density-ion-electron_all_interfaces} shows the ion and electron fluids at simulation end along with the Lorentz force in the x-direction. Charged particles within EM fields experience a Lorentz forces that provides a continuous forcing to drive the Rayleigh-Taylor instability (RTI). Characteristic RTI features (mushrooms) can be seen on the ion fluid bubbles of the SRMI-Li case in figure \ref{fig:dD_0p2_density-ion-electron_all_interfaces}. EM fields can similarly generate torques due to the gradient of Lorentz forces which acts as a source of vorticity that can amplify or diminish the KHI depending on the sign or the vorticity. The KHI rollers are seen along the neck of the SRMI-He, SRMI-Li, and TRMI cases, where it is more severe in the latter two due to the greater EM forces and consequent torques. The relative motion also produces interactions between the fluids that introduce additional interface perturbations, exacerbating the primary RMI and secondary instabilities. The secondary instabilities augment the primary interface perturbation growth rates, increasing them, as well as contributing to the emergence of high wavenumber secondary interface perturbations. 


Problem symmetry is obviously broken in the electron fluid and some ion fluid features also indicate this. The root cause is the interaction of complex systems of many vortices and inhomogeneous machine precission error. The small numerical deviations across the domain due to rounding error are not homogeneous, these differences are then amplified by the complex vortex field. 


The reader will note that the isotope interface is distinctly restrained despite the permitted relative motion between the fluids. This relative motion is important for generation of EM fields that drive the secondary ERTI and can augment the evolution of the secondary KHI instability. The Secondary instabilities occurring in 2D introduce translational motion in directions other than the shock propagation direction as well as rotational motion. These effects on relative motion are in addition to the influences discussed for 1D i.e. oscillatory motion, differing shock speeds, and fluid charge density discontinuities across the interface, see section \ref{sec:1D_base_flow}. Furthermore, the development timescales of the secondary instabilities in electron and ion fluids are more disparate than the shock timescales, therefore relative motion with the instabilities is greater than without. Since the electron timescales for instabilities is much smaller than ions, the presence of a significant (or indeed non-zero) initial electron fluid density ratio produces much greater relative motion than when only the ion fluid possesses a density ratio. 


Figure \ref{fig:dD_0p2_density-ion-electron_all_interfaces} shows a dramatic difference in both evolution and severity of the IRMI compared to the other interface types. The primary factor in this difference is the absence of an initial electron fluid interface in the IRMI. This absence results in greatly reduced EM field generation and effective removal of ion interface excitation by the electron fluid prior to the ion shock arrival. The effect of the ERTI is subsequently reduced due to its driving Lorentz force being proportional to the EM fields. The KHI is also reduced due to decreased shear and fewer seed points along the interface which would have been excited by energetic electron fluid perturbations. All these changes produce a slower amplitude growth rate that aligns closely with the single-fluid reference case (green dashed line labelled H-p=1), as seen in Figure \ref{fig:dD_0p2_interface_statistics_comparison_all}. 

\begin{figure}
	\centerline{\includegraphics[width=\linewidth]{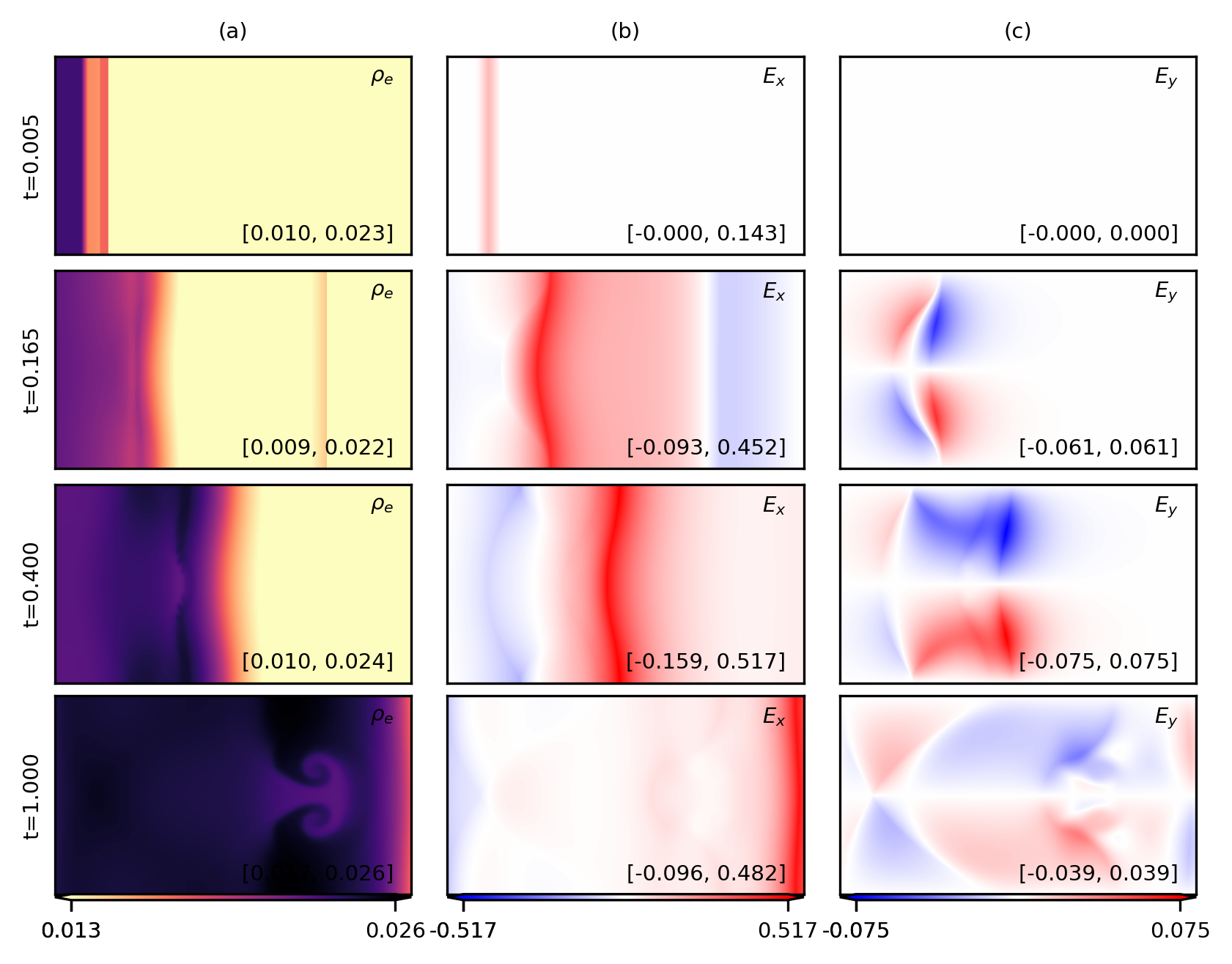}}
	\caption{Evolution of the IRMI (a) electron density contour, (b) x- and (c) y- electric field, left to right, for $\lambda_D = 2\times10^{-1}$. Bracketed tuples indicate the minimum and maximum each contour. }
	\label{fig:IRMI_0p2_electron_bubble}
\end{figure}

An interesting aspect of the isotope case is the bubble that forms on the x-symmetry line in the electron fluid DI, where all other interfaces produce a spike. The bubble is more clearly seen in figure \ref{fig:IRMI_0p2_electron_bubble} in the $\rho_e$ contour plot at $t=1$. Initially in the electron fluid there exists only a shock and a homogeneous and stationary density field. As the simulation begins, the breakdown of the electron shock and the differing propagation speed of the ion shock and the remaining electron waves produces a narrow-planar band of higher density in the electrons, figure \ref{fig:IRMI_0p2_electron_bubble} up to $t=0.005$. The acceleration of this narrow band of electrons at early times is due to the x-electric field. This acceleration is aligned with the band's density gradient and therefore does not produce any baroclinic torque. However, after the ion shock processes the perturbed ion-fluid DI, figure \ref{fig:IRMI_0p2_electron_bubble} $t=0.165$, appreciable y-Electric fields are produced, consequently misaligning the directions of the pressure and density gradients, generating baroclinic torque. The electric field, which dominates the Lorentz force in these simulations, is opposite signed either side of the x-symmetry line, producing opposite signed baroclinic torque either side of the x-symmetry line. The result is positive and negative baroclinic torque above and below the x-symmetry line, acting to contort the narrow band of electrons into a bubble. By $t=0.4$, after the initial contortion of the narrow band by the y-Electric field, the x-electric field is now misaligned with the density gradient along the interface. The long lived x-electric field continues to drive continual baroclinic torque generation that reinforces the bubble. The bubble continues to grow until simulation end since there is no significant counteracting torque.

The electron fluid flow field that evolves in the IRMI is not produced n the other interface types and may provide a unique seed for instabilities at later times. The formation of a bubble or spike on the x-symmetry line depends on the relative magnitude and orientation of the circulation produced by the electron shock passage and the ERTI (note the baroclinic torque is one source of circulation). In the TRMI and SRMI, the circulation deposited by the ERTI (at early simulation time) is lesser in magnitude and opposes the electron shock deposited circulation, and so it has no effect on the direction of the spike. However, in the IRMI the ERTI circulation does not encounter an appreciable opposing circulation in the electron fluid and so it generates a bubble where all other interface types form a spike. The IRMI bubble then grows and develops features along its extent which will evolve the opposite manner to the analogous cases in the TRMI and SRMI for the same continuous or impulsive forcing due to the direction of the density gradient. This distinctly opposite evolution may lead to unique and fundamentally different behaviours at late time. 

Figure \ref{fig:dD_0p2_density-ion-electron_all_interfaces} shows the Li-3 SRMI final state is closer to the TRMI results of \citet{bond2017richtmyer} than the He-3 SRMI case. The Li-3 species case exhibits three distinct qualities in the ion fluid when compared to the He-3 species case; a more severe RMI in terms of amplitude growth rate and final amplitude; a much more perturbed interface exhibiting more developed features (KHI rollers and ERTI spikes); and a narrower spike and neck region. To reiterate, the electron fluid density ratio greatly affects the overall plasma RMI since it governs the charge separation induced by the relative motion of the interfaces. The increase in electron density ratio in the Li-3 case to the levels of the thermal interface results of \cite{bond2017richtmyer} and subsequent aligning of RMI amplitude and growth rate suggests that the electron density ratio may play a dominant role in the plasma RMI development at large Debye lengths. This is a physically plausible mechanic when understood from the perspective of the secondary ERTI. The ERTI is driven by EM fields of which the electron fluid dynamics contribute to with their motion and separation from the ion fluid. Both of these factors are amplified in the case of a greater electron fluid density ratio. Likewise, the increased interface disturbance is due to the more energetic electron fluid exciting the ion fluid. However, the third distinct feature --- the narrowing of the spike in comparison to the thermal case --- must be a consequence of varying ion properties of temperature, number density, and charge between the initial parameters of the TRMI and Li-3 SRMI. 

\begin{figure}
	\centerline{\includegraphics[width=\linewidth]{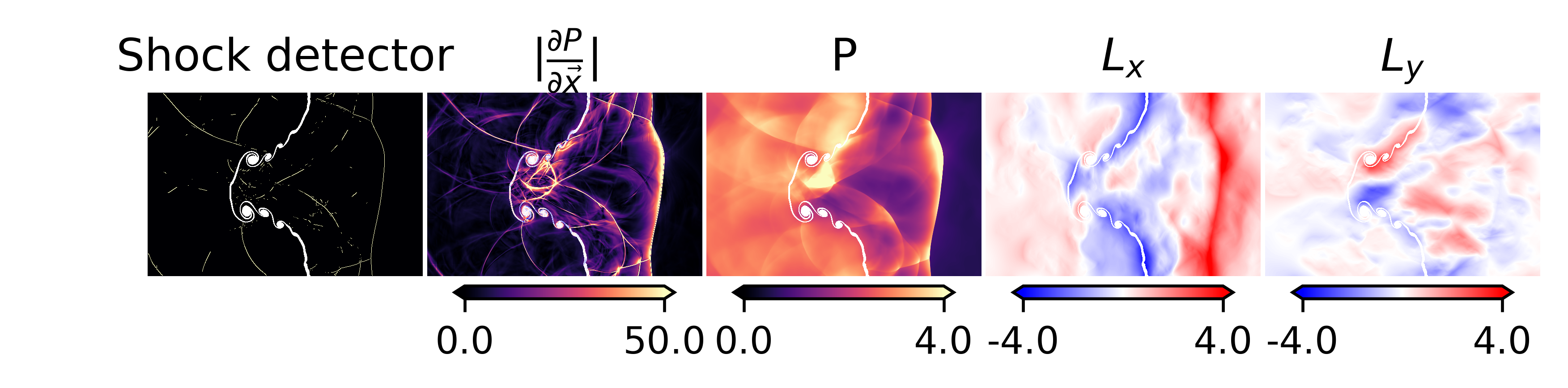}}
	\caption{TRMI (a) Canny shock detector results, (b) low filtered pressure gradient magnitude, (c) pressure, (d) x-Lorentz force, (e) y-Lorentz force for $\lambda_D = 2\times10^{-1}$ at $t-0.9$.}
	\label{fig:TRMI_dD_0p2_Canny_narrowing}
\end{figure}

\begin{figure}
	\centerline{\includegraphics[width=\linewidth]{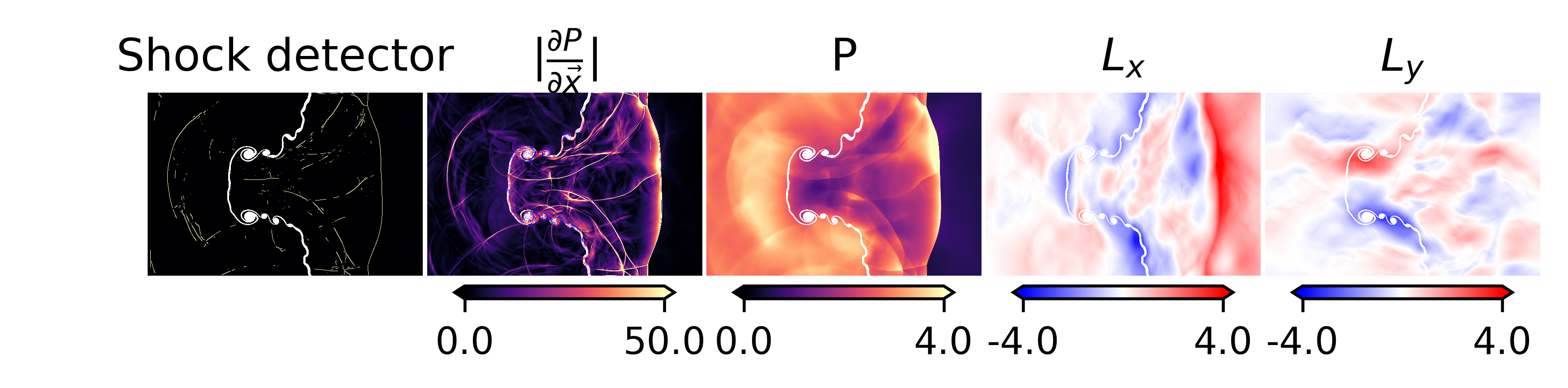}}
	\caption{SRMI-Li (a) Canny shock detector results, (b) low filtered pressure gradient magnitude, (c) pressure, (d) x-Lorentz force, (e) y-Lorentz force for $\lambda_D = 2\times10^{-1}$ at $t-0.9$}
	\label{fig:SRMI_Li_dD_0p2_Canny_narrowing}
\end{figure}



The thermal and lithium species cases are very similar in their evolution owing to their identical electron fluid and ion fluid density ratio. The TRMI does, however, produce a greater amplitude growth during simulation time, figure \ref{fig:dD_0p2_interface_statistics_comparison_all} $\eta$ time series. Figure \ref{fig:dD_0p2_density-ion-electron_all_interfaces} shows the SRMI-Li has a narrower neck than the TRMI at final time, the neck width of the TRMI (0.52) is 30\% greater than the SRMI-Li (0.4) at their narrowest points. Figures \ref{fig:TRMI_dD_0p2_Canny_narrowing} and \ref{fig:SRMI_Li_dD_0p2_Canny_narrowing} show the symmetry of the SRMI-Li case is coherent whereas the TRMI case has broken symmetry at large scale. 
The narrowing effect appears to be influenced by Lorentz forces and transverse shocks. 
The broken symmetry in the TRMI affects shock refraction and EM field evolution, thereby influencing the shock strength and direction, and Lorentz forces driving the fluid flow. 
In regard to a difference in shock refraction, the SRMI-Li (figure \ref{fig:SRMI_Li_dD_0p2_Canny_narrowing}), shows two coherent shocks having just traversed the interface (from both y-boundaries) and moving towards the line of symmetry, however, the TRMI (figure \ref{fig:TRMI_dD_0p2_Canny_narrowing}) shows only one shock. 
Additionally, the y-Lorentz force acting along the spike is positive (negative) when above (below) the axis of symmetry, leading up to and after the narrowing of the spike working to widen the spike. 
Beyond the Lorentz force, the only linear drivers of motion within the fluid are the existing pressure gradients. 
Inspection of the magnitudes of Lorentz forces show that the Li-3 SRMI experiences weaker y-Lorentz forces, which act to widen the spike, than the TRMI. Furthermore, there is a lower pressure behind the ion fluid DI in the Li-3 SRMI and greater pressure gradients, which mostly act to narrow the spike. 
These factors result in the much more pronounced narrowing in the Li-3 case, from the spike to the neck, as opposed to only the spike in the TRMI.




The instability evolution will change significantly when moving from the present 2D3V work to a full 3D3V simulation. 
The coherence observed in the 2D case e.g. KHI rollers will be subject to 3D instabilities and lose their 2D coherent structures. This type of interaction is likely for the primary instability spikes and bubbles also. In addition, turbulent mixing zones will likely evolve in the longer time scales. These effects will likely culminate in reduced instability growth. 

It would be practical, using the same effective resolution, to produce a 3D simulation of the ideal MFP RMI with Cerberus. It is estimated that the 3D simulation would require 33 M cells, increasing from approximately 2 M cells for the current work. This is only possible due to the AMR framework provided by AMReX. However, simulations with transport processes (collisions) seem to be out of reach for 3D simulations due to the severe time step constraint imposed by diffusive and conductive time scales, and the current configuration of the code (explicit and only second order accurate in time). Investigation of multi-mode perturbations are possible and could be easily implemented through designing the initial density interface geometry accordingly.

\subsubsection{Moderately coupled fluids $d_S=1$}
\begin{figure}
	\centerline{\includegraphics[width=\linewidth]{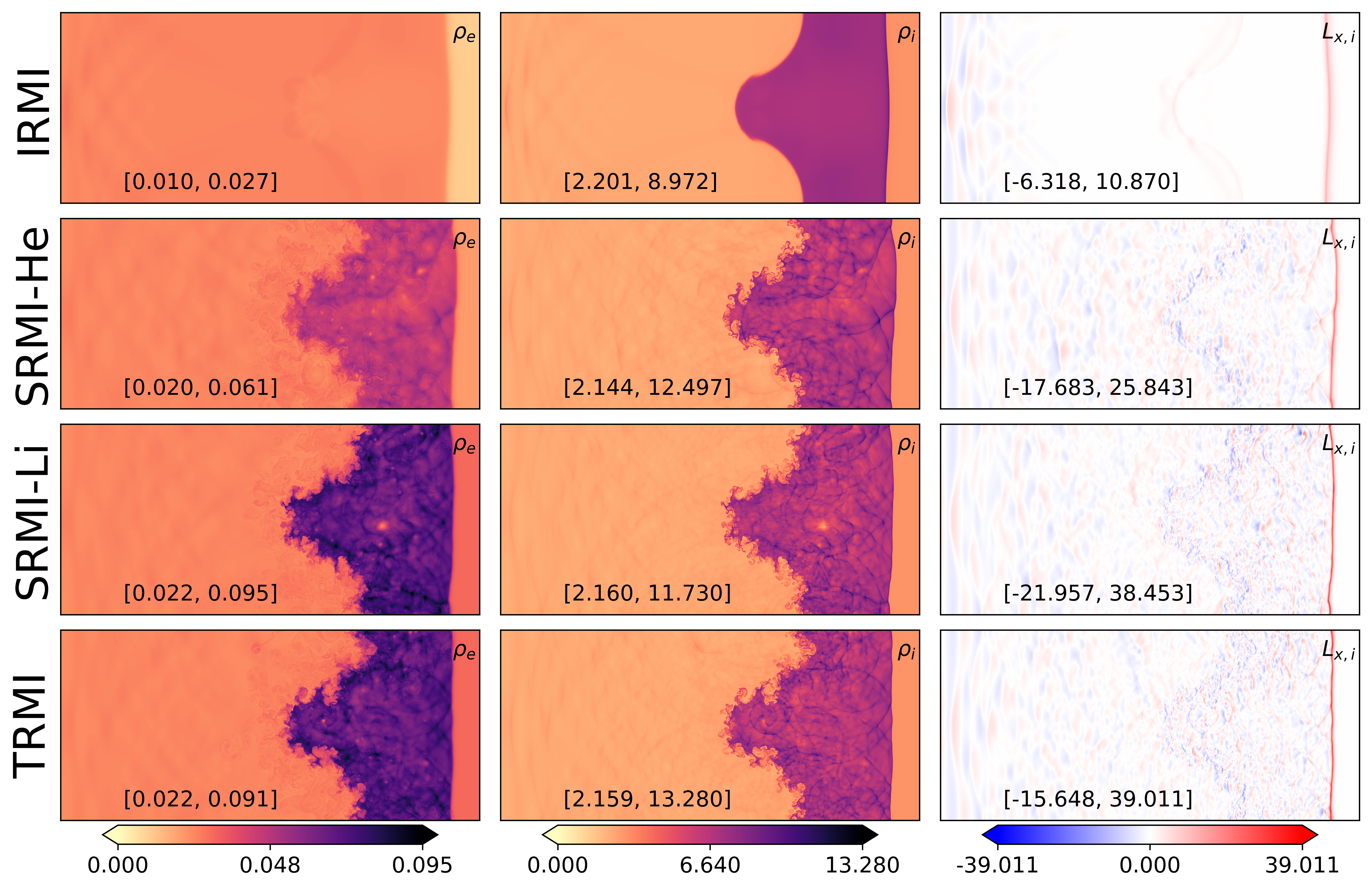}}
	\caption{Electron and ion fluid mass density contours at simulation end time $d_S = 1$ ($\lambda_D = 2\times10^{-2}$). Bracketed tuples indicate the minimum and maximum each contour. }
	\label{fig:dD_0p02_density-ion-electron_all_interfaces}
\end{figure}
\begin{figure}
	\centering
	\centerline{\includegraphics[width=\linewidth]{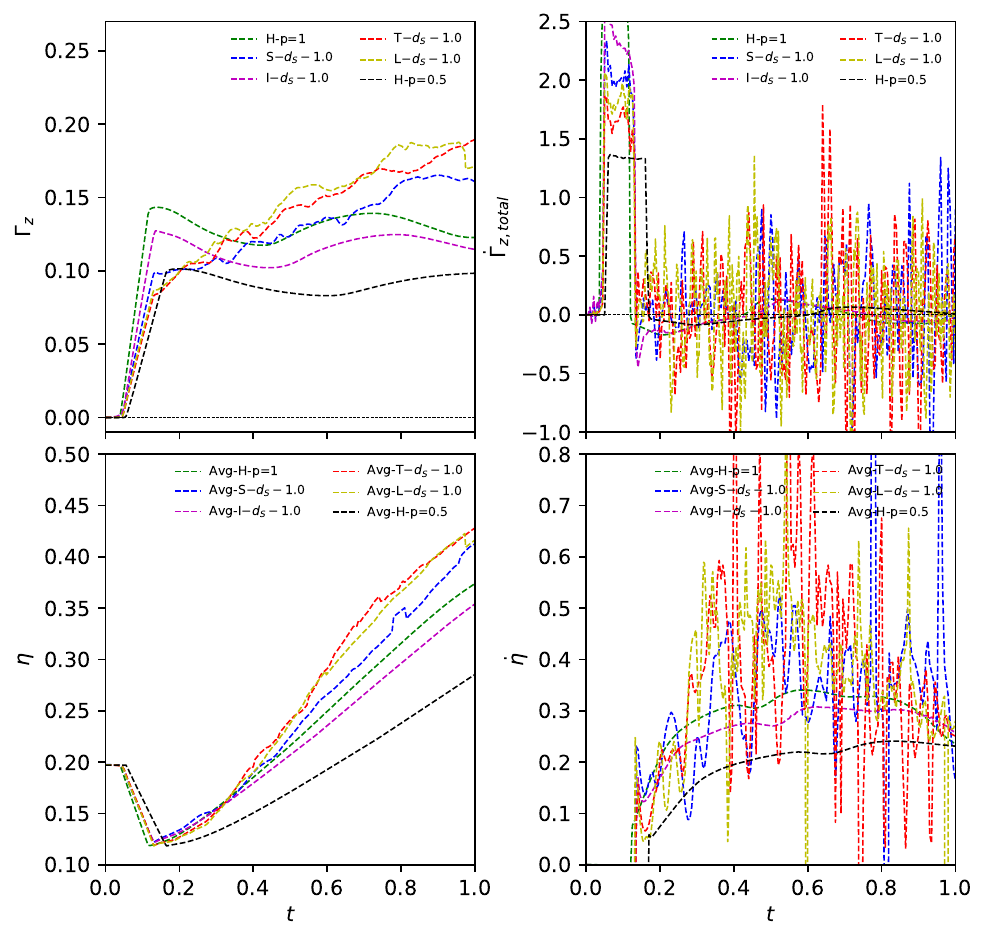}} 
	\caption{The evolution of the IFDI a) cumulative circulation $\Gamma_Z$, b) total instantaneous generation of circulation $\dot{\Gamma_{Z, total}}$, c) interface width $\eta$, and d) growth rate $\dot{\eta}$, for all scenarios with $d_S = 1$ ($\lambda_D = 2\times10^{-2}$).}
	\label{fig:dD_0p02_interface_statistics_comparison_all}
\end{figure}
The results of the reduction in skin depth, figures \ref{fig:dD_0p02_density-ion-electron_all_interfaces} and \ref{fig:dD_0p02_interface_statistics_comparison_all}, show the SRMI-He was the only scenario to produce an increase in interface amplitude when compared to the loosely couple fluid case, figures \ref{fig:dD_0p2_density-ion-electron_all_interfaces} and \ref{fig:dD_0p2_interface_statistics_comparison_all}. The loosely and moderately coupled cases develop a similar interface circulation deposition profile until $t\approx0.7$ where the moderately coupled case ($\lambda_D=2\times10^{-2}$) sees an increased deposition of circulation on the interface, where the loosely coupled case drops, compare figures \ref{fig:dD_0p2_interface_statistics_comparison_all} and \ref{fig:dD_0p02_interface_statistics_comparison_all}. 
The late time rally in circulation produces a final interface amplitude width that is greater than the preceding Debye length case, 0.412 compared to 0.402 non-dimensional lengths. The combination of increased amplitude width and development of high-wave number interface perturbations make the $\lambda_D=2\times10^{-2}$ SRMI-He case more severe than the loosely coupled case in terms of the detrimental effects for ICF experiments. 

\begin{figure}
	\centerline{\includegraphics[width=\linewidth]{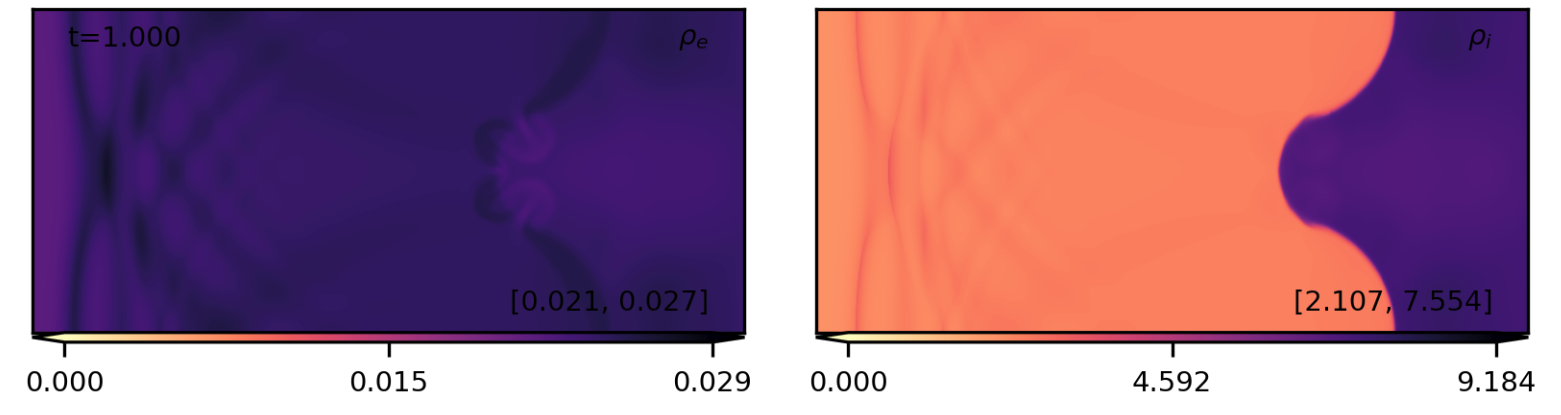}}
	\caption{IRMI $\lambda_D=2\times10^{-2}$ ion and electron density contours at simulation end showing the ERTI and TRS generated perturbations. Bracketed tuples indicate the minimum and maximum each contour. }
	\label{fig:dD_2e-2_IRMI_final_density_time_single} 
\end{figure}
The IRMI in the ion fluid produces a reduced interface amplitude, when compared to the loosely coupled case, but shares the same fluid and EM features. The electron fluid, however, is different, producing a greater number of ERTI features on the electron fluid DI. Figure \ref{fig:dD_2e-2_IRMI_final_density_time_single} shows the inverted primary electron fluid bubble centred on the x-symmetry line is still present (observed in the larger Debye length case for the IRMI) although diminished in size. Additionally, there is a secondary spike and bubble adjacent to the primary bubble on either side of the x-axis of symmetry. The secondary spike and bubble are caused by the same mechanism as discussed in section \ref{sec:dS10_discussion} but with the smaller and alternating EM field regions seeding the alternating pattern of bubble and spike that is amplified by the imprint of the ion fluid transverse-reflected shocks (TRS) on the electron fluid.
These ion fluid TRS imprints are not significant in the $\lambda_D=2\times10^{-1}$ IRMI case because of the weaker ion-electron coupling and the weak strength of the TRS. While the $\lambda_D=2\times10^{-2}$ case has some distinctions from the $\lambda_D=2\times10^{-1}$ case, these do not appear to be caused by a unique mechanism. In fact, had there been a greater density gradient in the electron fluid it is likely the inversions along the imprinted DI would not occur because its source, the TRS, does not seem strong enough in the cases studied thus far to produce enough circulation to overpower the initial deposition of circulation by incident ion and electron shocks. 

\begin{figure}
	\centering
	\begin{subfigure}[b]{.45\linewidth}
		\includegraphics[width=\linewidth]{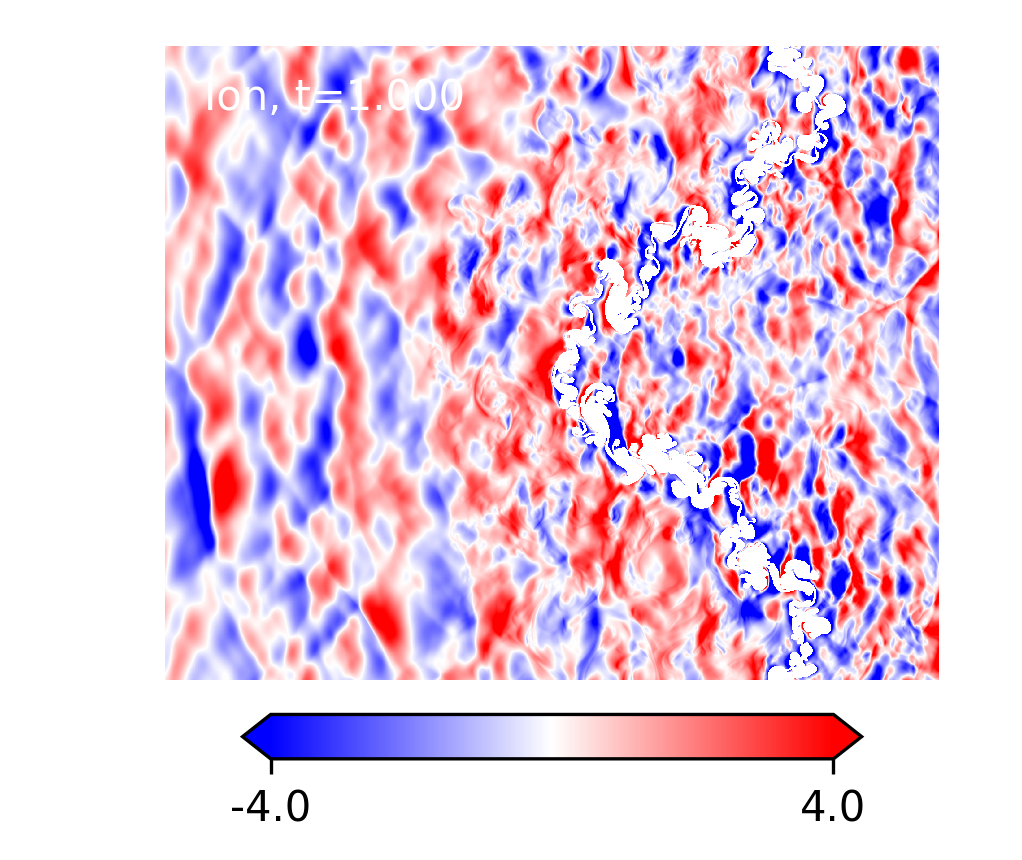}
		\caption{$\lambda_D=2\times10^{-2}$ He-3 SRMI x-Lorentz force at $t=1$ with interface contour whited out.}\label{fig:dS-1-SRMI-He-xLorentz-example} 
	\end{subfigure}
	\begin{subfigure}[b]{.45\linewidth}
		\includegraphics[width=\linewidth]{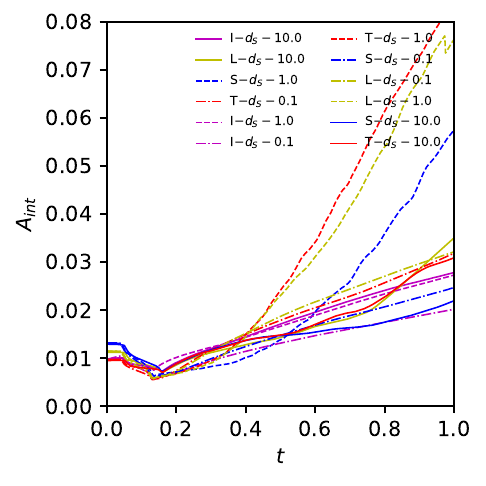}
		\caption{Interface area for all interface and skin depth scenarios. }\label{fig:interface_area_comparisons}
	\end{subfigure}
	\caption{An example of the intense and alternating regions of Lorentz force causing the erratic interface statistics for a smoothly evolving interface area.}
	\label{fig:dS_1_jagged_explanation}
\end{figure}

A reduction in skin depth shrinks the spatial scale of EM features and increases their intensity, overall reducing the influence on EM driven secondary instabilities. The ion x-Lorentz forces (directly proportional to the EM fields) shown in figure \ref{fig:TRMI_dD_0p2_Canny_narrowing} and figure \ref{fig:dS_1_jagged_explanation} are indicative of the EM fields for the $\lambda_D=2\times10^{-1}$ and $\lambda_D=2\times10^{-2}$ cases, respectively. In the $\lambda_D=2\times10^{-1}$ case (section \ref{sec:dS10_discussion}), the ion and electron fluids behave as if they were almost independent of each other, with their large separations giving rise to coherent and expansive regions of EM fields and forces. These EM forces compound the effects of the RMI to produce greater growth rates and amplitudes. In the $\lambda_D=2\times10^{-2}$ case, the self-generated EM fields exist in much smaller regions of alternating and intense strength. The net effect on interface growth with fine-grained regions of alternating EM fields and consequent Lorentz forces is lesser than the case of larger alternating regions of EM fields, even though the magnitudes of the Lorentz forces are greater. The ERTI for the moderately coupled fluids is thus reduced when compared to the tightly coupled case but is still a significant contributor to the interface growth rate. 

Increased coupling allows the electron fluid to excite the ion fluid interface with high wave number disturbances that are then amplified by the shock refraction and secondary instabilities. 
The electron fluid transfers (relative to the loosely couple case) more of its energy to the ion fluid, imparting high wavenumber content along the interface that is then amplified through deposition of circulation by the secondary ERTI and KHI. The high-wave number content is further amplified by shocks created from a refraction process not observed in the loosely coupled case. The numerous refracted shocks are observable in figure \ref{fig:dD_0p02_density-ion-electron_all_interfaces} behind the transmitted shock. These shocks result from the refraction of the incident ion-shock with the electron-excited ion fluid DI and the formation of triple points along the transmitted ion-shocks. The interaction of the transmitted shocks are clearly observable in the cellular patterns downstream of the ion fluid DI. The interface is multiply shocked, initially from the incident shock and then the shocks emanating from the triple point sweeping across the ion fluid DI.

The source of interface circulation deposition and growth rate appear to be highly transient phenomena. This is suggested by the time rate of change of circulation and interface amplitude in figure \ref{fig:dD_0p02_interface_statistics_comparison_all}. 2D Circulation contributions are given by the vorticity multiplied by relevant fluid area and the vorticity itself is governed by the equation
\begin{align}
\frac{\partial \mathbf{\omega}}{\partial t} = -\mathbf{\omega}\left(\div{\mathbf{u}}\right) -\left(\mathbf{u} \vdot \grad{}\right) \mathbf{\omega} + \frac{1}{\rho_\alpha^2}\left(\grad{\rho_\alpha}\cross\grad{p_\alpha}\right) + \curl{\left(\sqrt{\frac{2}{\beta}}\frac{q_\alpha}{d_S m_\alpha} \left(c\mathbf{E}+\mathbf{u}_\alpha \cross \mathbf{B}\right)\right)}.\label{eq:vorticity}
\end{align}
The sources of vorticity in the ideal case, ignoring stretching of vorticity due to velocity gradients and flow compressibility, are the EM fields and fluid pressure gradients (imposed by the initial shock and subsequent refracted shocks). Therefore the sources of vorticity and/or the interface area are highly transient. Figure \ref{fig:interface_area_comparisons} shows the time series of the interface area --- the interface area is defined as the sum of each cell's area where the cell contains between 0.05 and 0.95 volume of fluid tracer and is used to calculate the interface statistics. The simulation produces an interface area that varies smoothly with time, therefore, an incorrect identification of the interface is not the cause of the observed volatility in circulation. Rather, the cause is the highly energetic and variable nature of the $\lambda_D=2\times10^{-2}$ scenarios with regard to the shock refraction and EM field development. Figure \ref{fig:dS-1-SRMI-He-xLorentz-example} shows the highly variable x-Lorentz force acting on the ion fluid for the SRMI-He scenario, contributing to the variability in circulation generation on the interface. Figure \ref{fig:dD_0p02_density-ion-electron_all_interfaces} shows the numerous shocks generated from triple points, after refraction through the interface, which then interact with the interface once more. These numerous and incoherently propagating shocks provide a variable pressure and density along the interface that contributes to the highly variable generation rates of circulation in \ref{fig:dD_0p02_interface_statistics_comparison_all}. 

\subsubsection{Tightly coupled fluids $d_S=0.1$}

\begin{figure}
	\centerline{\includegraphics[width=\linewidth]{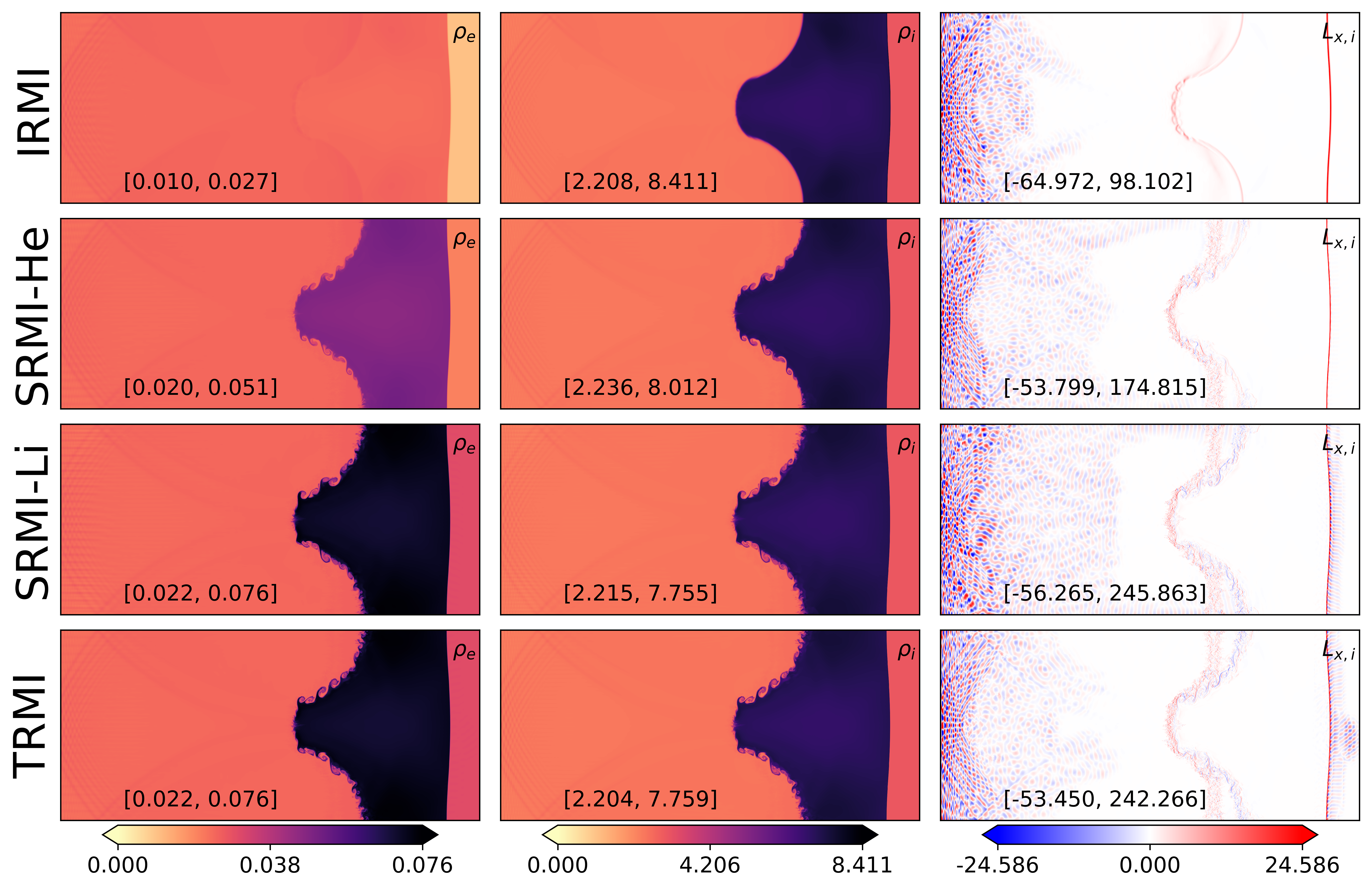}}
	\caption{Electron and ion fluid mass density contours at simulation end time $d_S = 0.1$ ($\lambda_D = 2\times10^{-3}$). Bracketed tuples indicate the minimum and maximum each contour. Note the contour plot scale is saturated due to the strong Lorentz forces surrounding the original transmitted shock front.} 
	\label{fig:dD_0p002_density-ion-electron_all}
\end{figure}

\begin{figure}
	\centerline{\includegraphics[width=\linewidth]{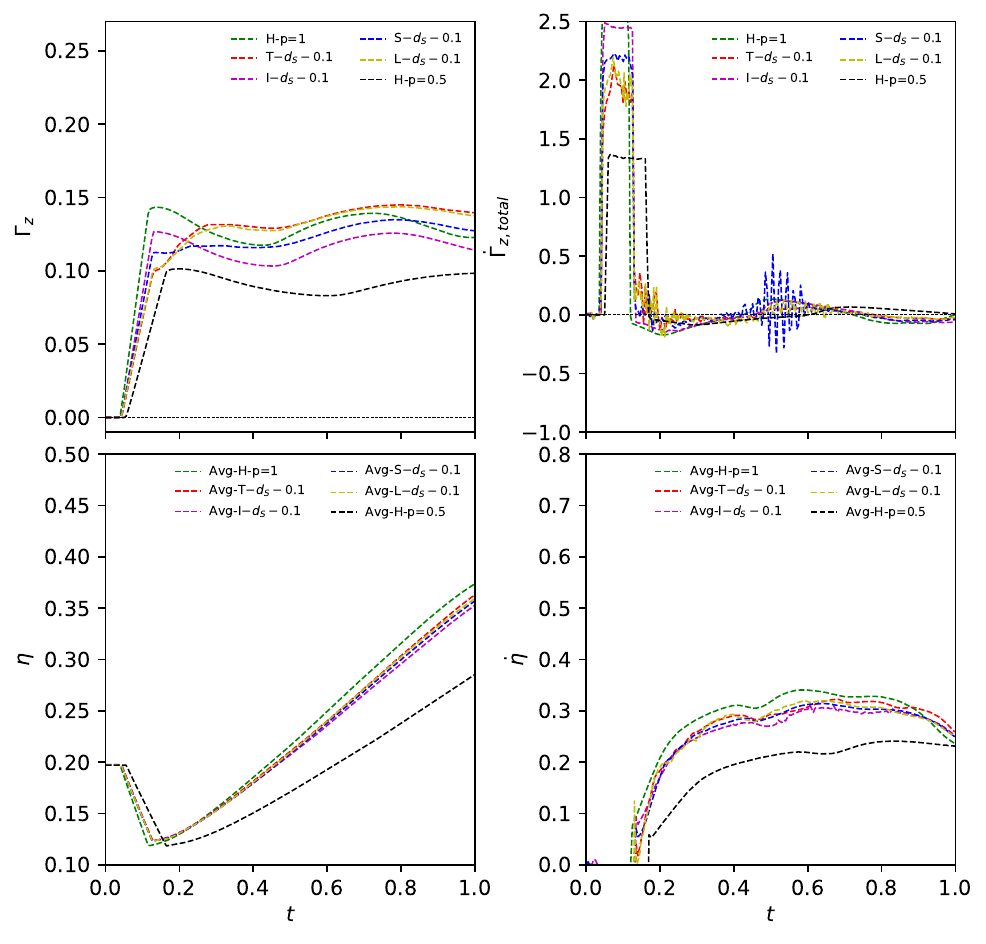}}
	\caption{The evolution of the IFDI a) cumulative circulation $\Gamma_Z$, b) total instantaneous generation of circulation $\dot{\Gamma_{Z, total}}$, c) interface width $\eta$, and d) growth rate $\dot{\eta}$, for all scenarios with $d_S = 0.1$ ($\lambda_D = 2\times10^{-3}$). }
	\label{fig:dD_0p002_interface_statistics_comparison_all}
\end{figure}
The smallest Debye length cases investigated appear to be approaching the single-fluid limit. Figure \ref{fig:dD_0p002_density-ion-electron_all} shows a high degree of resemblance between the ion and electron density contours in the TRMI and both SRMI cases. The IRMI also shows an imprint of the IFDI on the electron fluid. Indeed, the two fluids appear to be acting almost as one in all cases. The $\lambda_D=2\times10^{-3}$ produces TRMI and SRMI growth rate and interface amplitudes significantly below the reference single-fluid case, a significant difference from preceding $\lambda_D$ cases. The EM effects also appear to be reducing in magnitude across the interface. 

No new phenomena occur in the isotope RMI evolution in the case of tightly coupled fluids. The only modification, which is a common element of all interface cases is what seems to be an increase in the refraction frequency of the TRS. As mentioned above, the shock refraction process with a small Debye length produces many reflected waves at high frequency, in terms of emission from the interface rather than wave frequency, and in close spacing to each other. The combination of the ERTI and the imprint of the many closely spaced ion TRS on the electron fluid seems to produce many more perturbations that grow along the ion fluid DI spike and neck. Interestingly, the relative growth of high-wavenumber perturbations decreases from large to small Debye lengths, suggesting a decrease in the effectiveness of the ERTI and KHI. The decrease in secondary instabilities supports the idea that reducing differential movement between the fluids reduces the expanse of the EM fields, exposing the interface to many alternating regions of influence (consequent vorticity either reinforcing and opposing instabilities) instead of fewer regions of a definitive action. Overall the IRMI does not produce a large difference in ion fluid DI amplitude or growth rate across the Debye length trials, as seen from a comparison of figures \ref{fig:dD_0p2_interface_statistics_comparison_all}, \ref{fig:dD_0p02_interface_statistics_comparison_all}, and \ref{fig:dD_0p002_interface_statistics_comparison_all}.

\begin{figure}
	\centerline{\includegraphics[width=\linewidth]{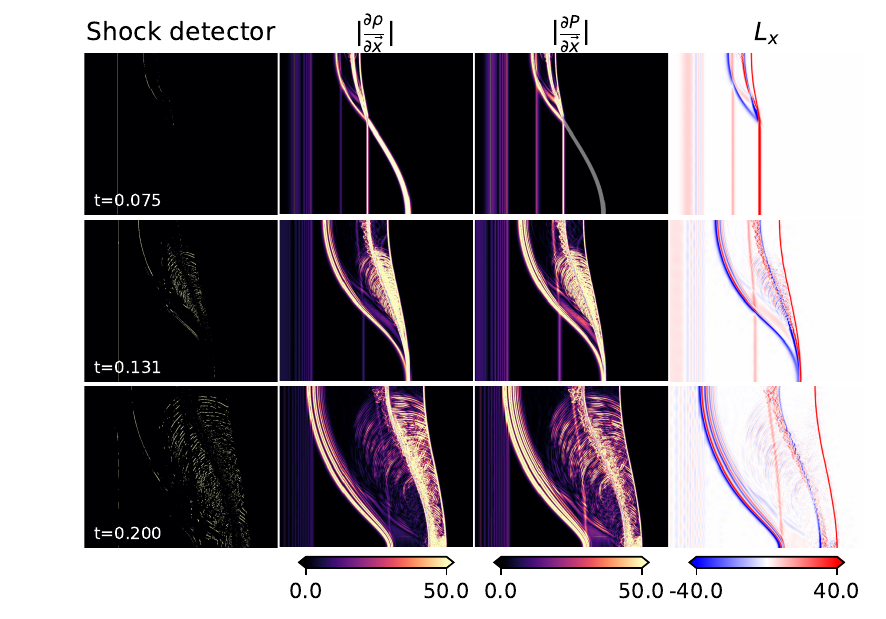}}
	\caption{SRMI-He (a) shock detector, (b) low filtered pressure gradients, and (c) x- and (d) y-Lorentz force components for $\lambda_D = 2\times10^{-3}$.}
	\label{fig:dD_0p002_CannyDetector_shock_refraction_SRMI}
\end{figure}

\begin{figure}
	\centerline{\includegraphics[width=\linewidth]{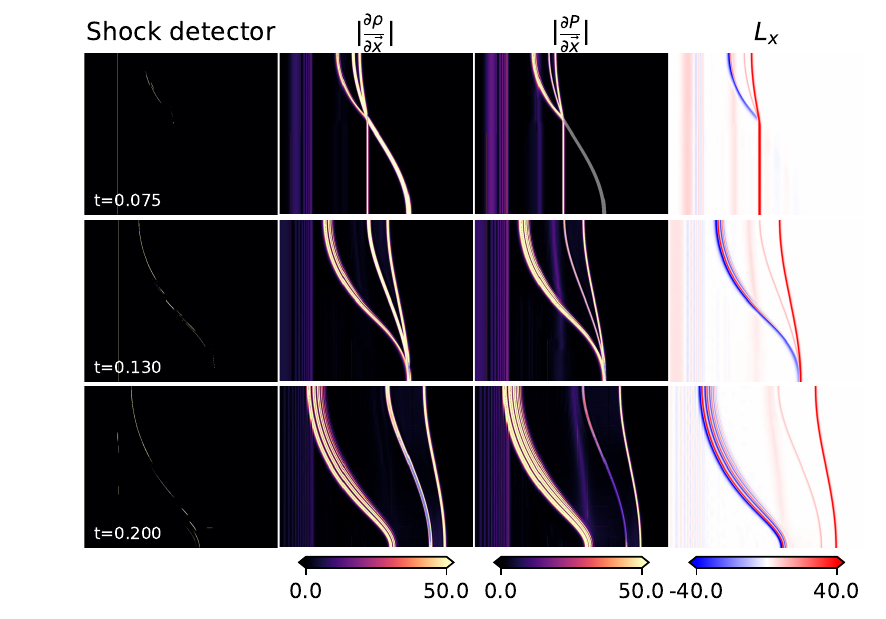}}
	\caption{IRMI (a) shock detector, (b) low filtered pressure gradients, and (c) x- and (d) y-Lorentz force components for $\lambda_D = 2\times10^{-3}$.}
	\label{fig:dD_0p002_CannyDetector_shock_refraction_IRMI}
\end{figure}
The $\lambda_D=2\times10^{-3}$ TRMI and both SRMI cases produce a significant MFP shock refraction effect despite approaching the single-fluid limit (p=1 reference case in figure \ref{fig:ref_HRMI_MHD_solution}).
The shock refraction of the incident shock on the density interface produces a reflected shock, transmitted shock, and high frequency waves. The high frequency waves, observed in figure \ref{fig:dD_0p002_CannyDetector_shock_refraction_SRMI}, are generated on the inclined portions of the interface, above and below the x-symmetry line, and propagate along the interface. 
The waves show strong structural resemblance in the pressure and density gradient contour plots.
As the high frequency waves pass \textit{along} the interface they seed high wavenumber instabilities. These waves then pass the interface and interact with the contact discontinuity --- generated from the spontaneous breakdown of the initial electron shock of which an imprint is made on the ion fluid due to high coupling of the fluids --- and excite the contact discontinuity. The contact discontinuity later impacts the interface, conforming around it but not penetrating during simulation time. The new interface perturbation development mechanism appears much less severe in size and number of features than the mechanism observed in the larger Debye length cases. However, the perturbations left behind are of a higher frequency. Note this interface perturbation development mechanism is different to that described for the moderately coupled case ($\lambda_D=2\times10^{-2}$) as that was driven by shocks formed from triple points after the initial refraction. Here in the $\lambda_D=2\times10^{-3}$ case, waves are generated directly from the refraction of the incident ion shock. The absence of this mechanism in the isotope RMI, see figure \ref{fig:dD_0p002_CannyDetector_shock_refraction_SRMI}  and \ref{fig:dD_0p002_CannyDetector_shock_refraction_IRMI}, and comparison to the single-fluid results suggests this is a phenomena unique to the two fluid model. 

\subsubsection{Qualitative hydrodynamic RMI features in the plasma RMI}\label{sec:qualitativePlasmaHydroRMI} 
\begin{figure}
	\centerline{\includegraphics[width=\linewidth]{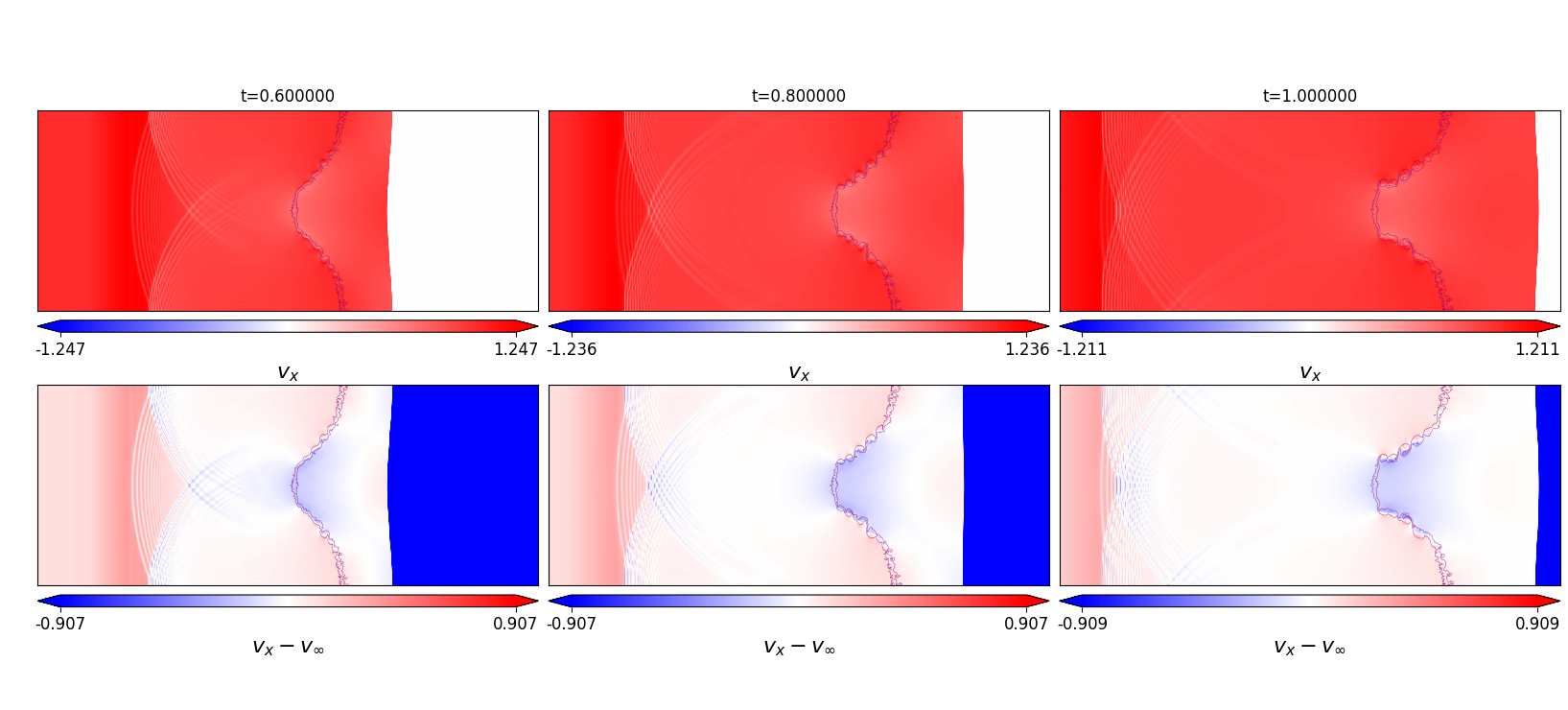}}
	\caption{The ion fluid x-velocity field in the stationary reference frame (top) and the background motion reference frame (bottom) for the TRMI small skin depth, $d_S = 0.1$.}
	\label{fig:dD_0p002_background_motion}
\end{figure}

The plasma RMI is subject to significant electromagnetic effects not present in the hydrodynamic case. 
Despite the additional Lorentz forcing terms, that provide additional sources of bulk fluid motion and direct circulation contributions, the typical qualitative features remain. 
We have observed already the plasma RMI instability evolves similarly to the hydrodynamic case, exhibiting the secondary KHI, but with the addition of Lorentz forcing and circulation generation, and the secondary ERTI. 
In the smallest skin depth case, an interesting parallel between the plasma and hydrodynamic RMI is that between the electromagnetic effects and the background motion. 
In the hydrodynamic RMI the fluid motion is most intense near to the interface and small away from it. 
A similar observation is made for the electromagnetic activity which is most intense near to the interface and negligible away from it (between the reflected and transmitted shocks), see figure \ref{fig:dD_0p002_density-ion-electron_all} x-Lorentz force. 
The behaviour of the background motion, in the small skin depth plasma RMI, follows that predicted by the hydrodynamic theory, figure \ref{fig:dD_0p002_background_motion}.

\begin{figure}
	\centerline{\includegraphics[width=\linewidth]{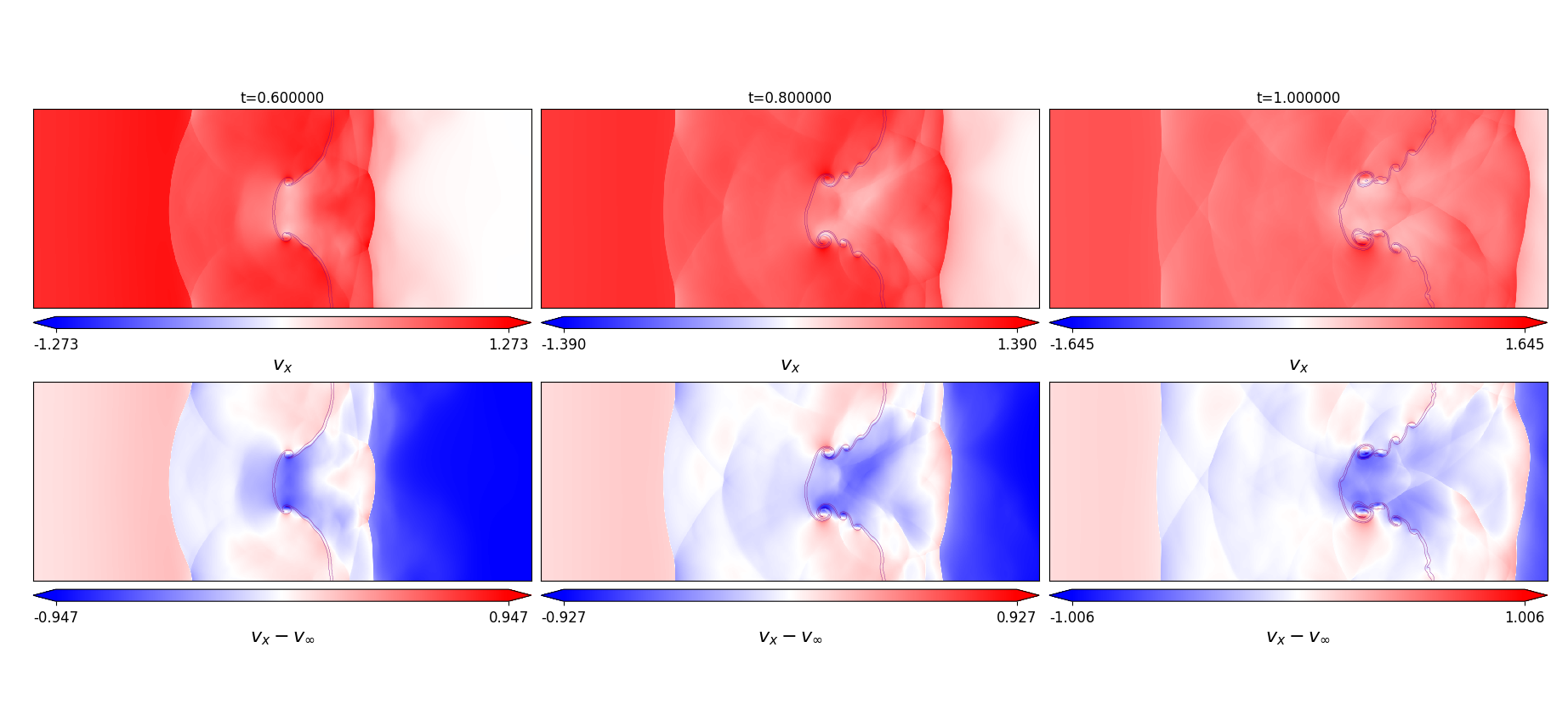}}
	\caption{The ion fluid x-velocity field in the stationary reference frame (top) and the background motion reference frame (bottom) for the TRMI large skin depth, $d_S = 10$, case.}
	\label{fig:dD_0p2_background_motion}
\end{figure}

The large skin depth behaviour of the plasma RMI diverges more from the hydrodynamic case due to the interaction of the electron and ion fluids which have loose electromagnetic coupling. 
The effect is akin to two miscible and separate RMI evolutions superposed with each other. 
The loose electromagnetic coupling allows expansive regions of electromagnetic fields to develop that affect both fluids. 
There is significant electromagnetic activity away from the interface, see the x-Lorentz force in figure \ref{fig:dD_0p2_density-ion-electron_all_interfaces}, and there is significant background motion away from the interface, figure \ref{fig:dD_0p2_background_motion}. 
In the SRMI-Li large skin depth case, shown in figure 	\ref{fig:dD_0p2_density-ion-electron_all_interfaces}, we see a fluid evolution that resembles bubble flattening that is much more pronounced than for the other interface types, raising doubts that the mechanism for this is the same in the hydrodynamic case \citep{wouchuk2001growth,pandian2017effect,abarzhi2003rayleigh}.

\begin{figure}
	\centerline{\includegraphics[width=\linewidth]{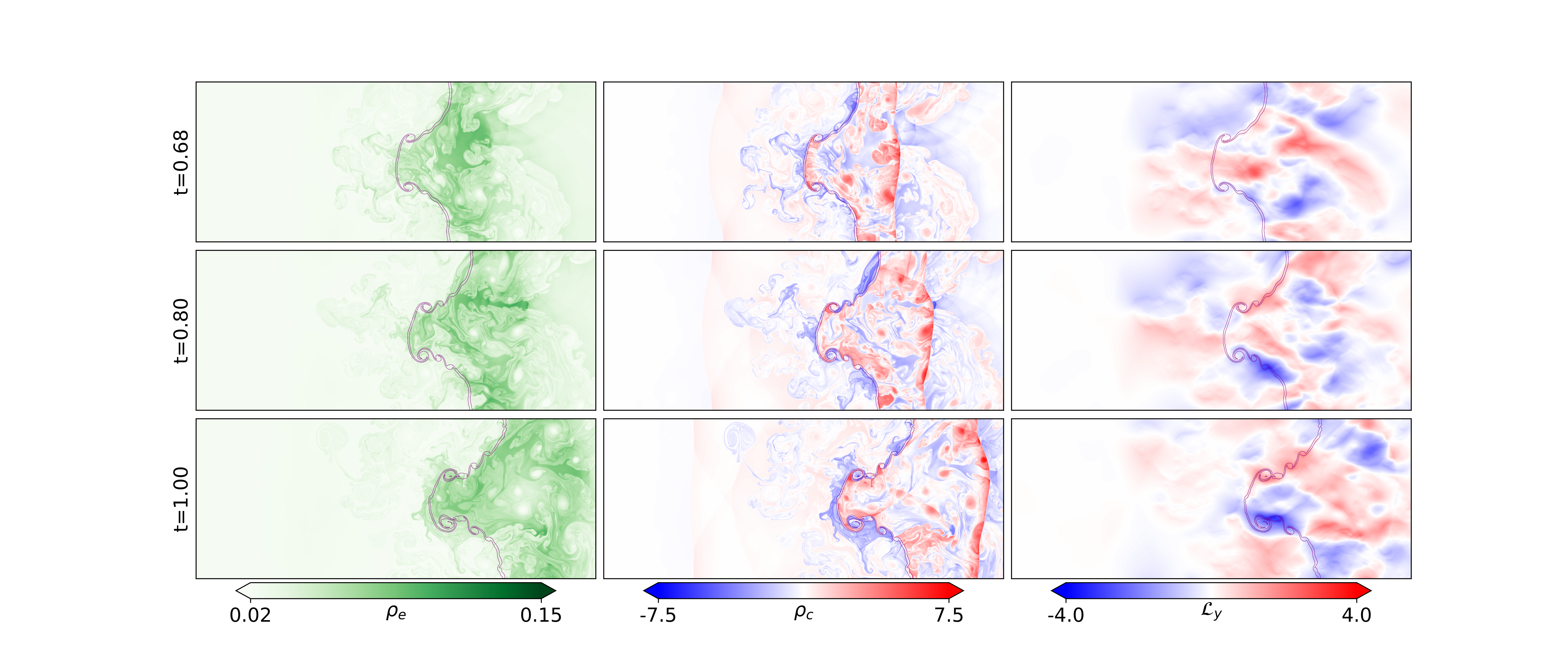}}
	\caption{The electron mass-density, charge-density, and ion y-Lorentz force TRMI large skin depth, $d_S = 10$, case. The ion interface contour is shown in purple. }
	\label{fig:TRMI_dD_0p2_bubbleFlattening}
\end{figure}

\begin{figure}
	\centerline{\includegraphics[width=\linewidth]{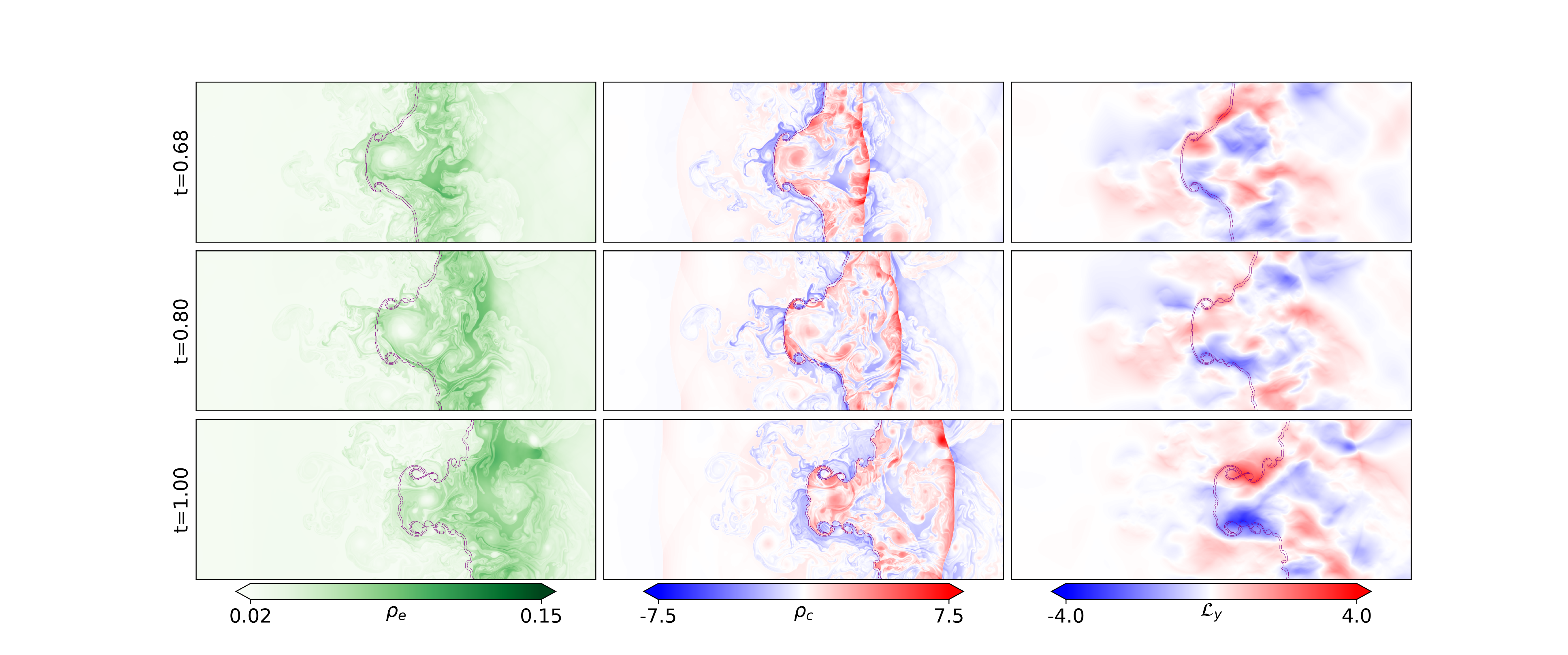}}
	\caption{The electron mass-density, charge-density, and ion y-Lorentz force SRMI-Li large skin depth, $d_S = 10$, case. The ion interface contour is shown in purple.}
	\label{fig:SRMI_dD_0p2_bubbleFlattening}
\end{figure}

The apparent bubble flattening in the SRMI-Li case is a result of the electromagnetic effects rather than the conventional RMI non-linear effect. 
We see from the hydrodynamic reference solutions (the $p=0.5$ being relevant due to the loose coupling), figure \ref{fig:ref_HRMI_MHD_solution}, and the isotope case, figure \ref{fig:dD_0p2_density-ion-electron_all_interfaces}, that the RMI is exhibiting early signs of bubble flattening. 
One might expect that the additional circulation generation from electromagnetic terms, in the plasma RMI with appreciable electron fluid-density interface, would drive the RMI to more quickly develop non-linear effects such as bubble flattening. 
However, the Lorentz forces within the plasma resist changes to the charge-distribution (such as that from a changing ion density-interface) thereby slowing the RMI evolution as seen in the TRMI and SRMI-He cases. 
The SRMI-Li case behaves differently, appearing to proceed with an evolution that resembles bubble flattening. 
The difference (consider a comparison of the TRMI and SRMI-Li cases) is the location of strongest vortex cores in the electron fluid, indicated by the low density spots in the mass-density contours of figures \ref{fig:TRMI_dD_0p2_bubbleFlattening} (TRMI) and \ref{fig:SRMI_dD_0p2_bubbleFlattening} (SRMI-Li). 
The strongest vortex core in the SRMI-Li and TRMI cases are closely behind the spike and far behind the spike, respectively. 
The location of these strongest vortices produces a significantly different electron fluid evolution and consequently a different electromagnetic field evolution and subsequent resisting Lorentz forces. 
Comparison of figures \ref{fig:TRMI_dD_0p2_bubbleFlattening} and \ref{fig:SRMI_dD_0p2_bubbleFlattening} (which have the same contour levels) at times before the bubble flattening illustrates these different evolutions. 
At $t=0.68$ the SRMI-Li case has large regions of ion y-Lorenz forcing toward the x-symmetry line (reinforcing flattening) in the bulk fluid adjacent to the bubble, whereas the the TRMI has fewer of these regions and more regions of ion y-Lorenz forcing away from the x-symmetry line. 
At $t=0.8$ as the bubble flattening is starting to become significant in the SRMI-Li case, the TRMI case has comparatively little change. At this point the SRMI-Li case shows weaker resisting y-Lorentz forces along the density-interface in comparison the TRMI case. 
These Lorentz forces in the SRMI-Li facilitate the flattening and diverge from the TRMI and SRMI-He cases. At later times, in response to the narrowing in the SRMI-Li case the Lorentz forces intensify to resist further narrowing of the region between the spike and bubble. 

There is potential for future work to investigate more closely the relation between the plasma RMI and the HRMI theory, particularly around the non-monotonic relationship between growth rate and initial amplitude. 
Direct quantitative comparison between simulation results of the plasma RMI to hydrodynamic theory would not be meaningful as the influence of electromagnetic effects has been shown to be substantial. 
A useful but challenging theoretical analysis would incorporate equations that can account for the coupling effect between the electron and ion fluids, for finite skin depth, that lead to variable generation and evolution of electromagnetic fields that drive the secondary ERTI. 
A possible approach is to model simplified (incompressible) sets of equation for each species, extract charge-density and current-density equations, and relate these to the electromagnetic fields, but this is beyond the scope of the present work.

\section{Discussion of dominant parameters}\label{sec:parameter_dicsussion}
Each interface type produces a unique evolution of the RMI that manifests as different interface amplitudes, growth rates, and imposed interface perturbations. These differences raises concern for the simplifications made in single-fluid models for specific plasma regimes. The key parameters influencing the evolution of the RMI under different interface types are those which significantly affect the severity of the RMI and secondary instabilities in regard to primary perturbation amplitude and growth. These key parameters are (1) the density ratio across the initial electron fluid DI and (2) the plasma Debye length. The ion charge may have a unique affect on the evolution of the RMI but this is not confirmed. The density ratio across the electron fluid DI and Debye length greatly influence the manifestation of the ERTI, TRS, and shock refraction process. All of these phenomena are uniquely MFP effects and are absent in the single-fluid case.

The influence of the EFDI density ratio is particularly observed through comparison of the isotope RMI and other scenarios. The IRMI simulations demonstrate a negligible interaction of the initial shock and the EFDI. Consequently the well-behaved electron fluid motion does not generate significant charge-separation and EM fields, hence it does not convert significant amounts of energy to the IFDI via the EM fields.
The electron fluid, in the case of low or zero EFDI density ratio, is unable to influence the evolution of the RMI and secondary instabilities. The electron fluid is unable to 1) perturb the IFDI prior to ion shock arrival and 2) generate significant EM fields that would drive Lorentz force oscillations and the secondary ERTI in the ion fluid. Without an initial EFDI, the single-fluid and IRMI cases, a distinctly single-fluid RMI evolution is observed. Conversely, the cases with significant EFDI density ratio, TRMI and SRMI cases, produce an energetic and erratic electron fluid, greater magnitude EM fields (EM energy density as well), and MFP effects that augment the ICF RMI. 

The expectation of approaching the magnetohydrodynamic limit with vanishingly small Debye length has been shown comprehensively in the literature \citep{chen2016introduction,bellan2008FundPlasma,shen2018magnetohydrodynamic} and is supported by the simulation results. Generally, decreasing only the Debye Length via skin depth results in significantly reduced amplitude and growth rates. The one exception is the SRMI-He $d_S=1$ where more expansive EM fields, relative to the $d_S=1$ SRMI-LI and TRMI, lead to a greater reinforcing of the RMI by the ERTI. However, reducing the Debye length further aligns the SRMI-He case with the single-fluid amplitude. There are many plasma regimes, both within ICF and other applications, that do contain a finite Debye length which requires the use of MFP modelling to capture relevant phenomena. 

Interestingly, all scenarios with an initial EFDI experience a maximum in interface perturbations at $d_S=1$. This suggests that excitation of the IFDI interface by the electron fluid relies on competing mechanisms. It seems that the scenario of $d_S=1$ may be close to a local maximum of multi-fluid effects, where the coupling of the ion and electron fluids is tight enough to allow significant excitation but the coupling is also flexible enough to allow for the necessary charge separation to produce strong enough fields to pump the electrons with energy to excite the ions. This effect is, however, diminished in the case of the $d_S=0.1$ scenarios where the excitation of the ion fluid by the electron fluid is visible in the high wavenumber perturbations of the interface, however, the ERTI is necessarily diminished by the reduced strength and expanse of EM fields. 

A comparison of the SRMI-Li and TRMI cases suggests the ion charge has a non-trivial effect on the interface evolution. Both cases have the same EFDI density ratio and despite having differing ion properties for S2 most fluid properties are identical. The only differing bulk property is the ion temperature --- 0.5 for SRMI-Li and 0.167 for the TRMI --- but shock propagation speeds are identical in the fluids due to the ratio of molar masses. The differing ion charge effect manifests as increased narrowing of the neck region and increased number of secondary perturbations along the bubble region of the ion interface in the loosely coupled case ($\lambda_D=2\times10^{-1}$). Some differences also manifest in the smaller Debye length cases but to a lesser extent. The SRMI-Li and TRMI show different shock refraction processes, which would explain the different effects seen as each can be attributed to the TRS interacting with the IFDI. 

\section{Conclusions}\label{sec:Conclusion}



The multi-fluid plasma (MFP) model produces a different RMI evolution for the isotope, species, and thermal density interface cases that are indistinguishable in the ideal MHD model and other single fluid models. MFP models are able to distinguish between ions, isotopes, and electrons allowing the modelling of charge separation and physical evolution of EM fields that is not possible with single-fluid models. The uniquely MFP effects that result affect the evolution and severity of the RMI in the interface cases studied which would otherwise be missed in the application of single-fluid models such as the ideal MHD model. The key phenomena influencing the interface scenarios are the 
\begin{enumerate}
	\item primary Richtmyer-Meshkov instability, 
	\item electromagnetically-driven Rayleigh-Taylor instability, 
	\item Kelvin-Helmholtz instability, 
	\item electron-fluid-excitation of the ion-fluid interface, 
	\item Lorentz force bulk fluid accelerations and vorticity deposition, 
	\item transverse-reflected shock-wave interaction with the ion-fluid interface, 
	\item a multi-fluid plasma shock refraction process.
\end{enumerate}
Points (i), (iii), and (vi) are observable in the hydrodynamic models of the RMI and MHD reductions of the plasma RMI, however, the remaining phenomena are beyond their grasp because of the required capacity to model self-generated EM fields and the evolution of electron fluid. All three interface types studied (isotope, species, and therm RMI) experience the phenomena above to varying degrees. The isotope case, lacking an initial DI in the electrons is the closest to the single-fluid limit. The TRMI and SRMI scenarios with their initial electron fluid mass density ratios experience significant multi-fluid effects and amplification of the RMI in the larger Debye length cases. As the Debye length is reduced (increasing coupling between the ion and electron fluids ) all three novel interface types tested experience a significantly reduced RMI growth rate and amplitude, approaching the single-fluid limit but retain multi-fluid phenomena. The results of the study show that MFP effects should be considered important when (i) Debye lengths are large, and (ii) distinct density interfaces are formed from isotope, species, and thermal discontinuities. 

In relation to ICF experiments, the phenomena discussed here present several points that pose a risk of degrading capsule implosion. As seen in the results, the species and thermal interfaces experience the worst instability growth (in comparison to hydrogen isotopes). A high atomic number species, such as carbon in the ICF fuel casing material, that is ionised during the implosion will produce a large density ratio in the electrons (when forming an interface with the ionised fuel). This could produce all the detrimental effects of a highly chaotic and energetic electron fluid RMI discussed in this paper. The shock refraction process at small Debye lengths results in significant interface perturbation. This increased surface perturbation is a concern for maintaining the carefully designed implosion geometry and coherence during implosions. The results represent only the early time evolution of the instability and under ideal conditions, introduction of transport processes is a vital next step in assessing the effect on ICF experiments.

\section*{Acknowledgements}
This research  was  supported  by  the  KAUST  Office  of  Sponsored  Research  under Award URF/1/2162-01. This work was supported by computational resources provided by  the  Australian  Government  under  the  National  Computational  Merit  Allocation Scheme.

\vspace{0.5cm}

Special thanks to Lachlan Whyborn for providing feedback on an early draft of this paper. 

\vspace{0.5cm}

Declaration of Interests. The authors report no conflict of interest.

\bibliographystyle{jfm}
\bibliography{RMI_interface_comparison}

\appendix

\end{document}